\documentclass{aa} 
\usepackage{graphicx}
\usepackage{txfonts}
\usepackage{subfig}
\usepackage{epstopdf}
\usepackage{lscape}

\usepackage[section] {placeins}
\usepackage{multirow}

\usepackage{natbib}

\usepackage{hyperref}
\hypersetup{
 colorlinks=true,
 citecolor=blue,
 linkcolor=magenta,
 urlcolor=black}

\bibpunct{(}{)}{;}{a}{}{,} 

\begin{document}

\newcommand {\methanol} {CH$_3$OH}
\newcommand {\cmethanol} {$^{13}$CH$_3$OH}
\newcommand {\acetonitrile} {CH$_3$CN}
\newcommand {\acetonitrilec} {CH$_3$$^{13}$CN}
\newcommand {\isocyanicacid} {HNCO}
\newcommand {\isocyanicacidc} {HN$^{13}$CO}
\newcommand {\methylformate} {HCOOCH$_3$}
\newcommand {\formaldehyde} {H$_2$CO}
\newcommand {\cformaldehyde} {H$_2$$^{13}$CO}
\newcommand {\oformaldehyde} {H$_2$C$^{18}$O}
\newcommand {\formamide} {NH$_2$CHO}
\newcommand {\methylether} {CH$_3$OCH$_3$}
\newcommand {\propionitrile} {C$_2$H$_5$CN}
\newcommand {\ethanol} {C$_2$H$_5$OH}
\newcommand {\ketene} {CH$_2$CO}
\newcommand {\acetaldehyde} {CH$_3$CHO}
\newcommand {\formicacid} {HCOOH}
\newcommand {\methylacetylene} {CH$_3$CCH}
\newcommand {\acetone} {CH$_3$COCH$_3$}
\newcommand {\Tex} {$T_\mathrm{ex}$}
\newcommand {\Trot} {$T_\mathrm{rot}$}
\newcommand {\Eup} {$E_\mathrm{up}$}

\authorrunning{Isokoski et al.}
\titlerunning{m09bn10}
\title{Chemistry of massive young stellar objects with a disk-like structure}


\author{K. Isokoski\inst{1}
\and
S. Bottinelli\inst{2,3}
\and
E.~F. van Dishoeck\inst{4,5}
}
\offprints{K. Isokoski;  \email{isokoski@strw.leidenuniv.nl}}

\institute{$^1$Raymond and Beverly Sackler Laboratory for Astrophysics, Leiden Observatory, Leiden University, PO Box 9513, 2300 RA Leiden, The Netherlands\\
$^2$ Universit\'{e} de Toulouse, UPS-OMP, IRAP, Toulouse, France \\
$^3$ CNRS, IRAP, 9 Av. colonel Roche, BP 44346, F-31028 Toulouse cedex 4, France \\
$^4$ Leiden Observatory, Leiden University, P.O. Box 9513, 2300 RA Leiden, The Netherlands\\
$^5$ Max-Planck Institut f\"ur Extraterrestrische Physik (MPE), Giessenbachstr. 1, 85748 Garching, Germany \\
}



\abstract
{}
{Our goal is to take an inventory of complex molecules in three well-known high-mass protostars for which disks or toroids have been claimed and to study the similarities and differences with a sample of massive YSOs without evidence of such flattened disk-like structures. With a disk-like geometry, UV radiation can escape more readily and potentially affect the ice and gas chemistry on hot-core scales.}
{ A partial submillimeter line survey, targeting \methanol, \formaldehyde, \ethanol, \methylformate, \methylether, \acetonitrile, \isocyanicacid, \formamide, \propionitrile, \ketene, \formicacid, \acetaldehyde, and \methylacetylene, was made toward three massive YSOs with disk-like structures, IRAS20126+4104, IRAS18089-1732, and G31.41+0.31. Rotation temperatures and column densities were determined by the rotation diagram method, as well as by independent spectral modeling. The molecular abundances were compared with previous observations of massive YSOs without evidence of any disk structure, targeting the same molecules with the same settings and using the same analysis method.}
{Consistent with previous studies, different complex organic species have different characteristic rotation temperatures and can be classified either as warm ($>$100~K) or cold ($<$100~K). The excitation temperatures and abundance ratios are similar from source to source and no significant difference can be established between the two source types. Acetone, CH$_3$COCH$_3$, is detected for the first time in G31.41+0.31 and IRAS18089-1732. Temperatures and abundances derived from the two analysis methods generally agree within factors of a few.}
{The lack of chemical differentiation between massive YSOs with and without observed disks suggest either that the chemical complexity is already fully established in the ices in the cold prestellar phase or that the material experiences similar physical conditions and UV exposure through outflow cavities during the short embedded lifetime.}


\maketitle

\section{Introduction}

Millimeter lines from complex organic molecules are widely associated with high-mass star forming regions and indeed form one of the signposts of the deeply embedded phase of star formation \citep[e.g.,][]{Blake1987, Hatchell1998, Gibb2000, Fontani2007,RequenaTorres2008, Belloche2009, Zernickel2012}.  Many studies of the chemistry in such regions have been carried out, either through complete spectral surveys of individual sources or by targetting individual molecules in a larger number of sources \citep[see][for reviews]{Herbst2009, Caselli2012}. In spite of all this work, only few systematic studies of the abundances of commonly observed complex molecules have been performed across a sample of massive YSOs, to search for similarities or differences depending on physical structure and evolutionary state of the object. Intercomparison of published data sets is often complicated by the use of different telescopes with different beams, different frequency ranges and different analysis techniques.

Chemical abundances depend on the physical structure of the source such as temperature, density and their evolution with time, as well as the amount of UV radiation impinging on the gas and dust. In contrast with the case of solar-mass stars, the physical structures and mechanisms for forming massive ($M > 8$ M$_{\odot}$) stars are still poorly understood.  Indeed, theoretically, the powerful UV radiation pressure from a high-mass protostellar object (HMPO) should prevent further accretion and so inhibit the formation of more massive stars \citep{Zinnecker2007}. However, a number of recent studies have claimed the presence of disk- or toroid-like `equatorial' structures surrounding a handful of high-mass protostars \citep{Cesaroni2007}. These data support theories in which high-mass star formation proceeds in a similar way as that of low-mass stars via a disk accretion phase in which high accretion rates and non spherically symmetric structures overcome the problem of radiation pressure. The best evidence so far is that for a $\sim$5000 AU disk in Keplerian rotation around IRAS20126+4104, claimed by \citet{Cesaroni1999} on the basis of the presence of a velocity gradient in the CH$_3$CN emission perpendicular to the direction of the outflow, as predicted by the disk-accretion paradigm. Surprisingly, even in the best case of IRAS20126+4104, the detailed chemistry of these (potential) disks has not yet been studied.

Most chemical models invoke grain surface chemistry to create different generations of complex organic molecules \citep{Tielens1997}. Hydrogenation of solid O, C, N and CO during the cold ($T_{\rm d}<20$~K) prestellar phase leads to ample production of CH$_3$OH and other hydrogenated species like H$_2$O, CH$_4$ and NH$_3$ \citep{Tielens1982}.  Exposure to UV radiation results in photodissociation of these simple ices, with the fragments becoming mobile as the cloud core heats up during the protostellar phase. First generation complex molecules result from the subsequent recombination of the photofragments, and will eventually evaporate once the grain temperature rises above the ice sublimation temperature of $\sim$100~K \citep{Garrod2006, Garrod2008}. Good examples are \ethanol, \methylformate~and \methylether~resulting from mild UV processing of CH$_3$OH ice \citep{Oberg2009b}. Finally, a hot core gas phase chemistry between evaporated molecules can drive further complexity in second generation species \citep[e.g.,][]{Millar1991, Charnley1992, Charnley1995}. 

\begin{figure}
   \centering
   \includegraphics[width=0.45\textwidth]{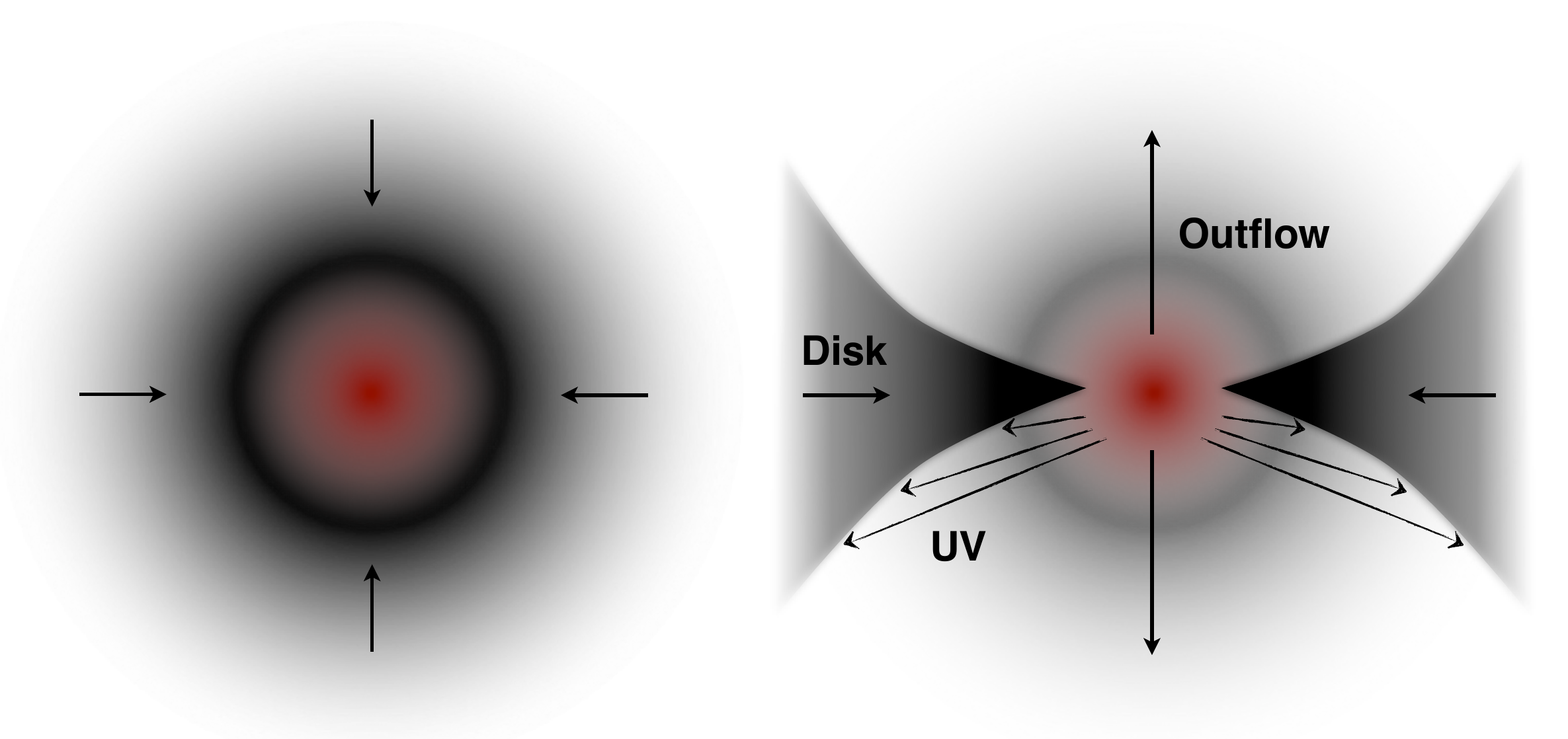}
      \caption{Illustration of a protostar with a spherical structure (left) and a protostar with a flattened disk-like structure (right) with enhanced UV photons illuminating the walls of the outflow cavity.}
         \label{fig:structure}
   \end{figure}

One of the most obvious consequences of an equatorial rather than spherical structure is that UV radiation can more easily escape the central source and illuminate the surface layers of the surrounding disk or toroid as well as the larger scale envelope \citep{Bruderer2009, Bruderer2010} (Fig.~\ref{fig:structure}). This can trigger enhanced formation of complex organic molecules in the ices relative to methanol. Another effect of UV radiation is that it drives increased photodissociation of gaseous N$_2$ and CO. The resulting atomic N and C would then be available for grain surface chemistry potentially leading to enhanced abundances of species like HNCO and NH$_2$CHO.

To investigate the effects of disk-like structures on the chemistry, we present here a single-dish survey using the James Clerk Maxwell Telescope of three HMPOs for which large equatorial structures (size$>2000$ AU) have been inferred, namely IRAS20126+4104, IRAS18089--1732, and G31.41+0.31. The results are compared with those of a recent survey of a sample of HMPOs by \citet{Bisschop2007} (hereafter BIS07), targeting many of the same molecules and settings. The use of the same telescope and analysis method allows a meaningful comparison between the two samples of sources. BIS07 found that the O-rich complex molecules are closely correlated with the grain surface product CH$_3$OH supporting the above general chemical scenario.  N-rich organic molecules do not appear to be correlated with O-rich ones, but overall, the relative abundances of the various species are found to be remarkably constant within one chemical family.  One of the main questions is whether this similarity in abundance ratios also holds for sources with disk-like structures. Although our observations do not spatially resolve these structures, they are sensitive enough and span a large enough energy range to determine abundances on scales of $\sim$1$''$ and thereby set the scene for future high-angular resolution observations with interferometers like the Atacama Large Millimeter/submillimeter Array (ALMA). Moreover, current interferometers resolve out part of the emission, which is why single-dish observations are still meaningful.

This paper is organized as follows. In \hyperref[sect:observations]{\S 2}, the observed sources and frequency settings are introduced and the details of the observations are presented. \hyperref[sect:data_analysis]{\S 3} focuses on the data analysis methods. Specifically, two techniques are used to determine excitation temperatures and column densities: the rotation diagram method employed by BIS07 and spectral modeling tools in which the observed spectra are simulated directly.  \hyperref[sect:results]{\S 4} presents the results from both analysis methods and discusses their advantages and disadvantages. Finally,  \hyperref[sect:discussion]{\S 5} compares our results with those of BIS07 to draw conclusions on similarities and differences in chemical abundances between sources with and without large equatorial structures.

\section{Observations}
\label{sect:observations}

\subsection{Observed sources}

Table \ref{tab:sources} gives the coordinates, luminosity $L$, distance $d$, galactocentric radius $R_\mathrm{GC}$, velocity of the local standard of rest $V_\mathrm{LSR}$ and the typical line width for the observed sources. The selected sources are massive young stellar objects (YSOs), for which strong evidence exists for a circumstellar disk structure. All sources are expected to harbor a \textit{hot core}: a compact, dense ($\ge$10$^{7}$ cm$^{-3}$) and warm ($\ge$100~K) region with complex chemistry triggered by the grain mantle evaporation (Kurtz et al. \cite{Kurtz2000}). \acetonitrile~emission from $\ge$100~K gas is present in all sources. Moreover, CH$_3$OH 7$_K$-6$_K$ transitions (338.5 GHz) with main beam temperatures of $\ge$1~K have been observed for these sources. Sources also needed to be visible from the James Clerk Maxwell Telescope (JCMT) \footnote{The James Clerk Maxwell   Telescope is operated by the Joint Astronomy Centre, on behalf of   the Particle Physics and Astronomy Research council of the United   Kindom, the Netherlands Organization for Scientific Research and the   National Research Council of Canada. The project ID is m09bn10.}.

\begin{table*}
\begin{center}
\caption{Coordinates, luminosity, distance, galactocentric radius, velocity of the local standard of rest, $^{12}$C/$^{13}$C isotope ratio and the typical line width for the observed sources.}
\label{tab:sources}
\begin{tabular}{l | c c c c c c c c}
\hline \hline
Sources  & $\alpha$(2000) & $\delta$(2000) &  $L$ & $d$ & $R_\mathrm{GC}$$^*$ & $V_\mathrm{LSR}$ & $^{12}$C/$^{13}$C & $\Delta V$ \\
 & & & [10$^5$ L$_{\odot}$] & [kpc] & [kpc] & [km s$^{-1}$] & &  [km s$^{-1}$] \\
\hline
IRAS20126+4104 & 20:14:26.04 & +41:13:32.5 & 0.13$^a$ & 1.64$^b$ & 8.3 & -3.5 & 70  &  6$^c$\\
IRAS18089-1732 & 18:11:51.40 & -17:31:28.5 & 0.32$^d$ & 2.34$^e$ & 6.2 & 33.8 & 54  &  5$^c$ \\
G31.41+0.31 & 18:47:34.33 & -01:12:46.5 & 2.6$^f$ & 7.9$^g$ & 4.5 & 97.0 & 41  & 6-10$^h$ \\
\hline
\end{tabular}
\end{center}
(a)~\citet{Cesaroni1997}, (b)~\citet{Moscadelli2011}, (c)~\citet{Leurini2007}, (d)~\citet{Sridharan2002}, (e)~\citet{Xu2011}, (f)~\citet{Cesaroni1994b}, (g)~\citet{Churchwell1990}, (h)~\citet{Fontani2007}. \\
$^*$ The galactocentric radii were calculated using distances $d$ in this table and a IAU recommended distance from the galactic center $R_0$ = 8.5 kpc. The $^{12}$C/$^{13}$C isotope ratios are calculated from Eq. \ref{eq:isotope_ratio}.
\end{table*}

\subsubsection{IRAS20126+4104}

IRAS20126+4104 (hereafter IRAS20126) is a luminous ($\sim$10$^4$ L$_\odot$) YSO located relatively nearby at a distance of 1.64 kpc \citep{Moscadelli2011}. It was first identified in the IRAS point source catalog by the IR colours typical of ultracompact HII regions and by H$_2$O maser emission characteristic of high-mass star formation \citep{Comoretto1990}. IRAS20126 features a $\sim$0.25-pc scale inner jet traced by H$_2$O maser spots in the SE-NW direction with decreasing velocity gradient \citep{Tofani1995}. Source and masers are embedded inside a dense, parsec-scale molecular clump \citep{Estalella1993, Cesaroni1999}. The inner jet feeds into a larger scale bipolar outflow with the two having reversed velocity lobes \citep{Wilking1990, Cesaroni1999}. The reversal is likely to be due to precession of the inner jet caused by a companion separated by a distance of $\sim$0.5$''$ (850 AU) \citep{Hofner1999, Shepherd2000, Cesaroni2005, Sridharan2005}.  
A rotating, flattened, Keplerian disk structure has been detected perpendicular to the inner jet. Observations of \acetonitrile~transitions show a Keplerian circumstellar disk (radius $\sim$1000~AU) with a velocity gradient perpendicular to the jet and a hot core with a diameter of $\sim$0.0082~pc and a temperature of $\sim$200~K at a geometric center of the outflow \citep{Cesaroni1997, Zhang1998, Cesaroni1999}. Direct near-infrared (NIR) observations show a disk structure as a dark line \citep{Sridharan2005}. The disk shows a temperature and density gradient and is going through infall of material with a rate of $\sim$2$~\times~10^{-3}$ M$_\odot~\mathrm{yr}^{-1}$ as expected for a protostar of this mass and luminosity \citep{Cesaroni2005}. Recent modeling of the Spectral Energy Distribution (SED) of IRAS20126 shows indeed a better fit when a disk is included \citep{Johnston2011}.

\subsubsection{IRAS18089-1732}

IRAS18089-1732 (hereafter IRAS18089) is a luminous ($\sim$10$^{4.5}$~L$_\odot$, \citealt{Sridharan2002}) YSO located at a distance of 2.34 kpc \citep{Xu2011}. It was identified based on CS detections of bright IRAS point sources with colours similar to ultracompact HII regions and the absence of significant free-free emission \citep{Sridharan2002}, and with H$_2$O and \methanol~maser emission \citep{Beuther2002}. The CO line profile shows a wing structure characteristic of an outflow, although no clear outflow structure could be resolved from the CO maps \citep{Beuther2002}. A collimated outflow in the Northern direction is however seen in SiO emission on scales of 5$''$ \citep{Beuther2004}.  IRAS18089 has a highly non-circular dust core of $\sim$3000~AU diameter ($\sim$1$''$), with optically thick \acetonitrile~at $\sim$350~K \citep{Beuther2005}. \methylformate~was found to be optically thin, with emission confined to the core, and showing a velocity gradient perpendicular to the outflow indicative of a rotating disk \citep{Beuther2004}. Also hot NH$_3$ shows a velocity gradient perpendicular to the outflow, although no Keplerian rotation was found, possibly due to infall and/or self gravitation \citep{Beuther2008}. Several hot-core molecules (\methylformate, \acetonitrile, \methylether, \isocyanicacid, \formamide, \methanol, \ethanol) were mapped by \citet{Beuther2005} but no column densities or abundances were reported.

\subsubsection{G31.41+0.31}

G31.41+0.31 (hereafter G31) is a luminous (2.6 $\times$ 10$^5$~L$_\odot$, \citealt{Cesaroni1994b}) YSO at a distance of 7.9~kpc \citep{Churchwell1990}. Preliminary evidence for a rotating disk with a perpendicular bipolar outflow was reported by \citet{Cesaroni1994a, Cesaroni1994b} and \citet{Olmi1996} showing a velocity gradient across the core in the NE-SW direction, similar to previously detected OH masers \citep{Gaume1987}. High-angular resolution \acetonitrile~observations by \citet{Beltran2005} could not detect Keplerian rotation typical for less luminous stars, nevertheless a toroidal structure undergoing gravitational collapse and fast accretion ($\sim$3 $\times$ 10$^{-2}$ M$_\odot$ yr$^{-1}$) onto the central object was found. The G31 hot molecular core (HMC) is part of a complex region in which multiple stellar sources are detected \citep{Benjamin2003}; indeed, it is separated from an ultracompact (UC) HII region by only $\sim$5$''$ and overlaps with a diffuse halo of free-free emission, possibly associated with the UC HII region itself \citep{Cesaroni1998}. More recent interferometric observations confirm the velocity gradient in the NH$_3$ (4,4) inversion transition and in \acetonitrile~data \citep{Cesaroni2010, Cesaroni2011}. Line profiles look like a rotating toroid with infall motion. Several complex hot-core molecules have been observed in G31, including glycolaldehyde CH$_2$OHCHO \citep{Beltran2009}, but again no abundance ratios have been presented.

\subsection{Observational details}

The observations were performed at the JCMT on Mauna Kea, Hawaii, between August 2007 and September 2009. The observations of the 338~GHz region covering \methanol~($7_K \rightarrow 6_K$) transitions were taken from JCMT archive. The front ends consisted of the facility receivers A3 (230~GHz region) and HARP-B (340~GHz region). The back-end was the digital autocorrelation spectrometer (ACSIS), covering 400 and 250 MHz of instantaneous bandwidth for A3 and HARP-B, respectively, with a channel width of 50 kHz. The noise level for both receivers was $T_{\rm rms}$$\sim$20~mK on a $T_{\rm A}^*$ scale when binned to 0.5 km s$^{-1}$. The integration time was $\sim$1 hr and 1.8 hr for A3 and HARP-B, respectively. The spectra were scaled from the observed antenna temperature, $T_{\rm A}^*$, to main-beam temperature, $T_{\rm MB}$, using main beam efficiencies $\eta_{\rm MB}$ of 0.69 and 0.63 at 230~GHz and 345~GHz, respectively. We adopt a $T_{\rm A}^*$ calibration error of 20$\%$.

The HPBW (half-power beam width, $\theta_{\rm{B}}$) for the 230 and 345~GHz band observations are 20-21$''$ and 14$''$, respectively. Emission from a volume with a source diameter $\theta_{\rm S}$ $\le$$\theta_{\rm{B}}$ undergoes beam dilution described by the beam-filling factor, $\eta_{\rm BF}$:

\begin{equation}
\label{eq:beamfillingfactor}
\eta_{\rm BF}=\frac{\theta_{\rm S}^2}{\theta_{\rm S}^2+\theta_{\rm B}^2}.
\end{equation}

Table~\ref{tab:frequencies} gives the observed frequency settings and the targeted molecular lines. The settings were taken from BIS07 and were chosen to cover at least one strong line for the target molecules as well as lines of other interesting species. Strong lines of target molecules were chosen due to their high main-beam temperatures and minimum line-confusion in line surveys of Orion-KL by \citet{sutton1985} and \citet{schilke1997} at 230 GHz and 345 GHz, respectively. In order to determine rotation temperatures, we covered at least two transitions for a given species with \Eup$<$100~K and \Eup$>$100~K each. BIS07 used the single pixel receiver B3, the predecessor of HARP-B, together with DAS (Digital Autocorrelation Spectrometer) as the back-end, covering a larger instantaneous bandwidth of 500 MHz. Only the central receptor of HARP-B array is analysed here as no significant off source emission was detected in the complex molecules.

\begin{table*}
\begin{center}
\caption{Observed frequency settings and molecular lines.}
\label{tab:frequencies}
\begin{tabular}{l l l l l l}
\hline \hline
Molecule  & Freq. & $E_{\rm up}$ & $\mu^2S$ & Transition & Additional molecules \\
 & [GHz] & [K] & [D$^2$] & & \\
\hline
\acetonitrile & 239.1195 & 144.77 & 811.86 & 13$_K$ -- 12$_K$ &  \acetonitrilec, \methylformate, \methylether \\
\isocyanicacid & 219.7983 & 58.02 & 28.112 & 10$_{0, 10, 11}$ -- 9$_{0, 9, 10}$ & \cformaldehyde, \propionitrile \\
 & 240.8809 & 112.53 & 30.431 & 11$_{1, 11, 12}$ -- 10$_{1, 10, 11}$ & \methylether, \methanol, \isocyanicacidc \\
\methylformate & 222.3453 & 37.89 & 42.100 & 8$_{5, 4}$ -- 7$_{4, 3}$ & \methylether, \formamide \\
 & 225.2568 & 125.50 & 33.070 & 18$_{6, 12}$ -- 17$_{6, 12}$ & \formaldehyde, \methylether, \cmethanol \\
\formaldehyde & 364.2752 & 158.42 & 52.165 & 5$_{3, 3}$ -- 4$_{3, 2}$ & \ethanol \\
\acetonitrile & 331.0716 & 151.11 & 513.924 & 18$_{K}$ -- 17$_{K}$ & \methylformate, \isocyanicacid, \acetonitrilec \\
\isocyanicacid & 352.8979 & 187.25 & 43.387 & 16$_{1, 15, 17}$ -- 15$_{1, 14, 16}$ & \propionitrile, \ethanol, \methylformate \\
\formamide & 345.1826 & 151.59 & 664.219 & 17$_{0, 17}$ -- 16$_{0, 16}$ & \methylformate, \ethanol, \cmethanol \\
\methylformate & 354.6084 & 293.39 & 87.321 & 33$_{2, 32}$ -- 32$_{2, 31}$ & \formamide, \propionitrile \\
\hline
\end{tabular}
\end{center}
\end{table*}

Data reduction and line fitting were done using the CLASS software package \footnote{CLASS is part of the GILDAS software package developed by IRAM.}.  Line assignments were done by comparison of observed frequencies corrected for source velocity with the JPL\footnote{\url{http://spec.jpl.nasa.gov/ftp/pub/catalog/catform.html}}, CDMS\footnote{\url{http://www.ph1.uni-koeln.de/vorhersagen}} and NIST\footnote{\url{http://physics.nist.gov/PhysRefData/Micro/Html/contents.html}} catalogs \citep{Pickett1998, Muller2005}. The line assignment/detection was based on Gaussian fitting with the following criteria: (i) the fitted line position had to be within $\pm$1~MHz of the catalog frequency, (ii) the FWHM had to be consistent with those given in Table~\ref{tab:sources} and (iii) a $S/N>$3 is required on the peak intensity.  Sect.~A.1 in the Appendix describes in more detail the error analysis. In general, our errors on the integrated intensities are conservative and suggest a lower $S/N$ than that on the peak intensity or obtained using more traditional error estimates.

\section{Data analysis}
\label{sect:data_analysis}

\subsection{Rotation diagrams}

Rotation temperatures and column densities were obtained through the rotation diagram (RTD) method \citep{Goldsmith1999}, when 3 or more lines are detected over a sufficiently large energy range. Integrated main-beam temperatures, $\int T_{\rm MB}dV$, can be related to the column density in the upper energy level by:

\begin{equation}
\label{eq:rotationdiagram1}
\frac{N_{\rm up}}{g_{\rm up}}=\frac{3k \int T_{\rm MB}dV}{8\pi^3\nu\mu^2S},
\end{equation}
where $N_{\rm up}$ is the column density in the upper level, $g_{\rm   up}$ is the degeneracy of the upper level, $k$ is Boltzmann's constant, $\nu$ is the transition frequency, $\mu$ is the dipole moment and $S$ is the line strength. The total \textit{beam-averaged} column density $N_{\rm T}$ in cm$^{-2}$ can then be computed from:

\begin{equation}
\label{eq:rotationdiagram2}
\frac{N_{\rm up}}{g_{\rm up}}=\frac{N_{\rm T}}{Q\left(T_{\rm rot}\right)} e^{-E_{\rm up}/T_{\rm rot}},
\end{equation}
where $Q(T_{\rm rot})$ is the rotational partition function, and $E_{\rm up}$ is the upper level energy in K.

Blended transitions of a given species with similar \Eup~($\Delta$\Eup$<$30~K) were assigned intensities according to their Einstein coefficients ($A$) and upper level degeneracies ($g_\mathrm{up}$):

\begin{equation}
\label{eq:flux_assignment}
\int T_{\mathrm {MB}}dV (i) = \int T_{\mathrm {MB}}dV \times \frac{A^i g_\mathrm{up}^i}{\sum_{i} A^i g_\mathrm{up}^i},
\end{equation}
where the summation is over all the contributing transitions.
Blended transitions with different \Eup, or with contamination from transitions belonging to other species, were excluded from the RTD fit.

The beam averaged column density, $N_{\rm T}$, is converted to the \textit{source-averaged} column density $N_{\rm S}$ using the beam-filling factor $\eta_{\rm BF}$:
\begin{equation}
\label{eq:N_S}
N_{\rm S}=\frac{N_{\rm T}}{\eta_{\rm BF}}.
\end{equation}

The standard RTD method assumes that the lines are optically thin. Lines with strong optical depth, determined from the arguments in \S 4.1 as well as the models discussed in \S 3.2, were excluded from the fit. For \methanol, also low-\Eup~lines arising from a cold extended component (see Sect. \ref{sect:methanol}) were excluded.

Differential beam dilution is taken into account by multiplying the line intensities in the 230~GHz range by $\eta_{\rm BF}$(340 GHz)$/\eta_{\rm BF}$(230 GHz). For warm and cold molecules, beam dilution is derived assuming source diameters $\theta_\mathrm {T=100 K}$ (see Eq.~\ref{eq:sourceradii} below) and 14'', respectively (see Eq.~\ref{eq:beamfillingfactor}). All emission is thus assumed to be contained within the smallest beam size. Same approach was used in BIS07.

The vibrational partition function was ignored assuming that all emission arises from the ground vibrational state so that the vibrational partition function can be set to unity. This approximation can cause an underestimation of the derived column densities for larger molecules, even at temperatures of 100--200~K. Indeed, \citet{widicus2005} show that the error can be up to a factor of 2 for temperatures up to 300~K. Since this approximation affects all complex molecules (albeit at different levels), the overall error in abundance ratios will be less than a factor of 2 and well within the other uncertainties.

\subsection{Spectral modeling}
\label{modeling}

The alternative method for analysing the emission is to model the observed spectra directly. For this purpose, we used the so-called `Weeds' extension of the CLASS software package \footnote{CLASS is part of the GILDAS software package developed by IRAM.}, developed to facilitate the analysis of large millimeter and submillimeter spectral surveys \citep{Maret2011}. In this model, the excitation of a species is assumed to be in LTE (Local Thermodynamic Equilibrium) at a temperature \Tex. The brightness temperature, $T_\mathrm{B}$, of a given species as a function of the rest frequency $\nu$ is then given by:

\begin{equation}
\label{eq:brightnesstemperature}
T_{\rm B}\left(\nu\right)=\eta_{\rm BF}\left[J_{\nu}\left(T_{\rm ex}\right)-J_{\nu}\left(T_{\rm bg}\right)\right]\left(1-e^{-\tau\left(\nu\right)}\right),
\end{equation}
where $\eta_{\rm BF}$ is the beam filling factor (see Eq.~\ref{eq:beamfillingfactor}), $J_\nu$ is the radiation field such that:
\[
J_\nu(T) = \frac{h\nu/k}{{\rm e}^{h\nu/kT}-1}
\]
and $T_{\rm bg}$ is the temperature of the background emission. The HPBW $\theta_{\rm   B}$ is calculated within the Weeds model as $\theta_{\rm B}=1.22{c}/{\nu D}$, where $c$ is the speed of light and $D$ is the diameter of the telescope \footnote{For JCMT with 15-m antenna diameter, the equation gives a $\theta_{\rm B}$ of 21.9$''$ and 14.8$''$ for the 230 GHz band and 345 GHz bands, respectively.}. The opacity $\tau(\nu)$ is:

\begin{equation}
\label{eq:opacity}
\tau\left(\nu\right)=\frac{c^2}{8\pi\nu^2}\frac{N_\mathrm{T}}{Q(T_{\rm ex})}\sum_{i} A^{i}g_{\rm up}^{i}e^{-E_{\rm up}^{i}/kT_{\rm ex}}\left(e^{h\nu_0^{i}/kT_{\rm ex}}-1\right)\phi^{i}
\end{equation}
where the summation is over each line $i$ of the considered species.  $\nu_0^{i}$ is the rest frequency of the line and $\phi^{i}$ is the profile function of the line, given by:

\begin{equation}
\label{eq:profilefunction}
\phi^{i}=\frac{1}{\Delta \nu \sqrt{2\pi}}e^{-\left(\nu-\nu_0^{i}\right)^2/2\Delta \nu^2},
\end{equation}
where $\Delta \nu$ is the line width in frequency units at $1/e$ height. $\Delta \nu$ can be expressed as a function of the line FWHM in velocity units $\Delta V$ by:

\begin{equation}
\label{eq:linewidth}
\Delta \nu=\frac{\nu_0^{i}}{c\sqrt{8\ln2}}\Delta V.
\end{equation}

The input parameters in the model are the column density
$N_\mathrm{S}$ in cm$^{-2}$, excitation temperature \Tex~in K, the
source diameter $\theta_{\rm S}$ in arcseconds, offset velocity from
the source LSR (Local Standard of Rest) in km s$^{-1}$ and the line
FWHM in km s$^{-1}$. All parameters excluding the source diameter are
optimized manually to obtain the best agreement with the observed
spectra. The source diameter for emission from cold species is allowed to
vary over a larger area; generally 14$''$ is used. The emission from
hot-core molecules targeted in our work is assumed to originate from
the central region with $T_\mathrm{dust}\ge$100~K. The source diameter
for the warm emission is calculated using a relation derived from dust
modeling of a large range of sources (BIS07):

\begin{equation}
\label{eq:sourceradii}
R_{\rm T=100~K}\approx 2.3\times10^{14}\left(\sqrt{\frac{L}{L_\odot}}\right) \ {\rm cm},
\end{equation}
where $L/L_\odot$ is the luminosity of the source relative to the solar luminosity. Table \ref{tab:sourceradii} gives the calculated $R_{\rm T=100~K}$ radii and diameters for the observed sources. 

\begin{table}
\begin{center}
\caption{Source radii and angular diameters for $T$=100~K.}
\label{tab:sourceradii}
\begin{tabular}{l | c c}
\hline \hline
Sources  & $R_{\rm T=100 K}$ & $\theta_{\rm S, T=100 K}$ \\
 & [AU] & [$''$] \\
\hline
IRAS20126+4104 & 1753 & 2.2 \\
IRAS18089-1732 & 2750 & 2.4 \\
G31.41+0.31 & 7840 & 2.0 \\
\hline
\end{tabular}
\end{center}
\end{table}

In the analysis for individual molecules, the initial values for \Tex~were based on the \Trot~from the RTD analysis in case of optically thin species. For optically thick species the value of \Tex~from the $^{13}$C-isotope was used. If \Tex~for the isotopologue could not be obtained, the initial temperature was guessed. The \Tex~value was then optimized visually based on the relative intensity of the emission lines. The simulated emission was not allowed to exceed the emission of optically thin lines in the spectrum in any of the observed spectral ranges.  Coinciding and blended transitions, which together contribute to an optically thick line, are excluded in the analysis.

Due to the visual optimization and the possibility of overlapping lines (particularly in the line-rich source G31), the resulting \Tex~values are only a rough estimate ($\pm$50-100~K) and do not
differ significantly from those from the RTD method. The column density for each species was constrained by optically thin, unblended lines, where available.

For the specific case of CH$_3$OH, which requires two temperature components for a proper fit, the analysis was also done with the
CASSIS analysis package\footnote{CASSIS has been developed by IRAP-UPS/CNRS (\url{http://cassis.irap.omp.eu}).}. CASSIS has the advantage that it can properly model the emission from overlapping optically thick lines, as well as from nested regions of emission. For \methanol, warm emission from the compact inner region may be absorbed by the surrounding colder gas, which can influence the derived hot core abundances.

\section{Results}
\label{sect:results}

\subsection{General results and comparison between sources}

\begin{figure*}
\centering
\includegraphics[width=0.95\textwidth]{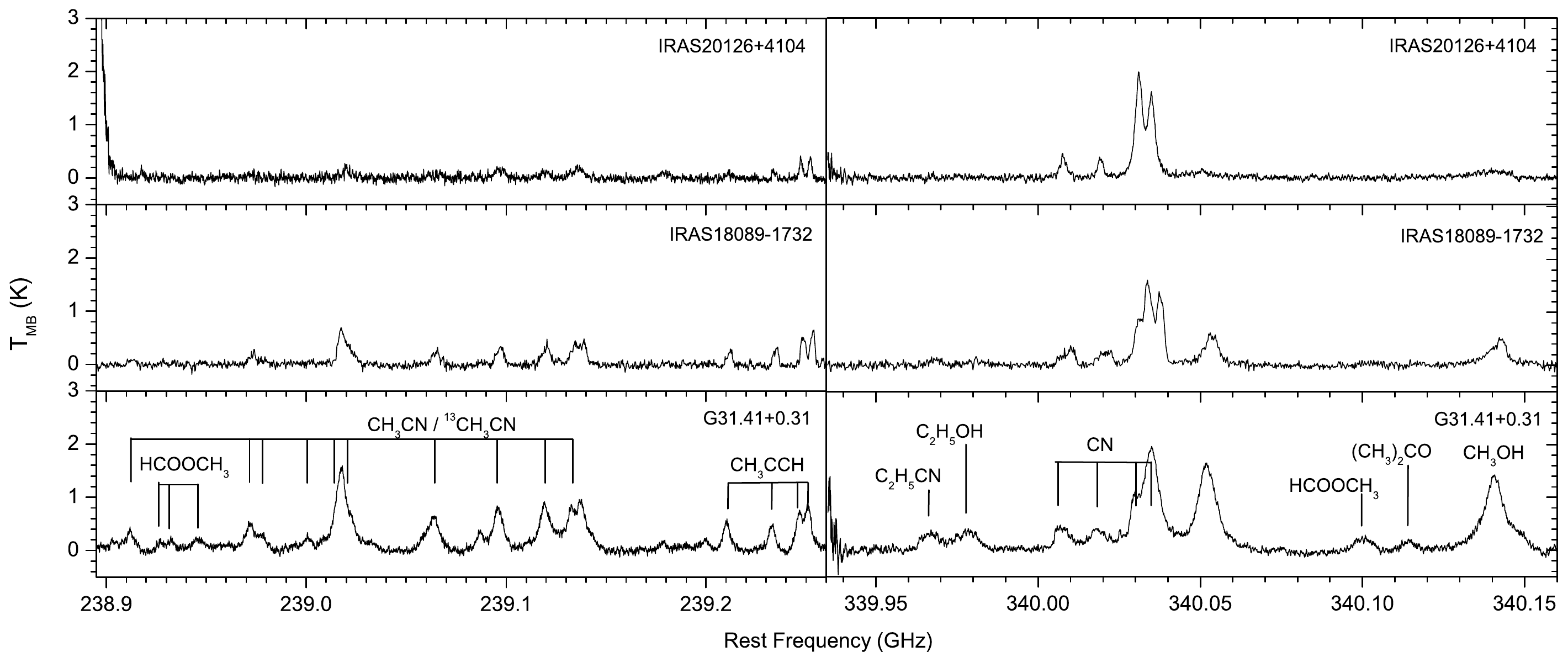}
\caption{Spectral ranges 238.83--239.26 GHz and 339.94--340.18 GHz with lines from several targeted species for the three sources.}
\label{fig:ranges}
\end{figure*}

The observed sources, IRAS20126, IRAS18089 and G31, differ from each
other significantly in the observed chemical richness in the JCMT
single-dish data. Fig.~\ref{fig:ranges} presents two frequency ranges
with lines from several observed species. For G31, strong lines of all
complex organic species are detected, whereas for IRAS20126 many
targeted lines are below the detection limit. Many complex molecules
are also found in IRAS18089 but with weaker lines than for G31. Among
the serendipitous discoveries acetone, CH$_3$COCH$_3$, is possibly
seen in G31 and IRAS18089 for the first time (see \S 4.5).  Integrated
line intensities for all detected lines and selected upper limits are
given in Tables \ref{fluxes_h2co}-\ref{fluxes_ch3cch}. The rotation
diagrams for the detected species are shown in Figures
\ref{fig:rtd_h2co}--\ref{fig:rtd_ch3cch} whereas the optimized
parameters in the Weeds model for each molecule and source are
available online in Tables \ref{tab:weeds_i20} -- \ref{tab:weeds_g31}.

\subsection{Optical depth determinations}
\label{subsect:optical_depth_determinations}

To assess the importance of optical depths effects, the ratio of lines
of different isotopologues are compared. The expected
$^{12}$C/$^{13}$C isotope ratio depends on the galactocentric radius,
$R$$_\mathrm{GC}$, according to Eq.~\ref{eq:isotope_ratio} (Wilson \&
Rood, 1994).

\begin{equation}
\label{eq:isotope_ratio}
^{12}$C$/^{13}$C$ = \left(7.5\pm 1.9 \right) R_{\mathrm{GC}} \left[ \mathrm{kpc} \right] +  \left(7.6\pm 12.9 \right)
\end{equation}
The isotope ratios derived for our sources using
Eq.~\ref{eq:isotope_ratio} are given in Table \ref{tab:sources}. The
galactocentric radii were calculated trigonometrically from the
galactic coordinates, 
using the IAU value for the distance to the Galactic center
$R_0=8.5$~kpc \citep{kerr1986}.

Table~\ref{tab:isotopologue_ratios} lists the observed isotopologue
intensity ratios for the most abundant species in our sources. The
\methanol/\cmethanol~ratios are derived from a low-energy transition
2$_{2,0,+0}$--3$_{1,3,+0}$ with \Eup=44.6~K.  No high \Eup~transitions
were reliably detected for \cmethanol~in our sources in our standard
settings and the low $E_{\rm up}$ ratios are therefore taken to apply
to cold methanol. For G31.41+0.31, one additional observation was
carried out to cover a transition with \Eup$>$100 K. A
\methanol/\cmethanol~intensity ratio of 4.0 is derived for a
transition with \Eup$\approx$210 K (13$_{1,12,-0}$ --
13$_{0,13,+0}$). This indicates that also warm methanol is optically
thick in this source.

The \acetonitrile/\acetonitrilec~ratios are derived from the 13$_{3}$--12$_{3}$ line intensities for G31 and IRAS18089 and indicate that CH$_3$CN is optically thick in these two sources, but not in IRAS20126. For H$_2$CO and HNCO, isotopologue lines are detected but for different transitions than the main isotopologues. Thus, a model is needed to infer the optical depths. Fits to each of the isotopologues independently at a fixed temperature of 150~K using the $R_\mathrm{T=100~K}$ source size gives column density ratios that are significantly smaller than the overall isotope ratios, suggesting that these species are also optically thick for G31 and IRAS18089.  In practice, we have excluded the optically thick lines (as indicated by the Weeds model) from the RTD fits for all species.

\begin{table*}
\begin{center}
\caption{Isotopologue line intensity ratios in the observed sources. Lower limits are those for which the $^{13}$C-isotopologue was not detected.}
\label{tab:isotopologue_ratios}
\begin{tabular}{ l | c c c }
\hline
\hline
Species & IRAS20126+4104 & IRAS18089-1732 & G31.41+0.31 \\
\hline
\methanol/\cmethanol  & $>$6 & $>$18 & 6 \\
\acetonitrile/\acetonitrilec  & $>$65 & 4.7 & 4.5 \\
\hline
\end{tabular}
\end{center}
\end{table*}

\subsection{Temperatures}

Table \ref{tab:temperatures} summarizes the derived temperatures from
the RTD fit for the various species.  As also found by BIS07,
molecules can be classified into \textit{cold} ($<$100~K) and
\textit{warm} ($>$100~K), and our categorization is similar to
theirs. The Weeds analysis is consistent with this grouping. There is
however variation in temperatures within the groups, with warm species
having rotation temperatures from 70 to 300~K, and cold species from
40 to 100~K. Some variation is seen in rotation temperatures of
individual species between different sources; the rotation
temperatures are generally higher in G31 than in IRAS18089, while
IRAS20126 has the lowest of the three.

\begin{figure}
   \centering
   \includegraphics[width=0.45\textwidth]{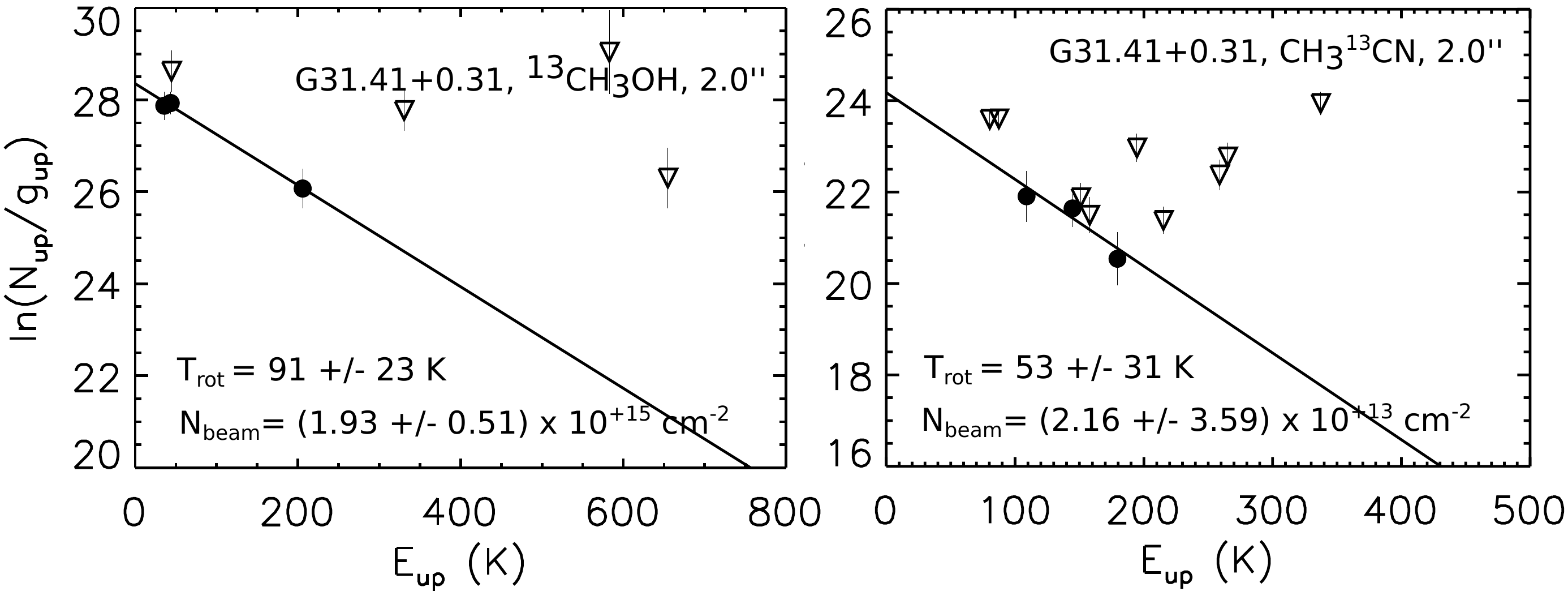}
      \caption{Rotation diagrams for $^{13}$CH$_3$OH~(left panel) and \acetonitrilec~(right panel) in G31. Triangles represent blended lines and are not included from the fit. }
         \label{fig:rtd_13isotopes}
   \end{figure}

The \Trot~value for \methanol~is $\sim$300~K for G31 and IRAS18089 and
$\sim$100~K for IRAS20126. For G31, several lines with \Eup$>$400~K
are detected, which makes the RTD fit more robust. For IRAS18089 and
IRAS20126 the accuracy of the RTD fits suffers from the small range of
\Eup~in the detected transitions. In addition to optically thick
lines, low-\Eup~lines have been excluded from the fit. These lines are
underestimated by the RTD fit and probably originate from a colder
extended region also seen in the $^{13}$C lines (Fig.~\ref{fig:rtd_13isotopes}). The \Trot~from the RTD
analysis therefore represents the warm \methanol~alone. 
See Sect.~\ref{sect:methanol} for a more detailed discussion on
the \methanol~emission.

For \acetonitrile, the \Trot~values range from $\sim$200 and $\sim$350~K
for the three sources.
The \acetonitrilec~rotation diagram gives a value of \Trot~of only
53~K, however.  (Fig. \ref{fig:rtd_13isotopes}). This illustrates the
large uncertainties at high temperatures for optically thick species
and the possibility of a cold component in addition to the warm one.

Contrary to the general trend, the \Trot~value for \formaldehyde~is
somewhat higher (204~K) in IRAS18089 than in G31 (157~K). The
discrepancy could be influenced by the small number of lines
used for the analysis. All fitted lines are those belonging to the
para-\formaldehyde, and the \Trot~fits are thus not affected by fluxes
from different spin states. A single transition of
\cformaldehyde~(3$_{1,2}$ -- 2$_{1,1}$) was covered, and no
information on the excitation temperature can therefore be obtained
from the minor isotopologue.

The \Trot~values of the other complex species, \isocyanicacid,
\ethanol, \propionitrile, \formamide~and \methylether\ are around 100~K
and are slightly higher in G31 than in IRAS18089. The \Trot~of
\methylformate~stands out in both sources, in G31 being closer to
200~K. No lines belonging to these species~were detected in IRAS20126.

The species classified as cold by BIS07, \ketene~and \methylacetylene,
indeed show cold rotation temperatures in all sources. Not enough
lines of \acetaldehyde~or \formicacid, which are also classified as
cold in BIS07, are observed in our sources for making rotation
diagrams.

\begin{table*}
\begin{center}
\caption{Temperatures (K) derived from the RTD analysis and the Weeds or CASSIS (CH$_3$OH) model. The species are classified  \emph{warm} and \emph{cold} according to BIS07. Square bracketed values are \Trot~values for optically thick species and round bracketed values are \Tex~values assumed based on temperatures derived from the other sources in this study.}
\label{tab:temperatures}
\begin{tabular}{ l | c c | c c | c c }
\hline
\hline
 & \multicolumn{2}{c}{IRAS20126+4104} & \multicolumn{2}{c}{IRAS18089-1732} & \multicolumn{2}{c}{G31.41+0.31} \\
\hline
 & RTD & model & RTD & model & RTD & model \\
\hline
& \multicolumn{6}{c}{warm species} \\
\hline
\formaldehyde & 	123$\pm$21 & 	150 & 	204$\pm$82 & 	(150) & 	157$\pm$44 &	 	(150) \\
\methanol & 		122$\pm$17 & 	300, 14$\pm$1 & 	291$\pm$37 & 		300, 15$\pm$2 & 	323$\pm$34 & 		200, 14$\pm$2 \\
\ethanol & 		-- & 			(100) & 	85$\pm$18 & 		150 & 	120$\pm$15 & 		100 \\
\isocyanicacid & 	-- & 			(200) & 	92$\pm$25 & 	200 & 	111$\pm$32 & 	200 \\
\formamide & 		-- & 			300 & 	72$\pm$28 & 		100 &  	94$\pm$50 & 	300 \\
\acetonitrile & 		217$\pm$352 & (200) &  	[346$\pm$106] & 	(200) & 	[311$\pm$68] & 	(300) \\
\propionitrile & 		-- & 			(80) & 	84$\pm$33 & 		80 & 		105$\pm$23 & 		80  \\
\methylformate & 	-- & 			(200) & 	118$\pm$20 & 		200 & 	174$\pm$11 & 		300 \\
\methylether & 		-- & 			(100) & 	66$\pm$11 & 		100 & 	90$\pm$6 & 		100 \\
\hline
& \multicolumn{6}{c}{cold species} \\
\hline
\ketene & 			-- & 			(50) & 	71$\pm$11 & 		50 & 		97$\pm$101 & 		50  \\
\acetaldehyde & 	-- & 			(50) & 	-- & 				(50) & 	-- & 				50  \\
\formicacid & 		-- & 			(40) & 	-- & 				(40) & 	-- & 				40  \\
\methylacetylene & 	40$\pm$10  & 	35 & 		46$\pm$12 & 		40 & 		67$\pm$14 & 		80  \\
\hline
\end{tabular}
\end{center}
-- means not enough lines were detected for a rotation diagram.\\
Typical uncertainties in the Weeds excitation temperatures are $\pm$50 K. \\
\end{table*}

\subsection{Column densities}

Table \ref{tab:column_densities} presents the column densities derived
from the RTD analysis, Weeds or CASSIS model, and from the $^{13}$C-isotopologues
for the optically thick species. Following BIS07, the column densities
for warm molecules are given as source-averaged values (see
Eq. \ref{eq:N_S}). The emission from cold molecules extends over a
larger volume and the values are given as beam-averaged column
densities. Typical uncertainty of the column densities derived from the RTD analysis is $\sim$40~\%.

\subsubsection{\methanol}
\label{sect:methanol}

An accurate determination of the \methanol~column density is essential for comparing the abundance ratios of complex organic species. For hot-core molecules, it is particularly important to quantify the warm \methanol~emission. The column densities of \methanol~in BIS07 were determined by the RTD method excluding the optically thick lines. The same is done in our analysis. Our rotation diagrams however also show emission from low-\Eup~transitions, which are strongly underestimated by the RTD fit on the warm lines, providing further evidence for the presence of a colder component. We have therefore also excluded these transitions. The fit to the higher \Eup, optically thin lines should give the warm \methanol~column density obtained in a similar way as BIS07.

\begin{figure}
\begin{centering}
\includegraphics[width=0.45\textwidth]{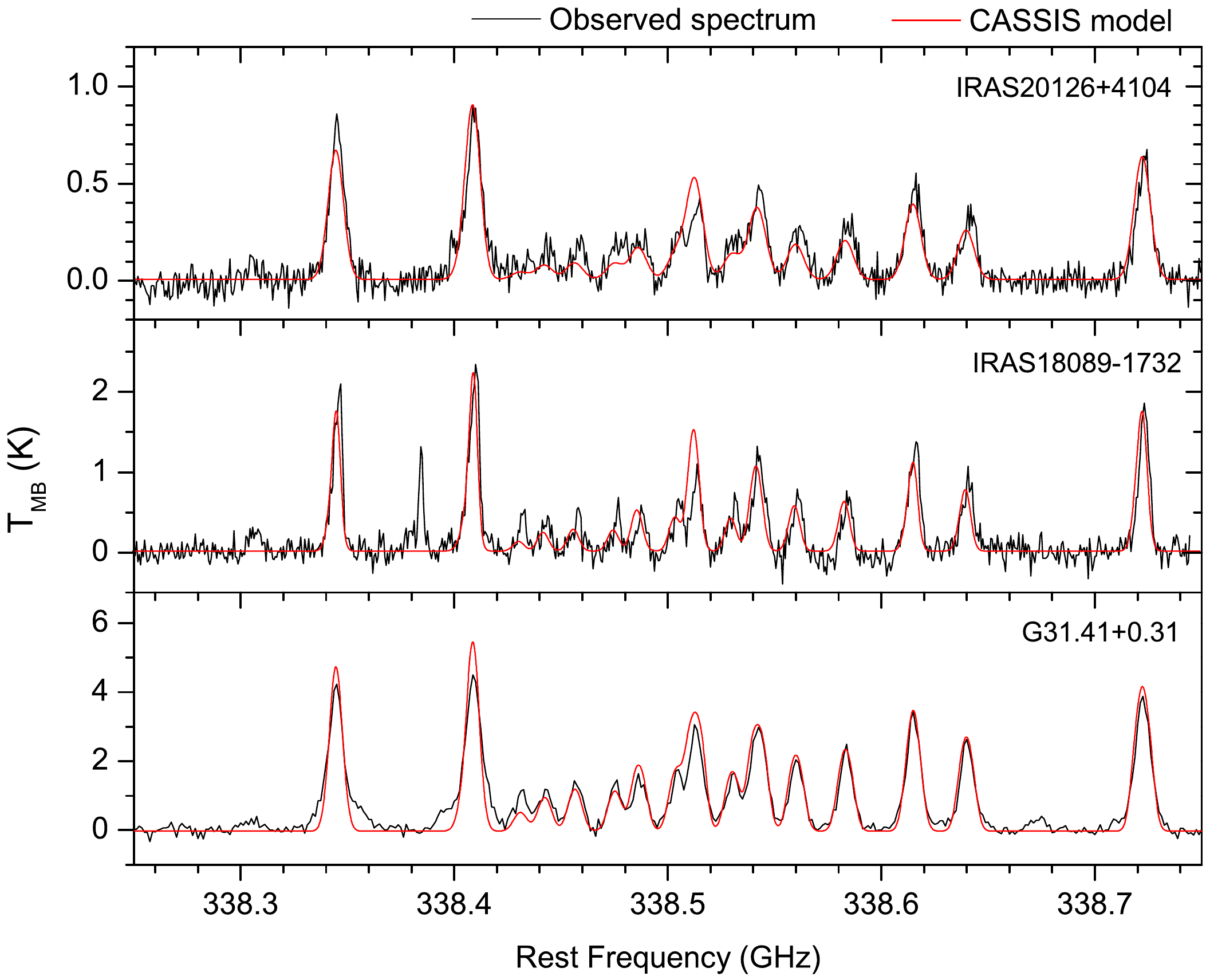}
\caption{CASSIS two-component models for \methanol~emission in the 338 GHz region covering the $7_K-6_K$ transitions with $K=0-6$.}
\label{fig:m_ch3oh}
\end{centering}
\end{figure}

\methanol~emission was also simulated using a two-component CASSIS model.  A single-component model is not able to simultaneously reproduce the warm and cold lines without overestimating the lines from intermediate energy levels. Indeed, a better agreement is obtained using a two-component model, consisting of a warm compact component and a cold extended component. Table~\ref{tab:ch3oh_cassis} shows the best model parameters obtained from the fitting. The warm components are fixed to $\theta_\mathrm{T=100K}$ while the cold component is allowed to extend beyond the beam diameter. The warm \methanol~column densities derived from the CASSIS fit are in agreement with the values derived from the RTD analysis. The best fits plotted onto the CH$_3$OH $7_K-6_K$ transitions (338.5 GHz) are shown in Fig.~\ref{fig:m_ch3oh}.

\begin{table*}
\begin{center}
\caption{CASSIS model parameters for \methanol~in the observed sources.}
\label{tab:ch3oh_cassis}
\begin{tabular}{l | c | c c c c c c}
\hline \hline
Source & Component$^a$ & $N_\mathrm{S}$ & $T_\mathrm{ex}$ & FWHM & $\theta_\mathrm{S}$ & $V_{\rm LSR}$ & $\chi^2$ \\
 & & [cm$^{-2}$] & [K] & [km s$^{-1}$] & [$''$] & [km s$^{-1}$] & \\
\hline
IRAS20126+4104
 & comp. & (1.1$\pm0.1$)E+17 & 300$^c$ & 8.15$\pm0.60$ & 2.2$^b$ & $-4.05\pm0.70$ & \multirow{2}{*}{1.9} \\
 & ext.  & (2.2$^{+0.9}_{-0.7}$)E+15 &  14$\pm1$ & 7.00$^{+0.60}_{-0.75}$ & $\gg$14.0 & $-4.25^{+0.80}_{-0.25}$ & \\
\hline
IRAS18089-1732 
 & comp. & (2.0$^{+0.1}_{-0.2}$)E+17 & 300$^c$ & 5.45$^{+0.40}_{-0.5}$ &  2.4$^b$ & 33.4$\pm0.2$ & \multirow{2}{*}{4.4}  \\
 & ext.  & (2.4$^{+1.4}_{-0.5}$)E+15 &  15$\pm2$ & 3.50$^{+0.50}_{-0.55}$ & $\gg$14.0 & 32.3$^{+0.4}_{-0.2}$ & \\
\hline
G31.41+0.31 
 & comp. & (1.0$\pm0.2$)E+18 & 200$^c$ & 6.35$\pm0.35$ &  2.0$^b$ & 97.2$\pm$0.2 & \multirow{2}{*}{7.4}  \\
 & ext.  & (1.2$\pm0.6$)E+16 &  14$\pm2$ & 3.95$\pm0.40$ & $\gg$14.0 & 97.2$^{+0.2}_{-0.3}$ & \\
\hline
\end{tabular}
\end{center}
$^a$ ``comp.'' for warm, compact component ; ``ext.'' for colder, more extended component.\\
$^b$ Fixed to $\theta_{\rm s,T=100K}$.\\
$^c$ Fixed.
\end{table*}

Several \cmethanol~lines are detected in G31. However, only
low-\Eup~lines are reliably detected since the high-\Eup~lines are
very weak or blended. Assigning these lines to the cold component and
assuming \Trot=20 K \citep{Oberg2011a, RequenaTorres2008}, the
beam-averaged \cmethanol~column density is $2.1\times 10^{15}$
cm$^{-2}$, corresponding to a \methanol~column density of $1.7\times
10^{17}$~cm$^{-2}$ for the 14$''$ volume. This is more than an order
of magnitude higher than found in the Weeds model for the cold
component, supporting the optically thick interpretation of the cold
component. Similarly, for the case of IRAS18089, 
the detected \cmethanol~lines may indicate a high optical depth.

As mentioned before, the \cmethanol~line with \Eup~=~211~K, covered in
additional observations for G31, reveals optical thickness in the warm
component as well. The \methanol~column density derived from this line
assuming temperature of 300 K is 2.7 $\times~10^{18}$~cm$^{-2}$ and
the CASSIS fit as well as the RTD analysis underestimate the warm
column density by a factor of $\sim$2.5. In principle, the CASSIS
analysis could be made consistent with the \cmethanol~value by letting
the warm source size vary as well. However, since such information is
not known for other molecules, as well as for consistency with BIS07,
we have chosen to keep the warm source size fixed at the 100~K radius.

In summary, it seems plausible that at least in our sources, methanol emission
arises from two temperature components, a warm (\Tex~$\approx$~300~K)
compact component and a cold (\Tex~$\approx$~20~K) significantly more
extended component. At least for G31, the \methanol~emission is optically
thick in both components.

\subsubsection{Other molecules}

Overall, the column densities of the various species follow the same
pattern in all sources, and hence seem to be well correlated relative
to each other. \methanol~is the most abundant complex molecule in all
sources. The other species have in general half to one order of
magnitude lower column densities.

The \acetonitrile~emission is optically thick for G31 and
IRAS18089. Column densities from the RTD analysis are therefore
underestimated. Due to a lack of sufficient optically thin lines, also
the Weeds analysis underestimates the column densities. Values derived
from the $^{13}$C-isotopologues are thus more reliable, even though some fraction may arise from a colder component. Indeed, the column density for \acetonitrile~derived from the
13-isotopologue is an order of magnitude larger than that from the RTD
of the main isotopologue alone. The same procedure was used by BIS07.

For \formaldehyde, the best estimates come from the RTD analysis on
the optically thin para-\formaldehyde~lines. 
The derived \formaldehyde~column
densities assuming \Trot~from Table~\ref{tab:temperatures}
and a statistical ortho-to-para ratio of 3 are given in Table~\ref{tab:column_densities}. The column densities derived
from the $^{13}$C-isotopologue of ortho-H$_2$CO 
are still larger than from RTD analysis
using optically thin lines, which indicates either larger ortho-to-para ratio in these sources (ratio of $\sim$3 to 5 has been predicted
for cold clouds by \citealt{Kahane1984}) or non LTE excitation. BIS07
derived the \formaldehyde~column densities from the
$^{13}$C-isotopologue.

The column density derived for \methylformate~from the RTD diagram is
significantly smaller than that derived from the Weeds model. The
\methylformate~emission is stronger in the 230 GHz beam (20-21$''$)
than in the 345 GHz beam (14$''$) probably indicating significant
extended emission. The RTD analysis was performed on the entire dataset,
while only lines in the 230 GHz band was used in the Weeds modeling.

\begin{figure*}
   \centering
   \includegraphics[width=0.90\textwidth]{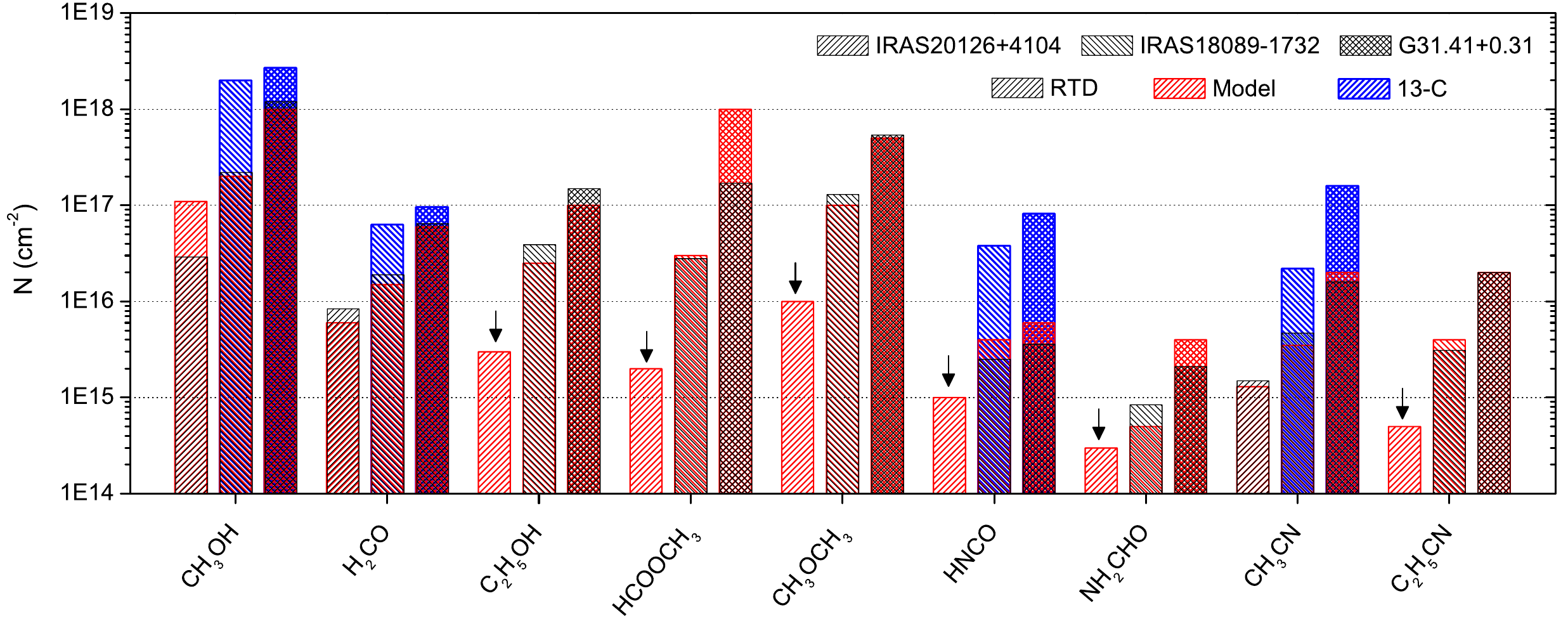}
      \caption{Source-averaged column densities for the targeted warm species in $\theta_\mathrm{T=100~K}$ volume. Column densities from the RTD analysis are marked in black bars. The red and blue bars show column densities from the Weeds or CASSIS (CH$_3$OH) models and from $^{13}$C-isotopologue, respectively. Arrows indicate upper limits.}
         \label{fig:column_densities_warm}
   \end{figure*}

\begin{figure}
   \centering
   \includegraphics[width=0.45\textwidth]{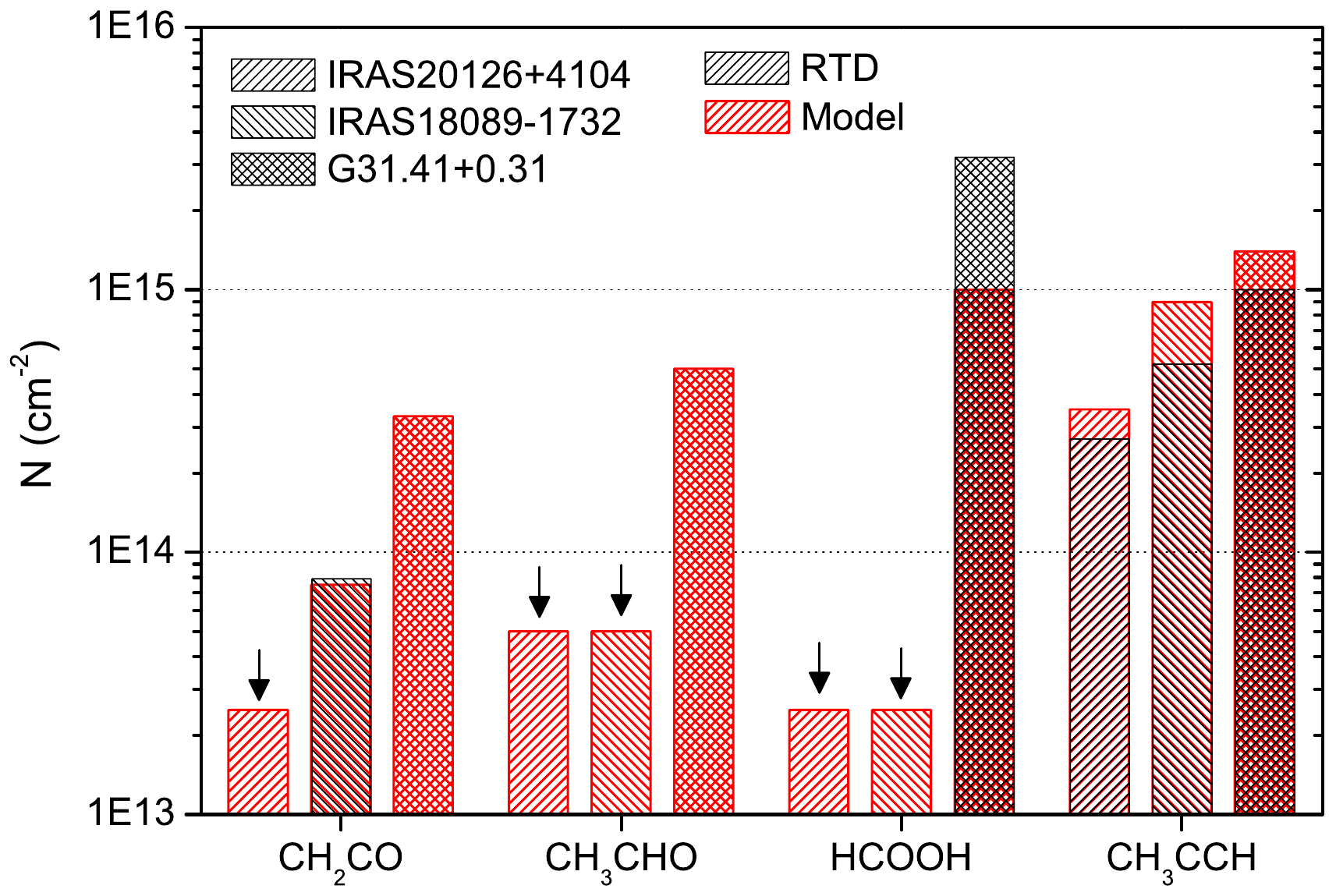}
      \caption{Beam-averaged column densities for the targeted cold species. Column densities from the RTD analysis are marked in black bars while the red bars show column densities from the Weeds model. Arrows indicate upper limits.}
         \label{fig:column_densities_cold}
   \end{figure}

\begin{table*}
\begin{center}
\caption{Column densities for the targeted species from the RTD analysis, Weeds or CASSIS (\methanol) models and those derived from $^{13}$C-isotopologues. Column densities for the warm molecules are source-averaged, while those for cold molecules are beam-averaged. Column densities where optical depth has a significant effect are labeled as lower limits.}
\label{tab:column_densities}
\begin{tabular}{ l | c c c | c c c | c c c }
\hline
\hline
 & \multicolumn{3}{c}{IRAS20126+4104} & \multicolumn{3}{c}{IRAS18089-1732} & \multicolumn{3}{c}{G31.41+0.31} \\
\hline
Species 			& RTD 		& Model 		& $N(^{12}$C)$^a$ 	& RTD 		& Model 		& $N(^{12}$C)$^a$ 	& RTD 		& Model 		& $N(^{12}$C)$^a$ \\
\hline
\formaldehyde$^b$ 		& 8.4E+15 	& 6.0E+15 	& $<$4.7E+16 		& 1.9E+16 	&  1.5E+16 	& 6.3E+16 		& 6.5E+16 	&  6.0E+16  	& 9.6E+16 \\
\methanol 			& 2.9E+16	& 1.1E+17 	& $<$4.0E+17$^c$ 		& 2.2E+17  	& 2.0E+17  	& 2.0E+18$^c$ 		& 1.2E+18  	& 1.0E+18 	& 2.7E+18$^d$ \\
\ethanol 				& -- 			& $<$0.3E+16 	& -- 				& 3.9E+16 	& 2.5E+16 	& -- 				& 1.5E+17 	& 1.0E+17 	& -- \\
\isocyanicacid 			& -- 			& 1.0E+15  	& $<$7.0E+15 		& 2.5E+15  	& 0.4E+16  	& $<$3.8E+16 		& 3.6E+15  	& 0.6E+16  	& $<$8.2E+16 \\
\formamide 			& -- 			& 0.3E+15 	& -- 				& 8.4E+14 	& 0.5E+15 	& -- 				& 2.1E+15 	& 0.4E+16 	& -- \\
\acetonitrile 			& 1.5E+15 	& 1.3E+15 	& $<$1.4E+15 		&  $>$4.7E+15  	& $>$3.5E+15  	& 2.2E+16 		& $>$1.6E+16 	& $>$2.0E+16  	& 1.6E+17 \\
\propionitrile 			& -- 			& $<$0.5E+15 	& -- 				& 3.1E+15  	& 0.4E+16 	& -- 				& 2.0E+16 	& 2.0E+16 	& -- \\
\methylformate 			& -- 			& $<$0.2E+16 	& -- 				& 2.8E+16  	& 3.0E+16 	& -- 				& 1.7E+17* 	& 1.0E+18 	& -- \\
\methylether 			& --  			& $<$0.1E+17 	& -- 				& 1.3E+17 	& 1.0E+17 	& -- 				& 5.4E+17 	& 5.0E+17 	& -- \\
\hline
\ketene 				& --  			& $<$2.5E+13 	& -- 				& 7.9E+13 	& 7.5E+13 	& -- 				& --		 	& 3.3E+14 	& -- \\
\acetaldehyde 			& -- 			& $<$0.5E+14 	& -- 				& -- 			& $<$0.5E+14 	& -- 				& -- 			&  5.0E+14 	& -- \\
\formicacid 			& -- 			& $<$2.5E+13 	& -- 				& -- 			& $<$2.5E+13 	& -- 				& 3.2E+15	& 1.0E+15 	& -- \\
\methylacetylene 		& 2.7E+14 	& 3.5E+14 	& -- 				& 5.2E+14 	& 0.9E+15  	& -- 				& 1.0E+15 	& 1.4E+15 	& -- \\
\hline
\end{tabular}
\end{center}
* Rotation diagram fit on \methylformate~on all lines, while Weeds model on the 230 GHz lines only.\\
$^a$ $N(^{12}$C) obtained from the $N(^{13}$C) adopting a $^{12}$C/$^{13}$C ratio equal to 70, 54 and 41 for IRAS20126, IRAS18089 and G31.41, respectively.\\
$^b$ Column density from $^{13}$C-isotopologue from ortho-\cformaldehyde, corrected using the statistical ortho to para ratio of 3:1.\\
$^c$ Calculated from the line intensity of transition 2$_{2,0,+0}$--3$_{1,3,+0}$  at 345.084 GHz with \Eup=44.6~K, assuming a \Trot~of 300~K.\\
$^d$ Calculated from the line intensity of transition 13$_{1,12,-0}$ -- 13$_{0,13,+0}$ at 341.132 GHz with \Eup=206~K, assuming a \Trot~of 300~K.\\
\end{table*}

\section{Discussion}
\label{sect:discussion}

\subsection{Comparison to massive YSOs without a disk structure}

In order to see the effect of a flattened circumstellar structure on
the chemistry around the YSO, we compare the temperatures and
abundances of the complex organic molecules in sources studied in this
work and those studied in BIS07. In addition to our sources
(IRAS20126, IRAS18089 and G31), AFGL 2591, NGC 7538 IRS1 and G24.78 from the
BIS07 sample are now known to have disk structures as well
\citep{vanderTak2006, Wang2012, Pestalozzi2004,
  Pestalozzi2009, Knez2009, Beltran2005}.  A disk-like structure may be present also
for other sources but current evidence is not as strong as for the
above sources. For the comparison with the BIS07 sample, we use the
results primarily from the RTD method to exclude method based
differences.

\subsubsection{Rotation temperatures}

\begin{figure*}
\centering
\includegraphics[width=0.9\textwidth]{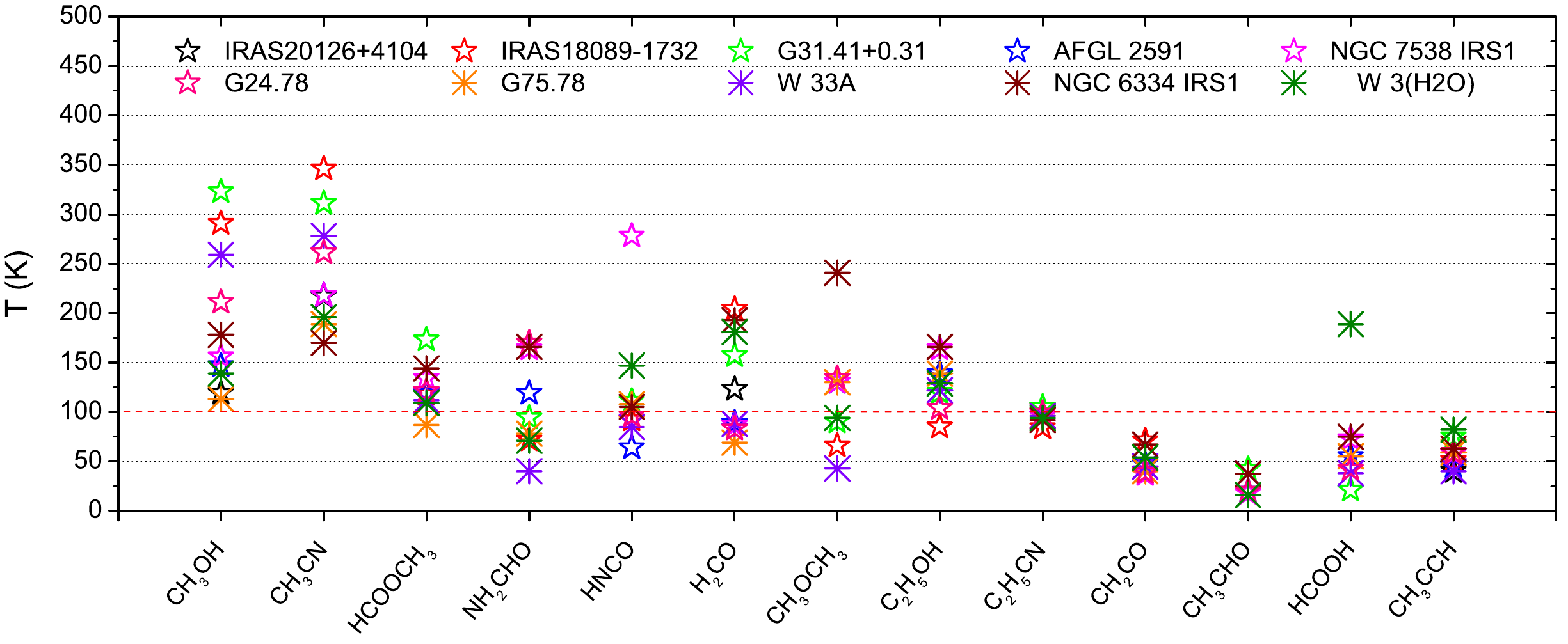}
\caption{Rotation temperatures for selected species in massive YSOs with (open stars) and without (asterisks) observed disk structure.}
\label{fig:temp_comparison}
\end{figure*}

Figure \ref{fig:temp_comparison} shows the rotation temperatures of the complex species for the observed sources together with those from BIS07. As in BIS07, we find that the complex molecules can be divided into warm and cold species based on their rotation temperatures. The rotation temperatures obtained for the molecules in our sources agree with the division; \methanol, \formaldehyde, \ethanol, \methylformate, \methylether, \isocyanicacid, \formamide, \acetonitrile~and \propionitrile~show rotation temperatures generally $\geq$100~K while \ketene, \acetaldehyde, \formicacid~and \methylacetylene~have rotation temperatures $<$100~K.

The rotation temperatures of the cold molecules show a very small scatter ($\pm$25~K) from source to source, with or without a disk structure, ignoring the one outlier for HCOOH. Of the warm species, \ethanol~has a consistent rotation temperature of 100--150 K from source to source.  The RTD analysis for these species is reliable due to low optical depths and lack of anomalous excitation.

The other warm species show more scatter in the derived rotation
temperatures. This is particularly the case for \methanol~and
\acetonitrile, which have rotation temperatures ranging from 100 to
350~K. No systematic difference is seen between the two source
types. Moreover, the \methylformate~rotation temperatures for AFGL
2591 and NGC 7538 IRS1 are similar to the diskless sources, disproving
any difference between the two source types.  The scatter in the
rotation temperatures could indicate that they exist in environments
with different temperatures, but it may also be caused by optical
depth effects (\formaldehyde, \isocyanicacid, \methanol~and
\acetonitrile) and (anomalous) radiative excitation. The latter has
been previously seen for, e.g., \formamide~and
\methylformate~\citep{nummelin2000} as well as HNCO
\citep{churchwell1986}. For species with significant radiative
excitation, e.g., \methylformate, the scatter of the data points in
the RTD plots is higher and results in additional uncertainty since
the linear fits do not properly capture this scatter.

\subsubsection{Column densities}

\begin{figure*}
   \centering
   \includegraphics[width=0.95\textwidth]{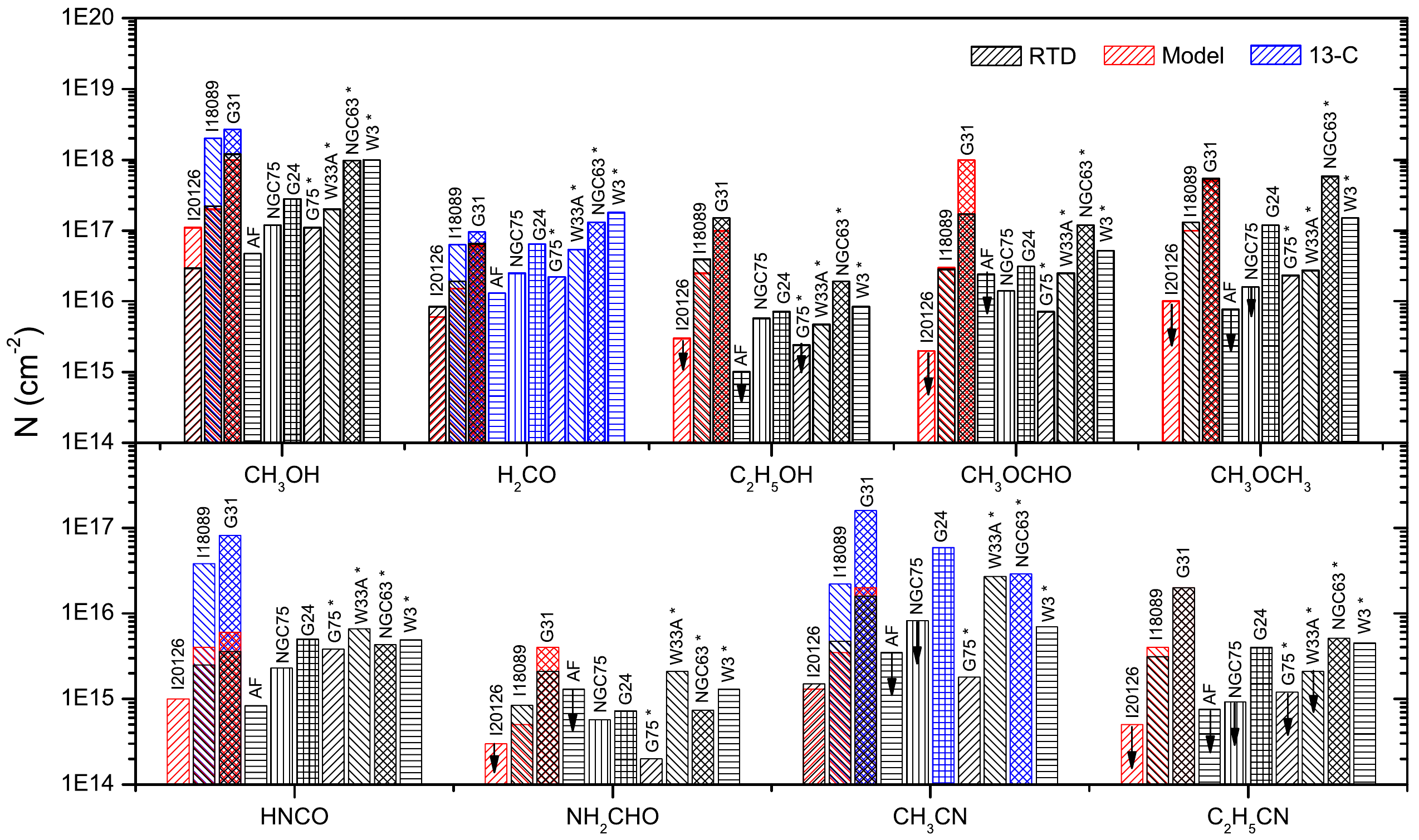}
      \caption{Source-averaged column densities for warm molecules. Column densities from the RTD analysis are marked in black bars. Sources without disk structure are marked with an asterisk. The red and blue bars show column densities from the Weeds or CASSIS (CH$_3$OH) models and from $^{13}$C-isotopologue, respectively. Upper limits are indicated with arrows.}
         \label{fig:column_densities_warm_all}
   \end{figure*}

\begin{figure*}
   \centering
   \includegraphics[width=0.95\textwidth]{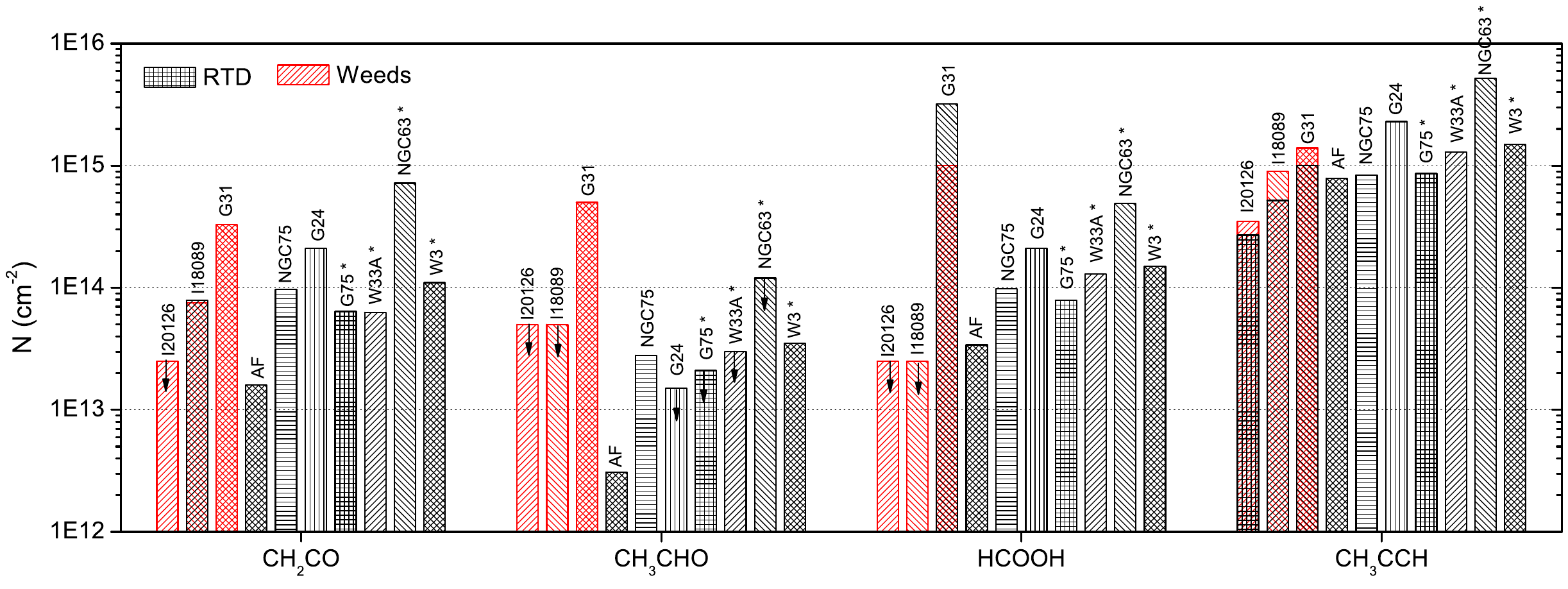}
      \caption{Beam-averaged column densities for cold molecules in sources observed in this study and those from BIS07. Sources without disk structure are marked with an asterisk. Column densities from the RTD analysis are marked in black bars while the red bars show column densities from the Weeds model. Upper limits are indicated with arrows.}
         \label{fig:column_densities_cold_all}
   \end{figure*}

Figures \ref{fig:column_densities_warm_all} and
\ref{fig:column_densities_cold_all} show the column densities for the
targeted species in our sources and those from BIS07 for warm and
cold molecules, respectively. For the warm species the column
densities within the sources vary by 1-2 orders of magnitude. There is
also a significant variation between sources. G31 is chemically
richest of the sources, with highest column densities of all the
targeted species, compared to other sources. IRAS20126 is among the
chemically poorest sources. The pattern of column densities is
remarkably similar, however, and the disk sources do not stand out.

\subsubsection{Abundance ratios}

Table \ref{tab:abundance_ratios} gives the abundance ratios for the
targeted species with respect to \methanol~ for oxygen-bearing species
and \isocyanicacid~for nitrogen-bearing species.  Column densities
from the RTD analysis are primarily used. In cases where column
densities are not available from the RTD analysis (mainly for
IRAS20126) the values are taken from the Weeds analysis.

\begin{table*}
\begin{center}
\caption{Abundance ratios of complex species in the observed sources and those in other chemically rich environments. The abundances are column densities with respect to \methanol~for oxygen-bearing species and with respect to \isocyanicacid~for nitrogen-bearing species. The \methanol~and \isocyanicacid~column densities are taken from the RTD analysis when available.}
\label{tab:abundance_ratios}
\begin{tabular}{ l | c c c c c c}
\hline
\hline
Source & \formaldehyde & \ethanol  & \formamide & \methylformate & \methylether & Refs. \\
\hline
\multicolumn{7}{l}{Massive protostellar objects with a disk structure} \\
\hline
IRAS20126+4104	&	0.29$^p$	& 	$<$0.10	& 	$<$0.30 & $<$0.34	&	$<$0.34	& this work\\
IRAS18089-1732	& 	0.09$^p$	& 	0.18		& 	0.34 	& 	0.13	&	0.60	& this work \\
G31.41+0.31		& 	0.06$^p$	&	0.12		&	0.59 	&	0.14 	&	0.46	& this work \\
AFGL 2591		&	0.28$^o$		&	$<$0.02		&	$<$1.00	&	0.51		&	$<$0.16	& 1\\
NGC 7538 IRS1	&	0.21$^o$		&	0.05			&	0.25		&	0.12		&	$<$0.13	& 1\\
G24.78			&	0.23$^o$		&	0.03			&	0.14	&	0.11			&	0.43	& 1	 \\
\hline
\multicolumn{7}{l}{Massive protostellar objects with no detected disk structure} \\
\hline
G75.78			&	0.20$^o$		&	$<$0.02		&	0.05	&	0.06			&	0.21	& 1 \\
W 33A			&	0.27$^o$		&	0.02			&	0.32	&	0.13			&	0.14	& 1	 \\
NGC 6334 IRS1	&	0.13$^o$ 		&	0.02			&	0.17	&	0.12			&	0.60	& 1	 \\
W 3(H2O)			&	0.18$^o$ 		&	0.01			&	0.27	&	0.05			&	0.15	& 1	 \\
\hline
\multicolumn{7}{l}{Low-mass outflows} \\
\hline
B1-b core 		& 1.2 	& $<$1.0 	& -- & 2.3 		& $<$0.8		& 2 \\
SMM4-W 		& 0.6 	& $<$0.4 	& -- & 3.5 		& 1.1			& 3\\
L1157 outflow 	& -- 		& 0.7 	& -- & 1.8 		& --			& 4, 5\\
\hline
\multicolumn{7}{l}{Low-mass protostars} \\
\hline
SMM1 				& 6.6 	& $<$3.4 	& -- & 10 		& 5.3		& 3\\
SMM4 				& 2.2 	& $<$0.6 	& -- & $<$1.0 	& 0.8		& 3\\
NGC1333 IRAS 4A env.	& -- 		& -- 		& -- & 56 		& $<$22	& 6, 7, 8\\
NGC1333 IRAS 4B env. 	& -- 		& -- 		& -- & 26 		& $<$19	& 6, 7, 8\\
IRAS 16293 env. 		& 4 		& -- 		& -- & 30 		& 20		& 8, 9, 10\\
\hline
\multicolumn{7}{l}{Low-mass hot corinos} \\
\hline
IRAS 16293 A 		& $<$0.02	& 1.4 	& -- 	& 0.6 	& 0.6		& 11, 12\\
IRAS 16293 B 		& 0.6 	& -- 		& -- 	& 0.8 	& 1.6		& 11, 12\\
NGC1333 IRAS 2A 	& -- 		& -- 		& -- 	& -- 		& 2		& 13, 14\\
\hline
\end{tabular}
\end{center}
$^p$ para-\formaldehyde.\\
$^o$ ortho-\formaldehyde.\\
\textbf{References.} (1) \citet{Bisschop2007}; (2) \citet{Oberg2010}; (3) \citet{Oberg2011a}; (4) \citet{Bachiller1997}; (5) \citet{Arce2008}; (6) \citet{Maret2005}; (7) \citet{Bottinelli2007}; (8) \citet{Herbst2009}; (9) \citet{vanDishoeck1995}; (10) \citet{Cazaux2003}; (11) \citet{Kuan2004}; (12) \citet{Bisschop2008}; (13) \citet{Huang2005}; (14) \citet{Jorgensen2005}.
\end{table*}

\begin{figure*}
   \centering
   \includegraphics[width=0.95\textwidth]{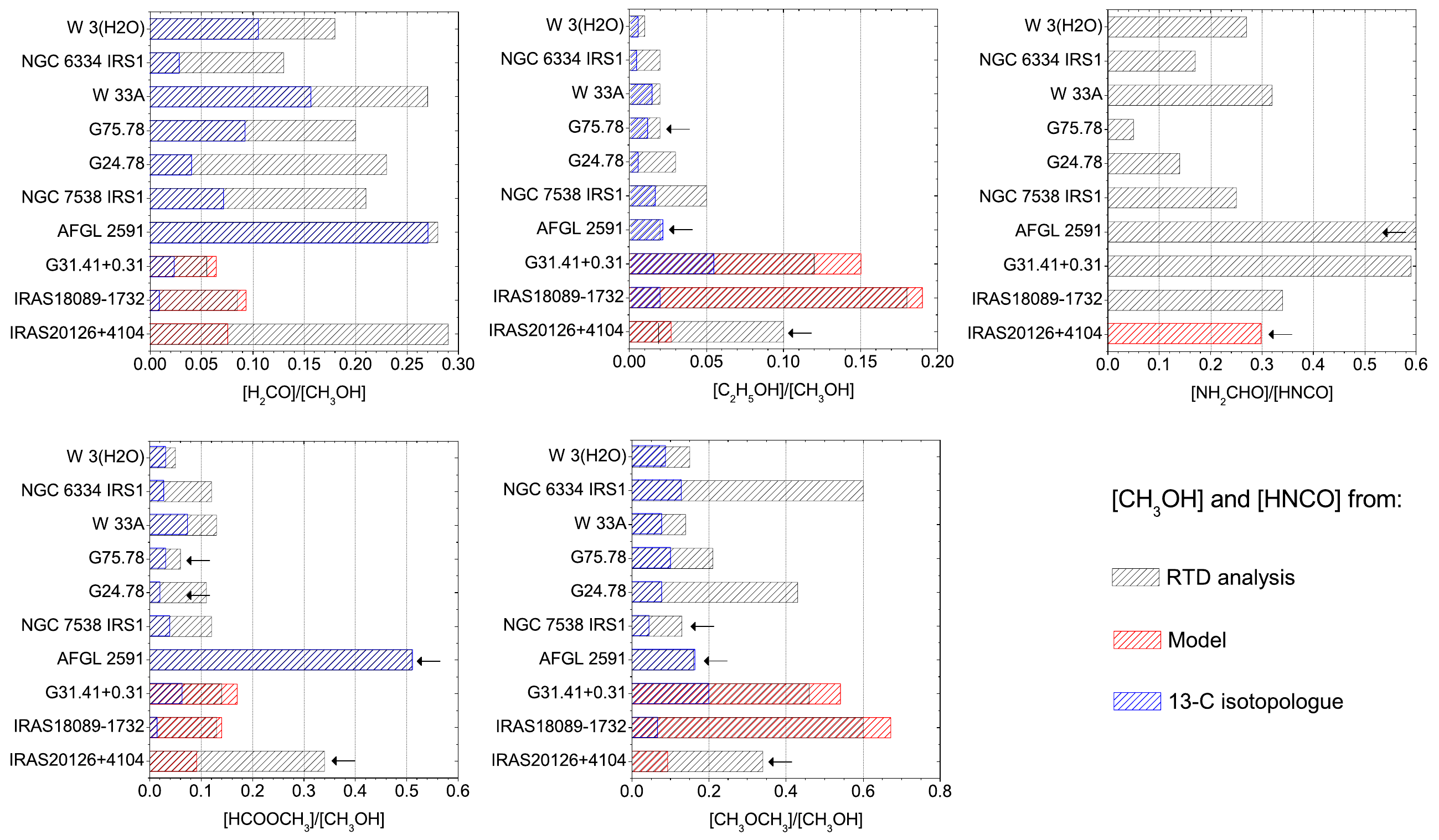}
      \caption{Abundance ratios of complex species with respect to \methanol~for oxygen-bearing species and \isocyanicacid~for nitrogen-bearing species. The black bars indicate abundance ratios calculated with the \methanol~and \isocyanicacid~column density derived from the RTD analysis similar to BIS07. The red and blue bars indicate abundance ratios where \methanol~and \isocyanicacid~column densities are derived from CASSIS or Weeds (\isocyanicacid) model, respectively. Upper limits are marked with arrows.}
         \label{fig:abundance_ratios}
   \end{figure*}

The resulting abundance ratios are presented in
Fig. \ref{fig:abundance_ratios}. Values 
from the RTD method are shown in black bars and can thus be
directly compared to BIS07, whereas those from the Weeds model and
$^{13}$C-isotopologue are in red and blue bars. In general, the
abundance ratios in different sources have larger variations within
the source types than between them. For example, \formaldehyde~has
both lowest and highest abundance ratio for two disk candidates,
G31 and IRAS20126, respectively. 
The \formaldehyde~results depend somewhat on the analysis method used.

\ethanol, \methylformate~and \methylether~abundances peak for G31 and
IRAS18089, and are generally lower for other
sources. \methylether~also peaks for some diskless sources. The
N-bearing species \formamide~has a large variation in the abundance
ratio with respect to HNCO. Among our sources, \formamide~peaks for
G31, the source with largest abundance of O-bearing species,
and the least clear disk structure.

\begin{figure*}
   \centering
   \includegraphics[width=0.95\textwidth]{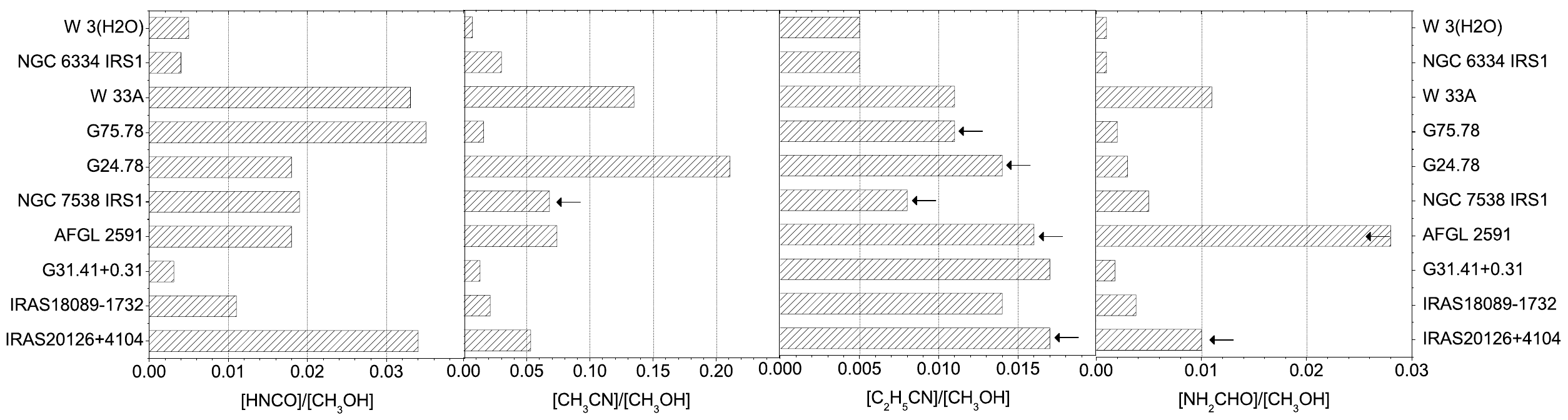}
      \caption{Abundance ratios of N-bearing species \isocyanicacid, \acetonitrile, \propionitrile~and \formamide~with respect to \methanol~(from RTD analysis). Arrows indicate upper limits.}
         \label{fig:abundance_ratios_nbearing}
   \end{figure*}

Figure \ref{fig:abundance_ratios_nbearing} shows the abundance ratios
of N-bearing species with respect to \methanol.  Again, a lot of
variation is seen within each source type, with no specific trend
found for sources with a disk-like structure.

We stress that the absolute inferred column densities and abundance
ratios are uncertain by a factor of a few up to an order of magnitude,
as already indicated by the different analysis techniques. An
independent assessment of the accuracy of the results can be made by
comparison with the inferred column densities and abundances of
\citet{Zernickel2012} for NGC 6334 I (=NGC 6334 IRS), who observed a
completely different set of lines of the same molecules with {\it
  Herschel}-HIFI and the SMA in beams ranging from
2--40$''$. Comparison with the results of BIS07 shows good agreement
within a factor of a few for the abundances of several species
(CH$_3$OH, CH$_3$OCH$_3$, CH$_3$OCHO, C$_2$H$_5$OH, C$_2$H$_5$CN)
whereas others differ by an order of magnitude (H$_2$CO,
CH$_3$CN). The species that show the largest discrepancy are those
with highly optically thick lines and without a large set of
isotopologue lines. Thus, a combination of differences in adopted
source sizes and optical depth effects can account for the
discrepancies. Because our approach is the same for all sources, the
relative values from source to source should be more reliable.

\subsection{Chemical and physical implications}

The main goal of this study is to investigate similarities and
differences between sources with and without a disk-like structures.
The presence of a flattened accretion disk should allow the UV
radiation from the central object to escape more readily and then
impinge on the gas and dust in the outer walls of a flared disk or
outflow cavity \citep{Bruderer2010}. Increased UV radiation could
manifest itself as enhancement in the complex organic species produced
through UV photodissociation from \methanol~in the solid state, such as
CH$_3$OCH$_3$, C$_2$H$_5$OH and CH$_3$OCHO \citep{Oberg2009b}. In this
scenario, \methanol\ is photodissociated into various radicals such as
CH$_3$, CH$_3$O and CH$_2$OH which can become mobile at higher ice
temperatures (20--40~K) and form the observed complex organic
molecules \citep{Garrod2006}.  Higher temperatures favor diffusion of
larger radicals resulting in the formation of larger complex organic
molecules compared with small molecules like H$_2$CO. Another related
parameter that plays a role is the CO content of ices, with
\formaldehyde, CHO- and COOH-containing species enriched in cold,
CO-rich \methanol~ices in which CO has not evaporated.  Indeed
\citet{Oberg2011a} show that the increased abundance of
\methylformate\ in low-mass YSOs compared with the high-mass sample of
BIS07 may be due to this effect. Although only limited laboratory
experiments exist on N-containing molecules, the abundance of
N-bearing complex organic molecules could be enhanced due to
photodissociation of N$_2$ or NH$_3$.

Our results show that there is no consistent enhancement of complex
organic species produced through \methanol~photochemistry, nor of
N-bearing species, that can be attributed to the presence of a
flattened disk structure on the scales probed by our data. The
variations within source types are larger than between the source
types. One explanation for this lack of differentation could be that
all high-mass sources have a source structure through which UV
radiation can impact the larger surroundings. In particular, just like
low-mass sources, all high-mass YSOs are expected to have outflows and
cavities through which UV radiation can escape and affect the
chemistry, whether or not they have a large disk-like structure. This
scenario can be tested with future high angular resolution data on
$<$1$''$ scale with ALMA which should then reveal the emission from
complex molecules coating the walls of the outflow cavities.

An alternative explanation is that enhanced temperature in the YSO
environment is not needed for the production of complex organic
molecules and that they are already formed in the prestellar stage
through UV- and cosmic-ray processing of cold ices, using just the
internal UV field produced by interaction of cosmic rays with H$_2$
\citep{Prasad1983} rather than that of the star. This scenario has
gained support over the last few years with the detection of
\methylformate~and \methylether~in cold low-mass cores away from
the YSO where the molecules are either released by shocks
\citep{Arce2008, Oberg2010} or by photodesorption \citep{Bacmann2012}
although the precise formation mechanisms are still unclear. Gas-phase
processes may also contribute in some cases. Again, high spatial
resolution and sensitivity such as ALMA will be key to testing this
scenario.

\section{Summary and conclusions}

We have carried out a partial submillimeter line survey, targeting
several complex organic species toward selected massive YSOs with
strong evidence of circumstellar accretion disks, IRAS20126+4104,
IRAS18089-1732 and G31.41+0.31. This is the first time that molecular
abundances are reported for these well-known sources. The analysis is
performed using both the rotation diagram method and spectral
modeling. The inferred rotation temperatures and molecular abundances
are compared to sources without reported disk structures analysed
using the same techniques. The molecules can be divided into two
different groups based on their rotation temperatures, independent of
source type.  In particular, the cold ($<$100~K) species have
remarkably constant rotation temperatures from source to source. The
warm ($>$100~K) species exhibit more scatter possibly due to optical
depth effects, non-LTE conditions and radiative excitation. The column
densities peak for the same sources, with G31.41+0.31 being chemically the
richest of the studied sources. 

The relative abundances of species follow the same pattern, and no
chemical differentiation could be established between the two source
types within the (considerable) uncertainties. The lack of chemical
differentiation between massive YSOs with and without observed
disk-like structure suggests similarity in the physical conditions and
level of UV exposure for all sources, for example through outflow
cavities. Alternatively, the complex molecules may already be formed
in the cold prestellar stage under similar conditions. Ultimately and
indirectly, these results imply that the mechanism of the formation of
massive stars does not differ significantly from source to
source. Future high angular and high sensitivity observations of
optically thin lines will be key to distinguishing the different
scenarios and derive accurate abundance ratios that can be directly
compared with chemical models.


\begin{acknowledgements}

  Astrochemistry in Leiden is supported by the Netherlands Research
  School for Astronomy (NOVA), by a Spinoza grant from the Netherlands
  Organisation for Scientific Research (NWO), and by the European
  Community's Seventh Framework Programme FP7/2007-2013 under grant
  agreement 238258 (LASSIE) and 291141 (CHEMPLAN).

\end{acknowledgements}


\bibliographystyle{aa} 
\bibliography{literature} 


\Online

\begin{appendix}

\section{Detected lines per species for all sources}
\label{detected}

The line assignment and detection is based on Gaussian fitting with
the following criteria: (i) the fitted line position has to be within
$\pm$1~MHz of the catalog frequency, (ii) the FWHM is consistent with
the those given in table \ref{tab:sources} and (iii) the peak
intensity has at least a $S/N=3$. The errors on the integrated
intensities are computed as follows.

The integrated main beam temperatures, $\int T_{\mathrm {MB}}dV$, were
obtained by Gaussian fits to the lines (Eq.~\ref{eq:TmbdV}).

\begin{equation}
\label{eq:TmbdV}
\int T_{\mathrm {MB}}dV = c^{st} T_0 \Delta V
\end{equation}
with
\[
c^{st}=\frac{\sqrt{\pi/{\rm ln}2}}{2}
\]
where $T_0$ is the peak intensity and $\Delta V$ is the FWHM of the line. The error, $d\int T_{\mathrm {MB}}dV$, is calculated from Eq.~\ref{eq:dTmbdV}.

\begin{equation}
\label{eq:dTmbdV}
d\int T_{\mathrm {MB}}dV = c^{st}\left[\Delta V d(T_0) + T_0 d(\Delta V)\right]
\end{equation}
with
\[
d(T_0)=\sqrt{\rm{rms}^2 + ({\rm cal} \times T_0)^2 + \sigma^2_{T0}}
\]
and
\[
d(\Delta V)=\sigma_{\Delta V}
\]
where rms is the root mean square amplitude of the noise in the
spectral bin $\delta {\rm v}$, cal is the calibration uncertainty of the
telescope, and $\sigma$'s are the statistical errors on $T_0$ and
$\Delta V$ from the Gaussian fits.

The errors on the integrated intensities derived from
Eq.~\ref{eq:dTmbdV} include all statistical errors from the Gaussian
fit.

\begin{equation}
\label{eq:error_2}
d\int T_{\mathrm {MB}}dV = c^{st} \Delta V \sqrt{\rm{rms}^2 + ({\rm cal}\times T_0)^2}
\end{equation}

For undetected transitions the upper limits were determined as 3$\sigma$ limits (Eq. \ref{eq:upper_limit}) using:
\begin{equation}
\label{eq:upper_limit}
\sigma=1.2\sqrt{\delta \mathrm{v} \Delta V} \mathrm{rms},
\end{equation}
where 1.2 is the coefficient related to the calibration uncertainty of
20~\%.

CLASS was used to determine the Gaussian fits and the uncertainties in
the individual parameters.  The formal errors on the integrated
intensities derived from Eq.~\ref{eq:error_2} in some cases yield a
$S/N< 2.5$. This is caused by: (i) a conservative estimate of
the statistical error on the FWHM parameter in the Gaussian fitter of
CLASS; (ii) all statistical errors are included into our error
calculation.  Considering the higher $S/N$ on the peak intensity, a more
traditional error estimate without the statistical errors from the
Gaussian fit (Eq.~\ref{eq:error_2}) would result in a $S/N>3$ for the
integrated intensity as well. All weak line fits were confirmed by visual inspection.

\begin{table*}
\caption{Observed line fluxes $\int T_{\rm MB}dV$ (K km s$^{-1}$) for \formaldehyde~and its isotopic species.}  
\begin{tabular}{l|lrc|ccc}
\hline
\hline
Frequency & Transition & $E_{\rm up}$ & $A$ & \multicolumn{3}{l}{Sources} \\
{[GHz]}     &            & {[K]} & {[s$^{-1}$]} & IRAS20126+4104 & IRAS18089-1732  & G31.41+0.31  \\
\hline
\multicolumn{7}{l}{\formaldehyde}\\
\hline
218.222 $^\emph{p}$	&	3$_{\rm 0, 3}$ -- 2$_{\rm 0, 2}$	& 21 & 2.8(-4) 		&	2.82 (0.65) 	&	-- 				&	$>$7.14 (1.71) 		\\
218.476 $^\emph{p}$	&	3$_{\rm 2, 2}$ -- 2$_{\rm 2, 1}$	& 68 & 1.6(-4) 		&	-- 			&	-- 				&	4.15 (1.03) 		\\
363.946 $^\emph{p}$	&	5$_{\rm 2, 4}$ -- 4$_{\rm 2, 3}$	&100 & 1.2(-3) 		&	4.65 (1.08)	& 	9.92 (2.15)		&	$>$20.29 (3.31) 	\\
364.103 $^\emph{p}$ 	&	5$_{\rm 4, 1/2}$ -- 4$_{\rm 4, 0/1}$&241 & 5.0(-4) 	&	1.90 (0.58)	&	4.25 (1.07)		&	10.97 (2.88) 		\\
364.275 $^\emph{o}$	&	5$_{\rm 3, 3}$ -- 4$_{\rm 3, 2}$	&143/158 &8.9(-4) 	&	$>$5.08 (1.50)	&	$>$12.62 (3.34) 	&	$>$21.56 (5.08) 	\\
364.289 $^\emph{o}$	&	5$_{\rm 3, 2}$ -- 4$_{\rm 3, 1}$	&143/158 &8.9(-4) 	&	$>$5.78 (1.71)	&	$>$10.68 (2.83)	&	$>$20.46 (4.85) 	\\
\hline
\multicolumn{7}{l}{\cformaldehyde}\\
\hline
219.909 $^\emph{o}$	&	3$_{\rm 1, 2}$ -- 2$_{\rm 1, 1}$	& 33 & 2.6(-4) &	$<$0.34	&	1.08 (0.60)	&	1.60 (0.55) \\
\hline
\end{tabular}

The notation $a(-b)$ stands for $a\times10^{-b}$\\
$>$ means lower limit (optically thick line)\\
-- means frequency not observed\\
$^\emph{p}$ para-\formaldehyde\\
$^\emph{o}$ ortho-\formaldehyde
\label{fluxes_h2co}
\end{table*}

\begin{table*}
\caption{Observed line fluxes $\int T_{\rm MB}dV$ (K km s$^{-1}$) for \methanol~and its isotopic species. }  
\begin{tabular}{l|lrc|ccc}
\hline
\hline
Frequency & Transition & $E_{\rm up}$ & $A$ & \multicolumn{3}{l}{Sources} \\
{[GHz]}     &            & {[K]} & {[s$^{-1}$]} & IRAS20126+4104 & IRAS18089-1732  & G31.41+0.31  \\
\hline
\multicolumn{7}{l}{\methanol}\\
\hline
218.440	&	 4$_{2,0}$ --  3$_{1,0}$	& 45 & 6.9(-5) &		--		&	--				&	10.10 (2.22)::\\
219.984	&	25$_{3,0}$ -- 24$_{4,0}$	&802 & 3.0(-5) &		$<$1.33	&	$<$0.24			&	$<$0.45 (0.24): \\ 
219.994	&	23$_{5,0}$ -- 22$_{6,0}$	&776 & 2.6(-5) &		$<$1.33	&	$<$0.24			&	$<$1.74 (0.56): \\
240.739	&	26$_{3,-0}$ -- 26$_{2,+0}$&864 & 1.1(-4) &		$<$0.20	&	$<$0.40 (0.32)		&	$<$0.78 (0.47): \\ 
240.818	&	5$_{1,2}$ -- 4$_{1,2}$	&834 & 8.5(-5) &		$<$0.20	&	$<$0.23			&	$<$3.58 (0.88): \\
240.861	&	5$_{-4,2}$ -- 4$_{-4,2}$	&779 & 3.2(-5) &		$<$0.20	&	$<$0.27			&	$<$0.81 (0.37): \\ 
240.870	&	5$_{0,2}$ -- 4$_{0,2}$	&769 & 8.9(-5) &		$<$0.20	&	$<$0.90 (0.71):		&	$<$3.26 (1.06):\\ 
240.916	&	5$_{3,-2/+2}$ -- 4$_{3,-2/+2}$	&693 & 5.7(-5) &	$<$0.20	&	$<$1.17 (0.46):		&	1.43 (0.45)\\ 
240.932	&	5$_{4,-2/+2}$ -- 4$_{4,-2/+2}$	&649 & 3.2(-5) &	$<$0.20	&	$<$0.57 (0.20)		&	$\sim$0.42 (0.21)\\
240.937	&	5$_{-2,2}$ -- 4$_{-2,2}$	&680 & 7.4(-5) &		$<$0.20	&	$<$1.03 (0.69):		&	$\sim$0.49 (0.24)\\
240.939	&	5$_{0,+2}$ -- 4$_{0,+2}$	&543 & 1.2(-4) &		$<$0.20	&	$<$1.03 (0.69):		&	2.54 (0.86)\\
240.948	&	5$_{3,2}$ -- 4$_{3,2}$	&656 & 5.6(-5) &		$<$0.20	&	$\sim$0.24 (0.12)*	&	$\sim$0.74 (0.30)*\\
240.952	&	5$_{2,2}$ -- 4$_{2,2}$	&621 & 7.5(-5) &		$<$0.20	&	$\sim$0.24 (0.12)*	&	$\sim$0.74 (0.30)*\\
240.959	&	5$_{-1,2}$ -- 4$_{-1,2}$  &567 & 8.5(-5) &  	     	$<$0.20	&	1.18 (0.42):		&	1.74 (0.61):\\ 
240.961	&	5$_{1,+1}$ -- 4$_{1,+1}$  &360 & 8.5(-5) &       		$<$0.20	&	1.18 (0.42):		&	1.74 (0.61):\\ 
241.043 	&	 $22_{ -6,   0} - 23_{ -5,   0}$ & 776 & 3.4(-5) & 		-- 		& $\sim$0.35 (0.48) 		&	$<$0.38 (0.54): \\
330.794	&	8$_{-3,0}$ -- 9$_{-2,0}$	&146 & 8.0(-5) &		$\sim$0.28 (0.57)	& 	3.23 (1.12)		&	8.06 (2.43)\\
331.220 	&	 $16_{ -1,   0} - 15_{ -2,   0}$ & 321 & 7.8(-5) & 		$<$0.39$^c$ 		& 	2.66 (0.79) 		&	 $>$3.19 (0.91)$^b$ \\ 
338.124	&	7$_{0,0}$ -- 6$_{0,0}$	& 78 & 2.5(-4) &		--				&	--				&	30.57 (11.11):: \\
338.345	&	7$_{-1,0}$ -- 6$_{-1,0}$	& 70 & 2.5(-4) &		7.29 (2.05)::		&	10.34 (2.58)::		&	38.03 (9.06)::\\
338.405	&	7$_{6,0}$ -- 6$_{6,0}$	&244 & 6.7(-5) &		$<$8.90 (2.54):		&	$<$12.95 (3.14):	&	$<$44.10 (10.87):\\
338.409	&	7$_{0,+0}$ -- 6$_{0,+0}$	& 65 & 2.5(-4) &		8.90 (2.54)::		&	12.95 (3.14)::		&	44.10 (10.87)::\\
338.431	&	7$_{-6,0}$ -- 6$_{-6,0}$	&254 & 6.7(-5) &		$\sim$0.72 (0.14)	&	$\sim$2.23 (1.06)	&	8.49 (2.90)\\ 
338.442	&	7$_{6,-0/+0}$ -- 6$_{6,-0/+0}$	&259 & 6.7(-5) &	$\sim$1.20 (0.79):	&	$\sim$2.29 (1.27)	&	8.80 (3.44)\\
338.457	&	7$_{-5,0}$ -- 6$_{-5,0}$	&189 & 1.2(-4) &		1.80 (0.66):		&	$\sim$2.55 (1.44)	&	10.12 (3.74)\\ 
338.475	&	7$_{5,0}$ -- 6$_{5,0}$	&201 & 1.2(-4) &		$\sim$1.43 (0.94):	&	$\sim$1.89 (1.21)	&	9.22 (3.46)\\
338.486	&	7$_{5,-0/+0}$ -- 6$_{5,-0/+0}$	&203 & 1.2(-4) &	$\sim$2.86 (0.94):	&	$>$3.40 (3.00)		&	$>$13.31 (5.28)\\
338.504	&	7$_{-4,0}$ -- 6$_{-4,0}$	&153 & 1.7(-4) &		$\sim$1.78 (0.85)	&	$\sim$3.39 (1.88)	&	$>$12.14 (4.99)\\
338.5126 &	7$_{4/4,-0/+0}$ -- 6$_{4/4,-0/+0}$ &145 & 1.7(-4) & 	$>$2.61 (0.94):		&	$>$5.87 (1.81):		&	$>$26.88 (9.59):\\
338.5129 &	7$_{2,-0}$ -- 6$_{2,-0}$	&103 & 2.3(-4) &		$>$2.61 (0.94):		&	$>$5.87 (1.81):		&	$>$26.88 (9.59):\\
338.530	&	7$_{4,0}$ -- 6$_{4,0}$	&161 & 1.7(-4) &		$\sim$1.34 (0.59)	&	$\sim$3.69 (2.22)	&	11.33 (4.40)\\
338.541	&	7$_{3,+0/-0}$ -- 6$_{3,+0/-0}$	&115 & 2.1(-4) &	$>$3.75 (1.24)		&	$>$8.71 (2.71)		&	$>$29.00 (8.01)\\
338.560	&	7$_{-3,0}$ -- 6$_{-3,0}$	&128 & 2.1(-4) &		$\sim$2.14 (0.95)	&	$>$4.50 (1.87)		&	14.94 (4.70)\\
338.583	&	7$_{3,0}$ -- 6$_{3,0}$	&113 & 2.1(-4) &		2.74 (1.17)		&	$>$4.02 (1.61)		&	18.35 (5.14)\\
338.615	&	7$_{1,0}$ -- 6$_{1,0}$	& 86 & 2.5(-4) &		4.48 (1.33)::		&	9.01 (2.57)::		&	28.49 (6.87)::\\
338.640	&	7$_{2,+0}$ -- 6$_{2,+0}$	&103 & 2.3(-4) &		$>$2.99 (1.07)		&	$>$6.46 (2.36)		&	21.12 (5.79)\\
338.722	&	7$_{2,0}$ -- 6$_{2,0}$	& 87 & 2.3(-4) &		2.60 (0.70)::		&	5.24 (1.36):: 		&	17.99 (8.33)::\\
338.723	&	7$_{-2,0}$ -- 6$_{-2,0}$	& 91 & 2.3(-4) &		2.60 (0.70)::		&	5.24 (1.36)::		&	17.99 (8.33)::\\
340.141	&	2$_{2,+0}$ -- 3$_{1,+0}$	& 45 & 4.1(-5) &		1.56 (0.71)::		&	3.31 (1.04)::		&	12.35 (2.83)::\\
342.730$^a$	&	13$_{1,12,-0}$ -- 13$_{0,13,+0}$ & 147 & 2.4(-5) & --			&	--				&	14.19 (3.54)\\
344.971	&	12$_{7,1}$ -- 11$_{6,1}$	&762 & 1.3(-4) &		$<$0.12			&	--				&	$\sim$2.27 (1.07)\\
364.159	&	9$_{3,1}$ -- 9$_{2,1}$	&522 & 6.5(-5) &		$\sim$0.20 (0.41)	&	2.97 (1.10):		&	2.66 (0.77)\\
\hline
\multicolumn{7}{l}{\cmethanol}\\
\hline
222.468 	&	21$_{ 1, 20, 0}$ -- 21$_{ 0, 21, 0}$ & 544 & 3.6(-5) & 	$<$0.23 & $<$0.32 & $<$0.59 \\ 
225.404	&	16$_{2,14,0}$ -- 15$_{3,12,0}$	&331 & 2.0(-5) &		$<$0.19	&	$<$0.45 (0.35):		&	$<$3.49 (1.55):\\
338.760	&	13$_{0,13,+0}$ -- 12$_{1,12,+0}$	&206 & 2.2(-4) &	$<$0.20	&	$<$0.39	& 			$\sim$5.55 (2.36) \\
341.132$^a$	&	13$_{1,12,-0}$ -- 13$_{0,13,+0}$ & 211 & 2.4(-5) & --		&	--			&	3.55 (2.22)\\
345.084	&	2$_{2,0,+0}$ -- 3$_{1,3,+0}$	& 45 & 2.9(-5) &		$<$0.29	&	$<$0.24	&	$<$1.69 (0.77):\\
345.133	&	4$_{0,4,0}$ -- 3$_{-1,3,0}$	& 36 & 8.2(-5) &		$<$0.29	&	1.81 (0.70)		&	4.06 (1.25)\\
354.446	&	4$_{1,3,0}$ -- 3$_{0,3,0}$	& 44 & 1.3(-4) &			$<$0.18	&	1.68 (0.65)		&	6.32 (1.59)\\
363.924	&	14$_{3,12,1}$ -- 14$_{2,13,1}$	&655 & 4.7(-5) &		$<$0.24	&	$<$2.06 (0.94):	&	$<$1.40 (0.92):\\
\hline
\end{tabular}

The notation $a(-b)$ stands for $a\times10^{-b}$\\
$^a$ from additional observations on G31 only\\
$^b$ line close to the edge of the frequency band\\
$^c$ used to constrain the RTD fit\\
$>$ means lower limit (optically thick line)\\
$<$ means upper limit\\
$\sim$ means uncertain detection, $S/N\lesssim2$\\
-- means frequency not observed\\
: means blended frequency\\
:: cold (\Eup$<$100 K) \methanol~line (treated as a blend with cold component)\\
~* calculated flux (see Eq.~\ref{eq:flux_assignment})
\label{fluxes_ch3oh}
\end{table*}

\begin{table*}
\caption{Observed line fluxes $\int T_{\rm MB}dV$ (K km s$^{-1}$) for \ethanol. }  
\begin{tabular}{l|lrc|ccc}
\hline
\hline
Frequency & Transition & $E_{\rm up}$ & $A$ & \multicolumn{3}{l}{Sources} \\
{[GHz]}     &            & {[K]} & {[s$^{-1}$]} & IRAS20126+4104 & IRAS18089-1732  & G31.41+0.31  \\
\hline
\multicolumn{7}{l}{\ethanol}\\
\hline
218.461	&  5$_{3,2,2}$  --  4$_{2,3,2}$	&  24 & 6.6(-5)&			--	&		--			&	$<$2.60 (0.52):\\
218.554	& 21$_{5,16,2}$ -- 21$_{4,17,2}$	& 226 & 6.2(-5)&		--	&		--			&	$\sim$0.77 (0.65)\\
222.217	& 20$_{5,15,2}$ -- 20$_{4,16,2}$	& 208 & 6.5(-5)&		$<$0.22	&	$\sim$0.56 (0.59)	&	$<$1.29 (0.26): \\
222.419	& 20$_{4,17,1}$ -- 19$_{5,15,0}$	& 256 & 1.1(-5)&		$<$0.22	&	$<$0.34 (0.30):	&	$<$1.18 (0.24): \\
222.519	& 26$_{4,23,2}$ -- 26$_{3,24,2}$	& 316 & 6.7(-5)&		$<$0.22	&	$<$0.27		&	$\sim$0.26 (0.24)\\
225.105	& 13$_{10,3/4,0}$ -- 12$_{10,2/3,0}$	& 255 & 4.2(-5)&	$<$0.19	&	$<$0.25		&	$\sim$0.66 (0.42)\\
225.108	& 13$_{11,2/3,0}$ -- 12$_{11,1/2,0}$	& 280 & 2.9(-5)&	$<$0.19	&	$<$0.25		&	$\sim$0.40 (0.42)\\
225.110	& 13$_{6,8/7,1}$ -- 12$_{6,7/6,1}$	& 181 & 8.0(-5)&		$<$0.19	&	$<$0.25		&	1.12 (0.42)\\
225.112	& 13$_{9,5/4,0}$ -- 12$_{9,4/3,0}$	& 231 & 5.3(-5)&		$<$0.19	&	$<$0.25		&	$\sim$0.74 (0.42)\\
225.116	&13$_{12,1/2,0}$ -- 12$_{12,0/1,0}$	& 308 & 1.5(-5)& 		$<$0.25	& 	$<$0.25 		& 	$<$0.55 (0.43): \\
225.131	& 13$_{8,5/6,0}$ -- 12$_{8,4/5,0}$	& 211 & 6.4(-5)&		$<$0.19	&	$<$0.25		&	$\sim$1.16 (1.12)\\
225.171	& 13$_{7,6/7,0}$ -- 12$_{7,5/6,0}$	& 192 & 7.3(-5)&		$<$0.19	&	$<$0.25		&	$\sim$1.86 (0.84)\\
225.210	& 19$_{5,14,2}$ -- 19$_{4,15,2}$	& 191 & 6.7(-5)&		$<$0.19	&	$<$0.25		&	$\sim$1.98 (1.05)\\
225.229	& 17$_{2,15,2}$ -- 16$_{3,14,2}$	& 137 & 2.5(-5)&		$<$0.19	&	$\sim$0.37 (0.24)	&	$<$2.76 (0.55):\\
225.249	& 13$_{6,8/7,0}$ -- 12$_{6,7/8,0}$	& 176 & 8.1(-5)&		$<$0.19	&	$\sim$0.40 (0.30)	&	$\sim$2.34 (0.93)\\
225.279	& 13$_{5,9,1}$ -- 12$_{5,8,1}$	& 168 & 8.7(-5)&			$<$0.19	&	$<$0.25		&	$\sim$1.61 (0.77)*\\
225.283	& 13$_{5,8,1}$ -- 12$_{5,7,1}$	& 168 & 8.7(-5)&			$<$0.19	&	$<$0.25		&	$\sim$1.61 (0.77)*\\
225.400	& 13$_{5,9,0}$ -- 12$_{5,8,0}$	& 163 & 8.7(-5)&			$<$0.19	&	$\sim$0.13 (0.15)	&	0.74 (0.32)*\\
225.404	& 13$_{5,8,0}$ -- 12$_{5,7,0}$	& 163 & 8.7(-5)&			$<$0.19	&	$\sim$0.13 (0.15)	&	0.74 (0.32)*\\
225.457	& 13$_{3,11,1}$ -- 12$_{3,10,1}$	& 148 & 9.7(-5)&		$<$0.19	&	$<$0.25		&	$\sim$0.95 (0.48)\\
238.841	& 21$_{2,19,0}$ -- 21$_{1,21,1}$	& 258 & 4.8(-5)&		--		&	$<$0.22		&	$<$2.28 (0.46):\\
239.020	& 28$_{4,25,2}$ -- 27$_{5,22,2}$	& 362 & 2.9(-5)&		$<$0.21	&	$<$0.22		&	$<$4.79 (0.96):\\
239.186	& 16$_{0,16,1}$ -- 15$_{1,14,0}$	& 171 & 4.4(-5)&		$<$0.21	&	$<$0.22		&	$<$0.92 (0.18):\\
240.654 	& 4$_{ 2, 2, 1}$ -- 3$_{ 1, 2, 0}$ &  75 & 4.1(-5)& 			-- 		& 	$<$0.22 			& 	-- \\
240.782	& 15$_{2,13,0}$ -- 14$_{1,13,1}$	& 163 & 2.9(-5)&		$<$0.2	&	$<$0.23		&	$<$3.38 (0.68):\\
240.839	& 14$_{1,13,0}$ -- 13$_{0,13,1}$	& 147 & 5.1(-5)&		$<$0.20	&	$\sim$0.20 (0.33)	&	$\sim$0.85 (0.52)\\
330.985	& 6$_{3,4,1}$ -- 5$_{2,4,0}$	&  90 & 1.2(-4)&			$<$0.31	&	--			&	$\sim$1.46 (0.84)\\
331.027	& 13$_{5,8,1}$ -- 13$_{4,10,0}$	& 168 & 1.1(-4)&	$<$0.29	&	$<$0.29		&	$<$1.84 (0.37):\\
331.079	& 43$_{ 7, 37, 2}$ -- 43$_{ 6, 38, 2}$& 860 & 2.2(-4)	& 		$<$0.29	&	$\sim$0.36 (0.39) 	& 	$\sim$0.15 (0.22) \\
331.095	& 27$_{7,20,2}$ -- 27$_{6,21,2}$	& 380 & 2.1(-4)&		$<$0.29	&	$<$0.29		&	$\sim$0.33 (0.55)\\
338.088 	& 25$_{ 1, 24, 1}$ -- 24$_{ 2, 22, 0}$ & 332 & 5.9(-5) & 		-- 		& 	-- 			& $\sim$3.64 (1.99) \\	
338.099 	& 18$_{ 7, 11, 2}$ -- 18$_{ 6, 12, 2}$ & 205 & 2.1(-4) & 		-- 		& 	-- 			& $\sim$2.05 (1.54)* \\	
338.110 	& 18$_{ 7, 12, 2}$ -- 18$_{ 6, 13, 2}$ & 205 & 2.1(-4) & 		-- 		& 	-- 			& $\sim$2.05 (1.54)* \\	
338.163 	& 10$_{ 2,  8, 1}$ --  9$_{ 1,  8, 0}$ & 113 & 8.6(-5) & 		-- 		& 	-- 			& 2.05 (0.98) \\	
338.412 	& 17$_{ 7, 10, 2}$ -- 17$_{ 6, 11, 2}$ & 190 & 2.1(-4) & 		$<$4.41 (1.31):  & $<$13.78 (3.29):	& $<$21.70 (5.13): \\	
338.417 	& 17$_{ 7, 11, 2}$ -- 17$_{ 6, 12, 2}$ & 190 & 2.1(-4) & 		$<$4.41 (1.31): 		& 	$<$0.71 			& $<$21.70 (5.13): \\	 
338.672 	& 16$_{ 7,  9, 2}$ -- 16$_{ 6, 10, 2}$ & 176 & 2.0(-4) & 		$<$0.32	& 	$<$0.83		& 1.63 (0.85)* \\ 
338.674 	& 16$_{ 7, 10, 2}$ -- 16$_{ 6, 11, 2}$ & 176 & 2.0(-4) & 		$<$0.32	& 	$<$0.83		& 1.63 (0.85)* \\
339.979	&	9$_{4,6,2}$ -- 8$_{3,5,2}$	&  58 & 2.2(-4)&			$<$0.15	&	$\sim$0.55 (0.47)	&	3.23 (0.93)\\
345.174	&	7$_{7,0/1,0}$ -- 6$_{6,0/1,1}$	& 140 & 2.5(-4)&		$<$0.26	&	$\sim$1.16 (0.70)	&	$\sim$2.28 (1.56)\\
345.229	&	21$_{1,21,0}$ -- 20$_{1,20,0}$	& 242 & 3.7(-4)&		$<$0.29	&	$\sim$0.58 (0.37)	&	$<$5.52 (1.10):\\
345.295	&	21$_{1,21,1}$ -- 20$_{1,20,1}$	& 246 & 3.7(-4)&		$<$0.29	&	$\sim$1.03 (1.00)	&	$\sim$2.91 (1.31)\\
345.312	&	22$_{3,19,1}$ -- 21$_{4,17,0}$	& 286 & 5.0(-5)&		$<$0.29	&	$<$0.19		&	$<$0.39\\
345.333	&	21$_{0,21,0}$ -- 20$_{0,20,0}$	& 242 & 3.7(-4)&			$<$0.29	&	$<$6.54 (1.31):	&	$<$17.35 (3.47): \\
345.408	&	21$_{0,21,1}$ -- 20$_{0,20,1}$	& 246 & 3.7(-4)&		$<$0.29	&	$\sim$0.75 (0.47)	&	$\sim$2.47 (1.13)\\
352.858	&	21$_{1,20,2}$ -- 20$_{2,19,2}$	& 196 & 1.2(-4)&		$<$0.18	&	$\sim$0.51 (0.43)	&	$\sim$1.56 (0.78)\\
353.034	&	12$_{3,9,2}$ -- 11$_{2,10,2}$	&  77 & 1.9(-4)&		$<$0.16	&	$\sim$0.61 (0.59)	&	3.12 (1.00)\\
354.758	&	20$_{3,17,1}$ -- 19$_{3,16,1}$	& 249 & 4.0(-4)&		$<$0.18	&	$\sim$0.59 (0.52)	&	3.96 (1.05)\\
363.968	& 21$_{7,15/14,1}$ -- 20$_{7,14/13,1}$	& 314 & 3.9(-4)&	$<$0.22	&	$\sim$0.64 (0.44)	&	2.66 (1.02)\\
364.001	& 21$_{8,14/13,0}$ -- 20$_{8,13/12,0}$	& 327 & 3.7(-4)&	$<$0.22	&	$\sim$0.70 (0.48)	&	2.28 (0.80)\\
364.233	& 21$_{7,15/14,0}$ -- 20$_{7,14/13,0}$	& 309 & 3.9(-4)&	$<$0.24	&	$\sim$0.64 (0.56)	&	$\sim$1.94 (1.48)\\
\hline
\end{tabular}

The notation $a(-b)$ stands for $a\times10^{-b}$\\
$<$ means upper limit\\
$\sim$ means uncertain detection, $S/N\lesssim2$\\
-- means frequency not observed\\
: means blended line\\
~* calculated flux (see Eq.~\ref{eq:flux_assignment})\\
\label{fluxes_c2h5oh}
\end{table*}

\begin{table*}
\caption{Observed line fluxes $\int T_{\rm MB}dV$ (K km s$^{-1}$) for
\isocyanicacid~and its isotopic species. }  
\begin{tabular}{l|lrc|ccc}
\hline
\hline
Frequency & Transition & $E_{\rm up}$ & $A$ & \multicolumn{3}{l}{Sources} \\
{[GHz]}     &            & {[K]} & {[s$^{-1}$]} & IRAS20126+4104 & IRAS18089-1732  & G31.41+0.31  \\
\hline
\multicolumn{7}{l}{\isocyanicacid}\\
\hline
219.657 & 10$_{3,8/7,9/11/10}$ -- 9$_{3,7/6,9/10/8}$ & 447 & 1.4(-4)& 		$<$1.33 & 		$<$0.73 (0.70): & 	$<$1.52 (0.63): \\ 
219.736 & 10$_{2,9/8,9/11/10}$ -- 9$_{2,8/7,9/10/8}$ & 231 & 1.4(-4)& 		$<$0.28 & 		$<$0.84 (0.75): & 	$\sim$0.64 (0.91) \\ 
219.798 & 10$_{0,10,9/11/10}$ -- 9$_{0,9,9/10/8}$ &  58 & 1.5(-4)& 		$<$0.28 & 		1.31 (0.50) & 	1.22 (0.55) \\
240.876 & 11$_{1,11,10/12/11}$ -- 10$_{1,10,10/11/9}$ & 113 & 2.0(-4)& 	$<$0.20 & 		$<$0.82 (0.54) & 	$<$3.10 (0.93): \\ 
330.849 & 15$_{1,14,14/16/15}$ -- 14$_{1,13,14/15/13}$ & 170 & 5.2(-4)& 	$\sim$0.32 (0.31) & 	$<$1.53 (0.71): & 	2.01 (0.71) \\ 
352.898 & 16$_{1,15,15/17/16}$ -- 15$_{1,14,15/16/14}$ & 187 & 6.3(-4)& 	$<$0.18 & 		$\sim$1.84 (1.00) & 	$<$3.12 (1.02): \\ 
\hline
\multicolumn{7}{l}{\isocyanicacidc}\\
\hline
219.664 & 10$_{3,8/7,10}$ -- 9$_{3,7/6,10}$ & 47 & 1.4(-4)& 				$<$1.32 & 	$<$0.21 & 	$<$0.41 \\
219.740 & 10$_{2,9,11/9/10}$ -- 9$_{2,8,10/8/9}$ & 31 & 1.4(-4)& 			$<$0.28 & 	$<$0.24 & 	$<$0.26 \\
219.744 & 10$_{2,8,11/9/10}$ -- 9$_{2,7,10/8/9}$ & 31 & 1.4(-4)& 			$<$0.28 & 	$<$0.24 & 	$<$0.26 \\
219.804 & 10$_{0,10,10/9/8}$ -- 9$_{0,9,10/9/8}$ & 58 & 1.5(-4)& 			$<$0.28 & 	$<$0.24 & 	$<$0.63 \\
240.881 & 11$_{11,11,12/11/10}$ -- 10$_{10,11,11/10/9}$ & 113 & 2.0(-4)& 	$<$0.20 & 	$<$0.23 & 	$<$0.98 \\
330.860 & 15$_{1,14,16/14/15}$ -- 14$_{1,13,15/13/14}$ & 170 & 5.2(-4)& 	$<$0.31 & 	$<$0.32 & 	$<$0.31 \\
\hline
\hline
\end{tabular}

The notation $a(-b)$ stands for $a\times10^{-b}$\\
$<$ means upper limit\\
$\sim$ means uncertain detection, $S/N\lesssim2$\\
-- means frequency not observed\\
: means blended line
\label{fluxes_hnco}
\end{table*}

\begin{table*}
\caption{Observed line fluxes $\int T_{\rm MB}dV$ (K km s$^{-1}$) for
\formamide. }  
\begin{tabular}{l|lrc|ccc}
\hline
\hline
Frequency & Transition & $E_{\rm up}$ & $A$ & \multicolumn{3}{l}{Sources} \\
{[GHz]}     &            & {[K]} & {[s$^{-1}$]} & IRAS20126+4104 & IRAS18089-1732  & G31.41+0.31  \\
\hline
\multicolumn{7}{l}{\formamide}\\
\hline
218.460	& 10$_{10,1,9}$ -- 9$_{9,1,8}$	& 61 & 7.5(-4)&				--			&	--			&	$<$2.45 (0.49):\\
339.716	& 16$_{16,8,8/9}$ -- 15$_{15,8,7/8}$	&329 & 2.2(-3)&	$<$0.15		&	$<$0.17		&	$<$4.04 (0.81): \\
339.781 	& 16$_{16,7,10/9}$ -- 15$_{15,7,9/8}$	&284 & 2.3(-3)&	$\sim$0.74 (0.42)	&	$\sim$0.29 (0.29)	&	$<$2.66 (1.20): \\ 
339.904 	& 16$_{16,6,11/10}$ -- 15$_{15,6,10/9}$	&246 & 2.5(-3)&	$<$0.89 (0.39):	&	$\sim$0.48 (0.30)	&	$\sim$1.41 (0.76)\\ 
340.135	& 16$_{16,5,12}$ -- 15$_{15,5,11}$	&213 & 2.6(-3)&		$<$0.78 (0.16):	&	$<$1.40 (0.28):	&	$<$1.11 (0.22): \\
340.139	& 16$_{16,5,11}$ -- 15$_{15,5,10}$	&213 & 2.6(-3)&		$<$0.78 (0.16):	&	$<$1.40 (0.28):	&	$<$1.11	(0.22): \\
345.183 	& 17$_{17,0,17}$ -- 16$_{16,0,16}$	&151 & 3.0(-3)&		$<$0.20		&	1.16 (0.53)	&	2.42 (0.94) \\
345.327	& 16$_{16,1,15}$ -- 15$_{15,1,14}$	&145 & 3.0(-3)&		$<$0.43			&	$<$1.87 (0.37):	&	$<$5.46 (1.09): \\
\hline
\end{tabular}

The notation $a(-b)$ stands for $a\times10^{-b}$\\
$<$ means upper limit\\
$\sim$ means uncertain detection, $S/N\lesssim2$\\
-- means frequency not observed\\
: means blended line
\label{fluxes_nh2cho}
\end{table*}

\begin{table*}
\caption{Observed line fluxes $\int T_{\rm MB}dV$ (K km s$^{-1}$) for
\acetonitrile~and its isotopic species. }  
\begin{tabular}{l|lrc|ccc}
\hline
\hline
Frequency & Transition & $E_{\rm up}$ & $A$ & \multicolumn{3}{l}{Sources} \\
{[GHz]}     &            & {[K]} & {[s$^{-1}$]} & IRAS20126+4104 & IRAS18089-1732  & G31.41+0.31  \\
\hline
\multicolumn{7}{l}{\acetonitrile}\\
\hline
238.844	&	13$_{8}$ -- 12$_{8}$	& 537 & 5.0(-4)&	--	&			$<$0.22	&	$<$0.17\\
238.913	&	13$_{7}$ -- 12$_{7}$	& 430 & 5.8(-4)&	$<$0.21	&		$\sim$0.37	 (0.27)	&	1.82 (0.62)\\
238.972	&	13$_{6}$ -- 12$_{6}$	& 337 & 6.4(-4)&	$\sim$0.33 (0.38)	&	$\sim$0.84 (0.37)	&	3.22 (1.11)\\
239.023	&	13$_{5}$ -- 12$_{5}$	& 259 & 6.9(-4)&	$\sim$0.93 (0.49)	&	$<$6.43 (1.29):	&	3.19 (0.88)\\
239.064	&	13$_{4}$ -- 12$_{4}$	& 195 & 7.4(-4)&	$\sim$0.51 (0.38)	&	1.17 (0.45)	&	5.25 (1.44)\\
239.096	&	13$_{3}$ -- 12$_{3}$	& 145 & 7.7(-4)&	1.04 (0.40)	&	1.64 (0.55)	&	5.21 (1.37)\\
239.120	&	13$_{2}$ -- 12$_{2}$	& 109 & 7.9(-4)&	$\sim$0.63 (0.38)	&	1.80 (0.61)	&	5.71 (1.37)\\
239.133	&	13$_{1}$ -- 12$_{1}$	&  87 & 8.1(-4)&	0.73 (0.26)*	&	1.72 (0.52)*	&	4.56 (1.19)\\
239.138	&	13$_{0}$ -- 12$_{0}$	&  80 & 8.1(-4)&	0.73 (0.26)*	&	1.72 (0.52)*	&	4.60 (1.15)\\
330.843	&	18$_{6}$ -- 17$_{6}$	& 408 & 1.9(-3)&	$\sim$0.31 (0.26)	&	4.63 (1.28)	&	7.44 (2.20)\\
330.913	&	18$_{5}$ -- 17$_{5}$	& 329 & 2.0(-3)&	--	&	3.59 (1.13)	&	5.91 (1.87)\\
330.970	&	18$_{4}$ -- 17$_{4}$	& 265 & 2.1(-3)&	--	&	2.81 (0.78)	&	6.70 (2.07)\\
331.014	&	18$_{3}$ -- 17$_{3}$	& 215 & 2.1(-3)&	--	&	4.66 (1.30)	&	8.26 (2.38)\\
331.046	&	18$_{2}$ -- 17$_{2}$	& 180 & 2.2(-3)&	--	&	4.51 (1.31)	&	10.24 (2.87)\\
331.065	&	18$_{1}$ -- 17$_{1}$	& 158 & 2.2(-3)&	--	&	4.01 (1.09)*	&	6.25 (1.96)\\
331.072	&	18$_{0}$ -- 17$_{0}$	& 151 & 2.2(-3)&	--	&	4.01 (1.09)*	&	7.86	(2.30)\\
\hline
\multicolumn{7}{l}{\acetonitrilec}\\
\hline
238.855	&	13$_{6}$ -- 12$_{6}$	& 337 & 9.2(-4)&	--	&		$<$0.22	&	$<$10.23 (2.58):\\
238.905	&	13$_{5}$ -- 12$_{5}$	& 259 & 1.0(-3)&	$<$0.21	&	$<$0.22	&	$<$1.00 (0.48):\\
238.946	&	13$_{4}$ -- 12$_{4}$	& 195 & 1.1(-3)&	$<$0.21	&	$<$0.22	&	2.21 (0.70)\\
238.978	&	13$_{3}$ -- 12$_{3}$	& 145 & 1.1(-3)&	$<$0.21	&	$\sim$0.35 (0.31)	&	1.51 (0.58)\\
239.001	&	13$_{2}$ -- 12$_{2}$	& 109 & 1.2(-3)&	$<$0.21	&	$<$0.22	&	$\sim$0.62 (0.35)\\
239.015	&	13$_{1}$ -- 12$_{1}$	&  87 & 1.2(-3)&	$<$0.21	&	$<$0.22	&	4.61 (0.97)*\\
239.020	&	13$_{0}$ -- 12$_{0}$	&  80 & 1.2(-3)&	$<$0.21	&	$<$0.22	&	4.61 (0.97)*\\
330.806	&	18$_{4}$ -- 17$_{4}$	& 265 & 3.0(-3)&	$<$0.31	&	$<$0.32	&	$<$8.08 (2.52):\\
330.851	&	18$_{3}$ -- 17$_{3}$	& 215 & 3.1(-3)&	$<$0.31	&	$<$2.66 (0.53):	&	$<$4.12 (1.23):\\
330.882	&	18$_{2}$ -- 17$_{2}$	& 179 & 3.1(-3)&	$<$0.31	&	$<$0.32	&	$\sim$0.69 (0.49)\\
330.901	&	18$_{1}$ -- 17$_{1}$	& 158 & 3.2(-3)&	$<$0.31	&	$<$1.49 (0.30):	&	2.37 (0.93)\\
330.908	&	18$_{0}$ -- 17$_{0}$	& 151 & 3.2(-3)&	$<$0.31	&	$<$1.49 (0.30):	&	3.48 (1.12)\\
\hline
\end{tabular}

The notation $a(-b)$ stands for $a\times10^{-b}$\\
$<$ means upper limit\\
$\sim$ means uncertain detection, $S/N\lesssim2$\\
-- means frequency not observed\\
: means blended line\\
~* calculated flux (see Eq.~\ref{eq:flux_assignment})\\
\label{fluxes_ch3cn}
\end{table*}

\begin{table*}
\caption{Observed line fluxes $\int T_{\rm MB}dV$ (K km s$^{-1}$) for
\propionitrile. }  
\begin{tabular}{l|lrc|ccc}
\hline
\hline
Frequency & Transition & $E_{\rm up}$ & $A$ & \multicolumn{3}{l}{Sources} \\
{[GHz]}     &            & {[K]} & {[s$^{-1}$]} & IRAS20126+4104 & IRAS18089-1732  & G31.41+0.31  \\
\hline
\multicolumn{7}{l}{\propionitrile}\\
\hline
218.390	&	24$_{3,21}$ -- 23$_{3,20}$	& 140 & 8.7(-4)&	$<$0.18	&	--		&	$\sim$1.03 (0.45)\\
219.903	&	12$_{3,10}$ -- 11$_{2,9}$	&  44 & 3.0(-5)&	$<$0.28	&	$<$0.24	&	$\sim$0.18 (0.13)\\
225.236	&	25$_{4,21}$ -- 24$_{4,20}$	& 158 & 9.4(-4)&	$<$0.00	&	$\sim$0.53 (0.36)	&	2.53 (0.85)\\
225.307	&	12$_{3,9}$ -- 11$_{2,10}$	&  44 & 3.2(-5)&	$<$0.19	&	$<$0.25	&	$<$1.46 (0.57):\\ 
225.317	&	23$_{2,22}$ -- 22$_{1,21}$	& 122 & 4.8(-5)&	$<$0.19	&	$<$0.25	&	$<$1.79 (0.65): \\ 
240.699	&	16$_{9,7/8}$ -- 17$_{8,10/9}$	& 148 & 7.9(-6)&	$<$0.20	&	$<$0.48 (0.10):	&	$<$0.89 (0.46): \\
240.861	&	28$_{1,28}$ -- 27$_{0,27}$	& 169 & 1.0(-4)&	$<$0.20	&	$<$0.27 (0.05):	&	$\sim$0.78 (0.37)\\
338.143	&	37$_{ 3,34}$ -- 36$_{ 3,33}$	& 317 &	3.3(-3)& 		-- & 			-- & 				5.74 (2.44) \\	
339.895	&	39$_{2,38}$ -- 38$_{2,37}$	& 334 & 3.3(-3)&	$<$0.15	&	$\sim$0.38 (0.68)	&	3.45 (1.05)\\
339.968	&	38$_{2,36}$ -- 37$_{2,35}$	& 327 & 3.3(-3)&	$<$0.15	&	$\sim$0.50 (0.32)	&	2.20 (0.69)\\
340.149	&	39$_{1,38}$ -- 38$_{1,37}$	& 334 & 3.3(-3)&	$<$0.15	&	$\sim$0.26 (0.17)	&	2.20 (0.82)\\ 
352.992 &	21$_{ 2,19}$ -- 20$_{ 1,20}$	& 105 & 1.4(-5)& 		$<$0.13 	& 	$<$0.24 &		$<$0.19 \\
353.089	&	23$_{3,20}$ -- 22$_{2,21}$	& 129 & 5.8(-5)&	$<$0.16	&	$<$0.24	&	$\sim$0.58 (0.29)\\
354.477	&	40$_{3,38}$ -- 39$_{3,37}$	& 361 & 3.8(-3)&	$<$0.18	&	$<$3.38 (0.68):	&	1.33 (0.65)\\
\hline
\end{tabular}

The notation $a(-b)$ stands for $a\times10^{-b}$\\
$<$ means upper limit\\
$\sim$ means uncertain detection, $S/N\lesssim2$\\
-- means frequency not observed\\
: means blended line\\
\label{fluxes_c2h5cn}
\end{table*}


\begin{table*}
\caption{Observed line fluxes $\int T_{\rm MB}dV$ (K km s$^{-1}$) for
\methylformate. }  
\begin{tabular}{l|lrc|ccc}
\hline
\hline
Frequency & Transition & $E_{\rm up}$ & $A$ & \multicolumn{3}{l}{Sources} \\
{[GHz]}     &            & {[K]} & {[s$^{-1}$]} & IRAS20126+4104 & IRAS18089-1732  & G31.41+0.31  \\
\hline
\multicolumn{7}{l}{\methylformate}\\
\hline
218.281	&	17$_{3,14,2}$ -- 16$_{3,13,2}$	& 100 & 1.5(-4)&		$<$0.18	&	--	&	1.38 (0.42)\\
218.298	&	17$_{3,14,0}$ -- 16$_{3,13,0}$	& 100 & 1.5(-4)&		$<$0.18	&	--	&	1.00 (0.430)\\
219.584	&	18$_{13,5/6,3}$ -- 17$_{13,4/5,3}$	&401 & 7.7(-5)&	$<$1.33	&	--	&	$\sim$0.17 (0.15)\\
219.592	&	28$_{9,19,2}$ -- 28$_{8,20,2}$	 & 295 & 1.6(-5)&		$<$1.33	&	--	&	$\sim$0.30 (0.22)\\
219.623	&	18$_{12,6/7,3}$ -- 17$_{12,5/6,3}$ & 384 & 8.9(-5) &	$<$1.33	&	--	&	$\sim$0.66 (0.32)\\ 
219.642	&	18$_{13,6,4}$ -- 17$_{13,5,4}$	& 401 & 7.7(-5) &		$<$1.33	&	$<$0.21	&	$\sim$0.55 (0.43)\\
219.696	&	18$_{11,8/7,3}$ -- 17$_{11,7/6,3}$	&369 & 1.0(-4) & 	$<$0.28	&	$<$0.21	&	0.69 (0.25)\\ 
219.705	&	18$_{4,15,3}$ -- 17$_{4,14,3}$	&300 & 1.5(-4)&		$<$0.28	&	$<$0.21	&	0.77 (0.27)\\ 
219.764	&	18$_{9,9,5}$ -- 17$_{9,8,5}$	& 342 & 1.2(-4)&		$<$0.28	&	$<$0.21	&	0.61 (0.24)\\
219.822	&	18$_{10,9/8,3}$ -- 17$_{10,8/7,3}$	& 355 & 1.1(-4)&	$<$0.28	&	$<$0.21	&	$<$1.10 (0.55):\\ 
222.149	&	18$_{6,12,3}$ -- 17$_{6,11,3}$	& 312 & 1.5(-4)&		$<$0.22	&	1.25 (0.25):	&	$<$3.05 (0.83):\\ 
222.177 	& 	18$_{ 6, 12, 5}$ -- 17$_{ 6, 11, 5}$  & 312 & 1.5(-4) & 	$<$0.18 	& 	$<$0.20 &	$\sim$1.14 (0.67) \\
222.421	&	18$_{8,10,2}$ -- 17$_{8,9,2}$ 	& 144 & 1.3(-4)&		$<$0.22	&	$<$1.32 (0.26):	&	1.62 (0.45)\\
222.438	&	18$_{8,11,0}$ -- 17$_{8,10,0}$	& 144 & 1.3(-4)&  		$<$0.22	&	0.35 (0.16)*	&	1.07 (0.39)*\\
222.440	&	18$_{8,10,0}$ -- 17$_{8,9,0}$	& 144 & 1.3(-4)&  		$<$0.22	&	0.35 (0.16)*	&	1.07 (0.39)*\\
222.442	&	18$_{8,11,1}$ -- 17$_{8,10,1}$	& 144 & 1.3(-4)&  		$<$0.22	&	0.35 (0.16)*	&	1.07 (0.39)*\\
225.372	&	20$_{21,9,3}$ -- 19$_{21,8,3}$	& 307 & 1.7(-4)&		$<$0.19	&	$<$0.31	&	1.38 (0.56)\\
225.449	&	20$_{11,9,3}$ -- 19$_{11,8,3}$	& 307 & 1.7(-4)&		$<$0.19	&	$<$0.18	&	$<$1.30 (0.55):\\ 
238.927	&	20$_{3,18,1}$ -- 19$_{2,17,2}$	& 128 & 2.1(-5)&		$<$0.21	&	$<$0.22	&	$\sim$0.40 (0.23)\\
238.933	&	20$_{3,18,0}$ -- 19$_{2,17,0}$	&128 & 2.1(-5)&		$<$0.21	&	$<$0.22	&	$\sim$0.47 (0.20)\\
238.947	&	19$_{3,16,5}$ -- 18$_{3,15,5}$	& 309 & 2.0(-4)&		$<$0.21	&	$\sim$0.29 (0.26)	&	$<$0.88 (0.38):\\ 
239.111	&	7$_{6,2,4}$ -- 6$_{5,2,4}$	& 227 & 2.7(-5)&			$<$0.21	&	$<$0.22	&	$<$0.15 (0.17)\\
241.059	&	30$_{3,27,0}$ -- 30$_{3,28,0}$		& 281 & 5.8(-6)&	--		&	$<$0.23	&	$\sim$0.33 (0.31)\\
241.068	&	30$_{3,27,0}$ -- 30$_{2,28,0}$		& 281 & 1.1(-5)&	--		&	$<$0.23	&	$\sim$0.76 (0.53)\\
330.941 	&	30$_{2/1,29,4/5}$ -- 29$_{2/1,28,4/5}$	& 443 & 5.5(-4)&	$<$0.31	&	$\sim$0.73 (0.38)	&	$\sim$0.75 (0.83)\\
331.021	&	25$_{5,21,0}$ -- 24$_{4,20,0}$		& 210 & 3.5(-5)&	$<$0.34	&	$<$0.40	& $<$8.29 (1.66): \\ 
331.036	&	25$_{5,21,1}$ -- 24$_{4,20,2}$		& 210 & 3.5(-5)&	$<$0.32	&		$<$0.34		& $<$10.01 (3.05): \\	
331.121	&	27$_{12,16,4}$ -- 26$_{12,15,4}$	&505 & 4.5(-4)&	$<$0.29	&	--	&	$<$0.74 (0.69)\\
331.149	&	28$_{4,25,1}$ -- 27$_{4,24,1}$		& 248 & 5.3(-4)&	$<$0.29	&	0.69 (0.31)	&	2.60 (1.39)\\
331.160	&	28$_{4,25,0}$ -- 27$_{4,24,0}$	& 248 & 5.3(-4)&		$<$0.29			&	$<$0.99 (0.42):	&	$<$2.24 (1.15): \\ 
331.161	&	27$_{11,17/16,3}$ -- 26$_{11,16/15,3}$	& 490 & 4.6(-4)&		$<$0.29	&	$<$0.99 (0.42):	&	$<$2.24 (1.15): \\ 
338.338	&  	27$_{ 8, 19, 2 }$ --  26$_{ 8, 18, 2}$ & 267 & 5.4(-4) & 	$<$0.26 	& 	$<$0.39 & 	6.26 (2.37) \\ 
338.356	&  	27$_{ 8, 19, 0 }$ --  26$_{ 8, 18, 0}$ & 267 & 5.4(-4) & 	$<$0.23 	& 	$<$0.44 & 	5.09 (2.08) \\ 	
338.393 	&  	28$_{ 5, 24, 3 }$ --  27$_{ 5, 23, 3}$ & 443 & 5.7(-4) & 	$<$0.23 	& 	$<$0.43 & 	$<$3.80 (2.05): \\
338.396 	&  	27$_{ 7, 21, 1 }$ --  26$_{ 7, 20, 1}$ & 258 & 5.5(-4) & 	$<$0.23 	& 	$<$0.45 & 	$<$3.80 (2.05): \\	
338.414 	&  	27$_{ 7, 21, 0 }$ --  26$_{ 7, 20, 0}$ & 258 & 5.5(-4) & 	$<$8.06 (2.04):	& 	$<$0.45 & 	$<$43.32 (9.59): \\
339.882	&	29$_{3,26,3}$ -- 28$_{3,25,3}$	& 450 & 5.8(-4)&		$<$0.22				&	$<$0.79 (0.16):	&	$\sim$0.73 (0.33)\\
340.044	&	29$_{4,26,4}$ -- 28$_{4,25,4}$	& 450 & 5.8(-4)&		$<$0.15			&	$<$0.12	&	$<$3.75 (0.75): \\
340.115	&	27$_{7,20,5}$ -- 26$_{7,19,5}$	& 444 & 5.5(-4)&		$<$0.15			&	$\sim$0.30 (0.31):	&	0.77 (0.32)\\ 
345.068	&	28$_{14,14,2}$ -- 27$_{14,13,2}$ & 370 & 4.7(-4)&		$<$0.26		&	0.28 (0.11)	&	$<$3.93 (1.62):\\ 
345.069	&	28$_{14,15/14,0}$ -- 27$_{14,14/13,0}$	& 370 & 4.7(-4)&		$<$0.26	&	0.56 (0.22)	&	$<$3.93 (1.62):\\ 
345.073	&	16$_{6,11,0}$ -- 15$_{5,10,0}$	& 104 & 3.9(-5)&		$<$0.26			&	$<$0.19	&	$<$5.62 (1.29):\\
345.085	&	19$_{13,6,2}$ -- 19$_{12,7,2}$		& 224 & 3.5(-5)&		--			&	$\sim$0.37 (0.20)	&	$<$1.73 (0.35):	\\
345.091	&	28$_{14,15,1}$ -- 27$_{14,14,1}$	& 370 & 4.7(-4)&		$<$0.26		&	$<$0.50 (0.32):	&	$<$4.14 (1.70):\\ 
345.148	&	28$_{6,23,3}$ -- 27$_{6,22,3}$	& 452 & 6.0(-4)&		$<$0.26	&	$\sim$0.55 (0.34)	&	$\sim$0.59 (0.54)\\
345.163	&	11$_{8,3,5}$ -- 10$_{7,3,5}$	& 269 & 7.1(-5)&		$<$0.26	&	$<$0.19			&	$\sim$0.50 (0.47)\\
345.230	& 	18$_{ 13, 5, 2}$ -- 18$_{ 12, 6, 2}$	& 213 & 3.2(-5)	& 	$<$0.25 	& 	$\sim$0.59 (0.40) 	& 	$<$8.18 (3.04): \\
345.242	& 	18$_{ 13, 6, 1}$ -- 18$_{ 12, 7, 1}$	& 213 & 3.2(-5)	& 	$<$0.25 	& 	$<$0.85 (0.83): 	& 	$<$8.18 (3.04): \\
345.248	&	28$_{10,19,4}$ -- 27$_{10,18,4}$	& 493 & 5.5(-4)&		$<$0.26		&	$<$0.19	&	$<$0.34 \\
352.817	&	33$_{0/1/0/1,33,3}$ -- 32$_{1/1/0/0,32,3}$ & 479 & 1.1(-4)	&	$<$0.18		&	$\sim$0.38 (0.28)	&	$<$2.68 (0.76):\\ 
352.841	&	33$_{1/0,33,4/5}$ -- 32$_{1/0,32,4/5}$	& 479 & 7.7(-4)&		$<$0.18	&	$<$0.50 (0.42):	&	1.72 (0.47)\\ 
352.912	&	31$_{2,29,2}$ -- 30$_{3,28,1}$	& 286 & 8.5(-5)&		$<$0.18			&	$<$0.26	&	0.30 (0.08)\\
352.918	&	31$_{3,29,1}$ -- 30$_{3,28,1}$	& 286 & 6.6(-4)&		$<$0.18			&	$<$1.10 (0.36):	&	2.32 (0.62)*\\ 
352.922	&	31$_{2,29,2}$ -- 30$_{2,28,2}$	& 286 & 6.6(-4)&		$<$0.18			&	$<$1.10 (0.36):	&	2.32 (0.62)*\\
352.926	&	31$_{3,29,0}$ -- 30$_{3,28,0}$	& 286 & 6.6(-4)&		$<$0.18			&	$<$1.10 (0.36):	&	2.32 (0.62)*\\
352.930	&	31$_{2,29,0}$ -- 30$_{2,28,0}$	& 286 & 6.6(-4)&		$<$0.18			&	$<$1.10 (0.36):	&	2.32 (0.62)*\\
\hline
\end{tabular}

The notation $a(-b)$ stands for $a\times10^{-b}$\\
$<$ means upper limit\\
$\sim$ means uncertain detection, $S/N\lesssim2$\\
-- means frequency not observed\\
: means blended line\\
~* calculated flux (see Eq.~\ref{eq:flux_assignment})\\
\label{fluxes_ch3ocho_1}
\end{table*}

\begin{table*}
\caption{Observed line fluxes $\int T_{\rm MB}dV$ (K km s$^{-1}$) for
\methylformate~continued. }  
\begin{tabular}{l|lrc|ccc}
\hline
\hline
Frequency & Transition & $E_{\rm up}$ & $A$ & \multicolumn{3}{l}{Sources} \\
{[GHz]}     &            & {[K]} & {[s$^{-1}$]} & IRAS20126+4104 & IRAS18089-1732  & G31.41+0.31  \\
\hline
\multicolumn{7}{l}{\methylformate}\\
\hline
354.427	&	29$_{16,13/14,3}$ -- 28$_{16,12/13,3}$	& 614 & 4.7(-4)  	&	$<$0.18	&	$<$0.23	&	$\sim$1.04 (0.64)\\
354.477 	&	29$_{ 15, 15, 4}$ -- 28$_{ 15, 14, 4}$ & 593 & 5.0(-4) 		&	$<$0.13	&	$<$0.23	&	$<$0.40 (0.58): \\
354.574	&	12$_{8,5/4,3}$ -- 11$_{7,4/5,3}$	& 276 & 7.4(-5)			&	$<$0.18	&	$<$0.23	&	$<$1.44 (0.78):\\ 
354.605	&	29$_{15,15/14,3}$ -- 28$_{15,14/13,3}$	& 593 & 5.0(-4)		&	$<$0.18	&	$<$0.23	&	1.96 (0.46)\\
354.608	&	33$_{1/0,33,1/2}$ -- 32$_{1/0,32,1/2}$ & 293 & 7.2(-4)		& 	$<$0.18 	&	1.11 (0.42)* &	3.14 (0.49)*\\
354.608	&	33$_{0/1,33,2/1}$ -- 32$_{1/0,32,1/2}$ & 293 & 6.8(-4)		& 	$<$0.18	&	1.05 (0.40)* &	2.75 (0.47)*\\
354.629	&	28$_{7,21,5}$ -- 27$_{7,20,5}$	& 461 & 6.4(-4)				&	$<$0.18	&	$<$0.23	&	$\sim$1.20 (0.68)\\
354.742	&	12$_{8,5,1}$ -- 11$_{7,5,1}$	&  88 & 7.3(-5)				&	$<$0.18	&	$<$0.23	&	1.19 (0.44)\\
354.759	&	12$_{8,4,2}$ -- 11$_{7,4,2}$		&  88 & 7.3(-5)			& 	$<$0.16 	& 	$<$0.17 	& 	$<$3.23 (0.65): \\
354.806	&	12$_{8,5/4,0}$ -- 11$_{7,4/5,0}$	&  88 & 7.3(-5)			&	$<$0.18	&	$<$0.23	&	3.90 (1.18)\\ 
354.839 	&	29$_{ 14, 15/16, 3}$ -- 28$_{ 14, 14/15, 3}$ & 574 & 5.2(-4) 	&	$<$0.15	&	$<$0.36 (0.46):	&	$<$4.91 (1.40): \\
364.297	&	33$_{1/2,32,2/1}$ -- 32$_{1,31,2}$	& 308 & 7.3(-4)			&	$<$0.24 	& 	$<$6.81 (1.36):	&	$<$3.67 (1.16):\\
364.302	&	33$_{2/1,32,0}$ -- 32$_{2/1,31,0}$	& 308 & 7.3(-4)			&	0.80 (0.91) &	$<$6.81 (1.36):	&	$<$5.00 (1.42):\\
\hline
\end{tabular}

The notation $a(-b)$ stands for $a\times10^{-b}$\\
$<$ means upper limit\\
$\sim$ means uncertain detection, $S/N\lesssim2$\\
-- means frequency not observed\\
: means blended line\\
~* calculated flux (see Eq.~\ref{eq:flux_assignment})\\
\label{fluxes_ch3ocho_2}
\end{table*}

\begin{table*}
\caption{Observed line fluxes $\int T_{\rm MB}dV$ (K km s$^{-1}$) for
\methylether. }  
\begin{tabular}{l|lrc|ccc}
\hline
\hline
Frequency & Transition & $E_{\rm up}$ & $A$ & \multicolumn{3}{l}{Sources} \\
{[GHz]}     &            & {[K]} & {[s$^{-1}$]} & IRAS20126+4104 & IRAS18089-1732  & G31.41+0.31  \\
\hline
\multicolumn{7}{l}{\methylether}\\
\hline
218.490	&	23$_{3,21,2/3}$ -- 23$_{2,22,2/3}$& 264 & 3.4(-5)&		$<$0.18	&	--	&	$<$1.19 (0.75):\\
218.492	&	23$_{3,21,1}$ -- 23$_{2,22,1}$	& 264 & 3.4(-5)&		$<$0.18	&	--	&	$<$1.19 (0.75):\\
218.495	&	23$_{3,21,0}$ -- 23$_{2,22,0}$	& 264 & 3.4(-5)&		$<$0.18	&	--	&	$<$1.19 (0.75):\\
222.239	&	4$_{3,2,2}$ -- 3$_{2,1,2}$	& 22 & 2.8(-5)&			$<$0.22	&	0.82 (0.56)	&	$\sim$0.93 (0.43)\\
222.248	&	4$_{3,2,3/1}$ -- 3$_{2,1,3/1}$	& 22 & 4.2(-5)&			$<$0.22	&	$\sim$0.43 (0.20)	&	$<$1.15 (0.35):\\ 
222.255	&	4$_{3,2,0}$ -- 3$_{2,1,0}$	& 22 & 4.9(-5)&				$<$0.22	&	$\sim$0.62 (0.21)	&	1.11 (0.98)\\
222.259	&	4$_{3,1,1}$ -- 3$_{2,1,1}$	& 22 & 1.6(-5)&				$<$0.22	&	$\sim$0.30 (0.21)*	&	2.37 (0.74)*\\
222.260	&	4$_{3,1,2}$ -- 3$_{2,1,2}$	& 22 & 2.1(-5)&				$<$0.22	&	$\sim$0.11 (0.21)*	&	0.83 (0.26)*\\
222.326	&	25$_{3,23,0}$ -- 24$_{4,20,0}$	& 308 &7.9(-6)&		$<$0.22	&	$<$0.27	&	$\sim$0.42 (0.19)*\\
222.327	&	25$_{3,22,1}$ -- 24$_{4,20,1}$	& 308 &7.9(-6)&		$<$0.22	&	$<$0.27	&	$\sim$1.12 (0.52)*\\
222.329	&	25$_{3,22/23,2/3}$ -- 24$_{4,20/20,2/3}$	& 308 &7.9(-6)&	$<$0.22	&	$<$0.27	&	$\sim$0.42 (0.21)*\\
222.414	&	4$_{3,2,2}$ -- 3$_{2,2,2}$	& 22 & 2.1(-5)&					$<$0.22	&	$\sim$0.20 (0.10)	&	$<$0.73 (0.15):\\
222.423	&	4$_{3,2,1}$ -- 3$_{2,2,1}$	& 22 & 1.6(-5)&				$<$0.22	&	$\sim$0.56 (0.30)	&	$<$2.09 (0.42):\\
222.427	&	4$_{3,1,3}$ -- 3$_{2,2,3}$	& 22 & 4.9(-5)&					$<$0.22	&	$\sim$0.23 (0.12)	&	$<$0.85 (0.11):\\
222.434	&	4$_{3,1,1/0}$ -- 3$_{2,2,1/0}$	& 22 & 4.2(-5)&			$<$0.22	&	1.68 (0.63)	&	$<$1.86 (0.52):\\ 
222.435	&	4$_{3,1,2}$ -- 3$_{2,2,2}$	& 22 & 2.8(-5)&					$<$0.22	&	$\sim$0.22 (0.08)	&	$<$0.34 (0.10):\\
225.202	&	24$_{4,21,2/3}$ -- 24$_{3,22,2/3}$	& 296 & 4.7(-5)&		$<$0.19	&	$<$0.30		&	0.55 (0.18)*\\
225.204	&	24$_{4,21,1}$ -- 24$_{3,22,1}$	& 296 & 4.7(-5)&			$<$0.19	&	$\sim$0.15 (0.14)*	&	0.87 (0.28)*\\
225.205	&	24$_{4,21,0}$ -- 24$_{3,22,0}$	& 296 & 4.7(-5)&			$<$0.19	&	$\sim$0.09 (0.08)*	&	0.55 (0.18)*\\
238.975	&	29$_{5,25,2/3/1/0}$ -- 28$_{6,22,2/3/1/0}$ & 432 & 1.6(-5)&	$<$0.21	&	$<$0.22	&	$<$1.64 (0.64):\\
239.020	&	24$_{5,19,5/3}$ -- 24$_{4,20,5/3}$	& 309 & 6.1(-5)&		$<$0.21	&	$<$0.63 (0.12):	&	$<$0.68 (0.14):\\
239.021	&	24$_{5,19,1}$ -- 24$_{4,20,1}$	& 309 & 6.1(-5)&			$<$0.21	&	$<$1.67 (0.33):	&	$<$1.82 (0.36):\\
239.021	&	24$_{5,19,0}$ -- 24$_{4,20,0}$	& 309 & 6.1(-5)&			$<$0.21	&	$<$0.63 (0.13):	&	$<$0.68 (0.14):\\
240.978	&	5$_{3,3,2}$ -- 4$_{2,2,2}$	&  26 & 4.5(-5)&				$<$0.20	&	$\sim$0.11 (0.07)*	&	2.16 (0.66)*\\
240.983	&	5$_{3,3,3}$ -- 4$_{2,2,3}$	&  26 & 5.4(-5)&				$<$0.20	&	$\sim$0.06 (0.04)*	&	0.19 (0.06)*\\
240.985	&	5$_{3,3,1}$ -- 4$_{2,2,1}$	&  26 & 5.1(-5)&				$<$0.20	&	$\sim$0.49 (0.30)*	&	1.49 (0.46)*\\
240.990	&	5$_{3,3,0}$ -- 4$_{2,2,0}$	&  26 & 5.4(-5)&				$<$0.20	&	$\sim$0.19 (0.12)	&	$<$0.64 (0.28):\\ 
\hline
\end{tabular}

The notation $a(-b)$ stands for $a\times10^{-b}$\\
$<$ means upper limit\\
$\sim$ means uncertain detection, $S/N\lesssim2$\\
-- means frequency not observed\\
: means blended line\\
~* calculated flux (see Eq.~\ref{eq:flux_assignment})\\
\label{fluxes_ch3och3}
\end{table*}

\begin{table*}
\caption{Observed line fluxes $\int T_{\rm MB}dV$ (K km s$^{-1}$) for
\ketene. }  
\begin{tabular}{l|lrc|ccc}
\hline
\hline
Frequency & Transition & $E_{\rm up}$ & $A$ & \multicolumn{3}{l}{Sources} \\
{[GHz]}     &            & {[K]} & {[s$^{-1}$]} & IRAS20126+4104 & IRAS18089-1732  & G31.41+0.31  \\
\hline
\multicolumn{7}{l}{\ketene}\\
\hline
222.120	&	11$_{4,8/7}$ -- 10$_{4,7/6}$	& 273 & 1.1(-4)&	$<$0.22	&	$<$0.27	&	$<$0.54 (0.10):\\
222.198	&	11$_{0,11}$ -- 10$_{0,10}$	&  65 & 1.2(-4)&	$<$0.22	&	$<$0.81 (0.26):	&	$<$2.34 (0.75):\\
222.201	&	11$_{3,9/8}$ -- 10$_{3,8/7}$	& 181 & 1.1(-4)&	$<$0.22	&	$<$0.81 (0.26):	&	$<$2.34 (0.75):\\
222.229	&	11$_{2,10}$ -- 10$_{2,9}$	& 116 & 1.2(-4)&	$<$0.22	&	$<$0.28 (0.06):	&	$<$2.22 (0.32):\\
222.315	&	11$_{2,9}$ -- 10$_{2,8}$		& 116 & 1.2(-4)&	$<$0.22	&	$\sim$0.24 (0.22)	&	$\sim$0.91 (0.57)\\
363.937!	&	18$_{2,16}$ -- 17$_{2,15}$	& 218 & 5.5(-4)	&	$<$4.65 (1.08):	&	$<$0.22	&	$<$20.29 (3.31):\\ 
\hline
\end{tabular}

The notation $a(-b)$ stands for $a\times10^{-b}$\\
$<$ means upper limit\\
$\sim$ means uncertain detection, S/N $\lesssim$2\\
-- means frequency not observed\\
: means blended line\\
! frequency from CDMS
\label{fluxes_ch2co}
\end{table*}

\begin{table*}
\caption{Observed line fluxes $\int T_{\rm MB}dV$ (K km s$^{-1}$) for
\acetaldehyde. }  
\begin{tabular}{l|lrc|ccc}
\hline
\hline
Frequency & Transition & $E_{\rm up}$ & $A$ & \multicolumn{3}{l}{Sources} \\
{[GHz]}     &            & {[K]} & {[s$^{-1}$]} & IRAS20126+4104 & IRAS18089-1732  & G31.41+0.31  \\
\hline
\multicolumn{7}{l}{\acetaldehyde}\\
\hline
330.822	&	5$_{3,3,1}$ -- 4$_{2,3,1}$	&  34 & 1.2(-4)&		$<$0.31			&	$\sim$0.71 (0.80)	&	$\sim$0.97 (0.52)	\\
331.039	&	17$_{3,14,3}$ -- 16$_{3,13,3}$	& 367 & 1.2(-3)&	$<$1.18 (0.88):		&	$<$0.42 (0.29):		&	$<$2.50 (1.47):		\\
354.458	&	18$_{2,16,3}$ -- 17$_{2,15,3}$	& 375 & 1.5(-3)&	$<$0.18			&	$<$3.11 (1.11):		&	$<$3.25 (1.26):		\\
354.525	&	19$_{0,19,5}$ -- 18$_{0,18,5}$	& 377 & 1.5(-3)&	$<$0.17			&	$<$0.19			&	$<$0.34	\\ 
354.813	&	18$_{2,16,2}$ -- 17$_{2,15,2}$	& 170 & 1.5(-3)&	$<$0.18			&	$<$0.23			&	$\sim$1.54 (0.92)	\\
354.844	&	18$_{2,16,0}$ -- 17$_{2,15,0}$	& 170 & 1.5(-3)&	$<$0.18:			&	$<$1.13 (0.55):		&	$<$4.92 (1.34):		\\
\hline
\end{tabular}

The notation $a(-b)$ stands for $a\times10^{-b}$\\
$<$ means upper limit\\
$\sim$ means uncertain detection, $S/N\lesssim2$\\
-- means frequency not observed\\
: means blended line
\label{fluxes_ch3cho}
\end{table*}

\begin{table*}
\caption{Observed line fluxes $\int T_{\rm MB}dV$ (K km s$^{-1}$) for
\formicacid~and its isotopic species. }  
\begin{tabular}{l|lrc|ccc}
\hline
\hline
Frequency & Transition & $E_{\rm up}$ & $A$ & \multicolumn{3}{l}{Sources} \\
{[GHz]}     &            & {[K]} & {[s$^{-1}$]} & IRAS20126+4104 & IRAS18089-1732  & G31.41+0.31  \\
\hline
\multicolumn{7}{l}{\formicacid}\\
\hline
222.110	&	7$_{2,6}$ -- 7$_{1,7}$	& 43 & 1.9(-6)	&	$<$0.22	&	$<$0.27			&	$<$0.24			\\
225.086 & 10$_{ 4, 7}$ --  9$_{ 4, 6}$ &	110 & 1.0(-4)	&	$<$0.18	&	$<$0.18	&	$<$1.97 (1.70): \\
225.091 & 10$_{ 4, 6}$ --  9$_{ 4, 5}$ &	110 & 1.0(-4)	&	$<$0.18	&	$<$0.18	&	$<$1.97 (1.70): \\
225.238	&	10$_{3,8}$ -- 9$_{3,7}$	& 88 & 1.1(-4)&		$<$0.19	&	$<$0.39 (0.24):		&	$<$2.59 (0.78):		\\
330.931	&	4$_{3,1}$ -- 4$_{2,2}$	& 39 & 5.1(-6)&		$<$0.31	&	$<$0.32			&	$\sim$0.47 (0.63)	\\
331.145	&	3$_{3,0}$ -- 3$_{2,1}$	& 35 & 3.6(-6)	&	$<$0.29	&	$<$0.29:			&	$<$2.77 (1.80):		\\
338.109 & 15$_{ 1, 15}$ -- 14$_{ 0, 14}$ & 127 & 1.1(-5) &	--	&	--	&	$<$5.11 (3.10): \\
338.202 & 15$_{ 3, 13}$ -- 14$_{ 3, 12}$ & 158 & 4.1(-4) &	--	&	--	&	$<$13.55 (6.28): \\
345.031	&	16$_{0,16}$ -- 15$_{0,15}$	& 143 & 4.5(-4)&	$<$0.26	&	$\sim$0.30 (0.30) 	&	$\sim$1.18 (0.70)	\\
345.253	&	14$_{3,12}$ -- 14$_{2,13}$	& 142 & 8.3(-6)&	$<$0.29	&	$<$0.19:			&	$<$1.59 (1.01):		\\
354.448	&	17$_{0,17}$ -- 16$_{1,16}$	& 161 & 1.3(-5)&	$<$0.18	&	$<$1.70 (0.57):		&	$<$6.42 (1.60):		\\
\hline
\end{tabular}

The notation $a(-b)$ stands for $a\times10^{-b}$\\
$<$ means upper limit\\
$\sim$ means uncertain detection, S/N $\lesssim$2\\
-- means frequency not observed\\
: means blended line
\label{fluxes_hcooh}
\end{table*}


\begin{table*}
\caption{Observed line fluxes $\int T_{\rm MB}dV$ (K km s$^{-1}$) for \methylacetylene.}  
\begin{tabular}{l|lrc|ccc}
\hline
\hline
Frequency & Transition & $E_{\rm up}$ & $A$ & \multicolumn{3}{l}{Sources} \\
{[GHz]}     &            & {[K]} & {[s$^{-1}$]} & IRAS20126+4104 & IRAS18089-1732  & G31.41+0.31  \\
\hline
\multicolumn{7}{l}{\methylacetylene}\\
\hline
222.099	&	13$_{4}$ -- 12$_{4}$	& 190 & 3.4(-5)&	$<$0.22		&	$<$0.65 (0.35):			&	$<$2.79 (0.56):\\
222.129	&	13$_{3}$ -- 12$_{3}$	& 140 & 3.6(-5)&	$\sim$0.29 (0.17)	&	1.14 (0.43)	&	3.15 (0.94)\\
222.150	&	13$_{2}$ -- 12$_{2}$	& 104 & 3.7(-5)&	$\sim$0.41 (0.19)	&	1.25 (0.46)	&	$<$4.25 (0.85):\\
222.163	&	13$_{1}$ -- 12$_{1}$	&  82 & 3.8(-5)&	0.74 (0.29)	&	2.20 (0.68)		&	3.14 (1.01)\\
222.167	&	13$_{0}$ -- 12$_{0}$	&  75 & 3.8(-5)&	0.93 (0.35)	&	2.13 (0.70)		&	3.27 (0.97)\\
239.088	&	14$_{6}$ -- 13$_{6}$	& 346 & 3.9(-5)&	$<$0.21		&	$<$0.22 (0.05):		&	$<$2.40 (0.48):\\
239.138	&	14$_{5}$ -- 13$_{5}$	& 267 & 4.1(-5)&	$<$0.21		&	$<$2.38 (0.48):		&	$<$6.39 (1.28):\\
239.179	&	14$_{4}$ -- 13$_{4}$	& 202 & 4.3(-5)&	0.71 (0.14)	&	$<$0.68 (0.14):		&	$\sim$0.39 (0.21)\\
239.211	&	14$_{3}$ -- 13$_{3}$	& 151 & 4.5(-5)&	$\sim$0.40 (0.24)	&	$\sim$0.75 (0.41)	&	2.47 (0.64)\\
239.234	&	14$_{2}$ -- 13$_{2}$	& 115 & 4.6(-5)&	0.36 (0.16)	&	$\sim$0.89 (0.89)	&	2.06 (0.63)\\
239.248	&	14$_{1}$ -- 13$_{1}$	&  93 & 4.7(-5)&	0.79 (0.23)	&	1.62 (0.52)	&	2.91 (0.74)\\
239.252	&	14$_{0}$ -- 13$_{0}$	&  86 & 4.7(-5)&	0.87 (0.24)	&	1.68 (0.43)	&	3.46 (1.07)\\
\hline
\end{tabular}

The notation $a(-b)$ stands for $a\times10^{-b}$\\
$<$ means upper limit\\
$\sim$ means uncertain detection, $S/N\lesssim2$\\
-- means frequency not observed\\
: means blended line
\label{fluxes_ch3cch}
\end{table*}

\section{Rotation diagrams}
\label{rtds}

RTD diagrams for \formaldehyde, \methanol, \ethanol, \isocyanicacid, \formamide, \acetonitrile, \propionitrile, \methylformate, \methylether, \ketene~and \methylacetylene. Lines from different frequency bands are corrected for differential beam dilution assuming a R$_{T=100K}$ source size for the warm species and 14'' source size for the cold species, as indicated in the plots. Optically thin, unblended lines with S/N$\gtrsim$2 are marked with filled circles. Lines with high S/N ratio ($\lesssim$2) are marked with diamonds, those included in the fit with filled diamonds. Optically thick lines and blended lines are marked with open squares and triangles, respectively. Upper limits are marked with arrows. Upper limits used to constrain the fit are marked with filled stars. The error bars are calculated using Eq. \ref{eq:dTmbdV}.

\begin{figure}
\begin{centering}
   \includegraphics[width=0.45\textwidth]{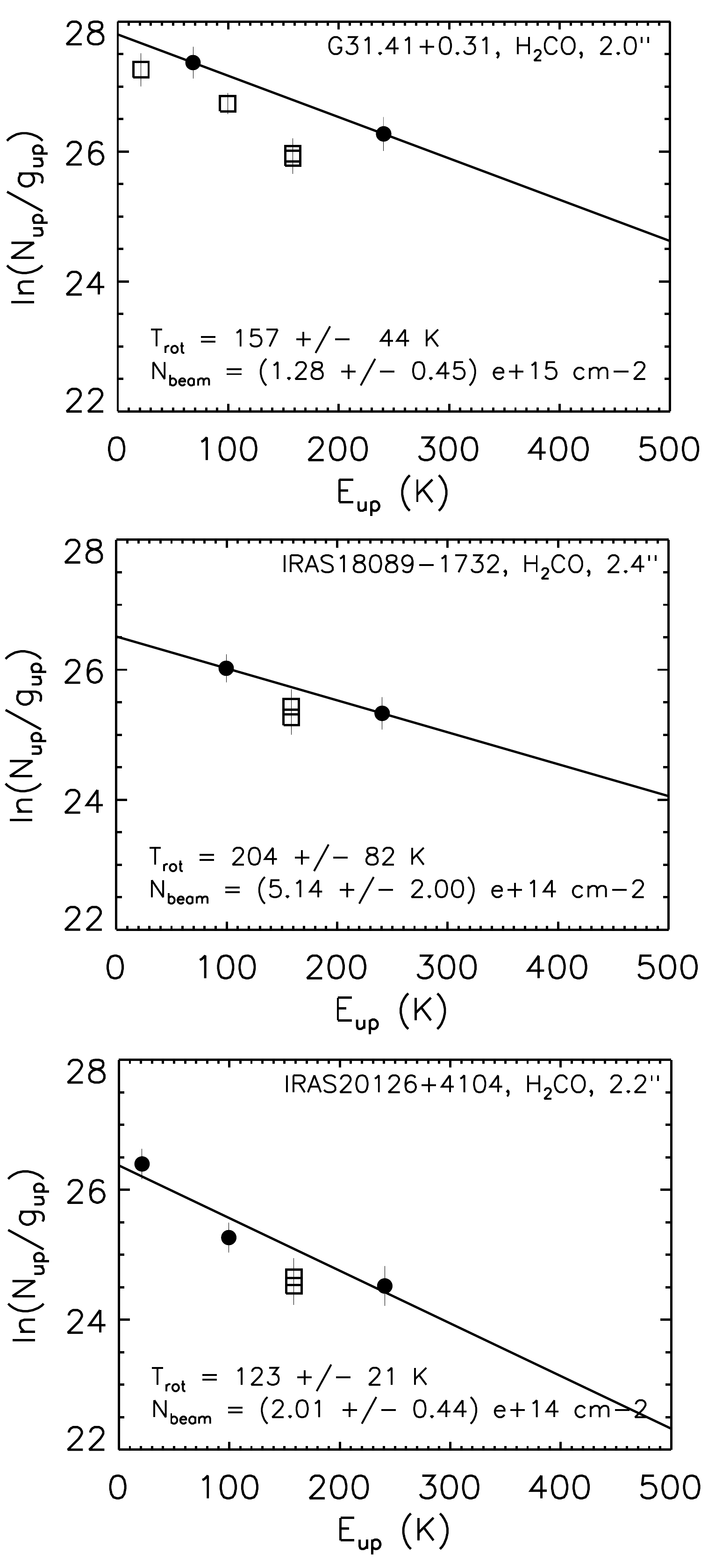}
      \caption{RTD fits for \formaldehyde. All lines included into the fit belong to para-\formaldehyde.}
         \label{fig:rtd_h2co}
\end{centering}
   \end{figure}
\begin{figure}
\begin{centering}
   \includegraphics[width=0.45\textwidth]{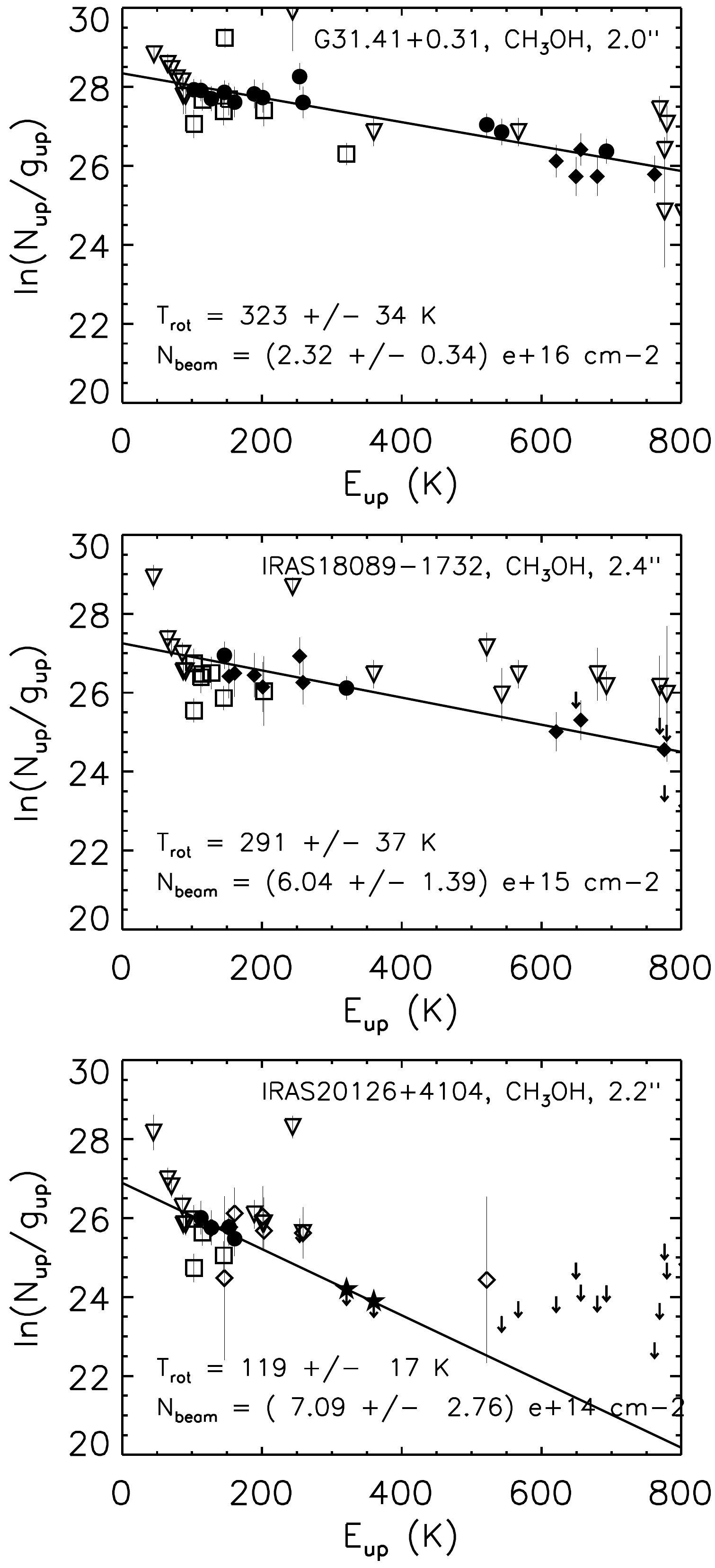}
      \caption{RTD fits for \methanol. Lines with \Eup$<$100 K are considered contaminated with the cold \methanol. Lines with S/N$\gtrsim$1 are included in the fit.}
         \label{fig:rtd_ch3oh}
\end{centering}
   \end{figure}
\begin{figure}
\begin{centering}
   \includegraphics[width=0.45\textwidth]{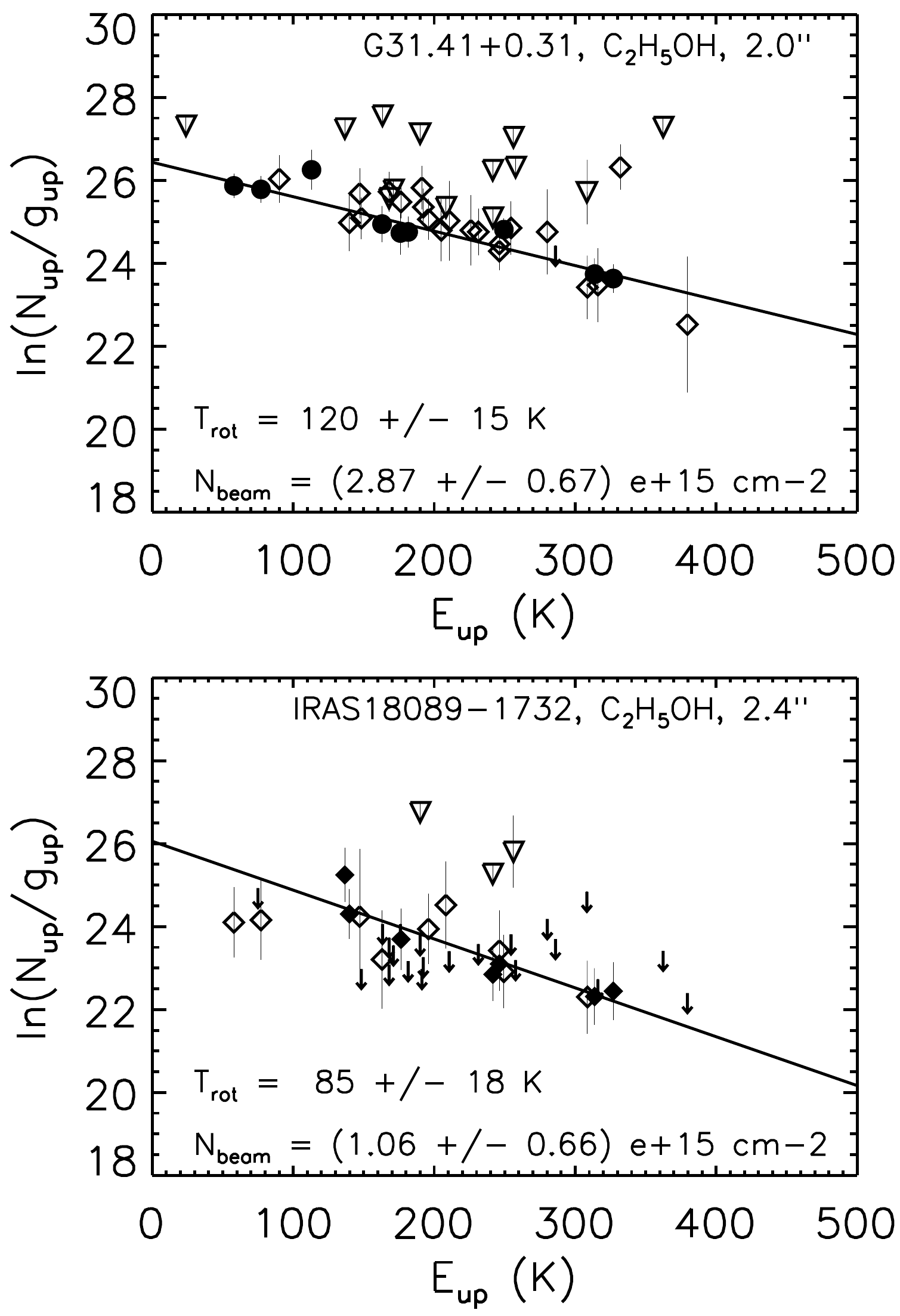}
      \caption{RTD fits for \ethanol. No \ethanol~was detected in IRAS20126+4104. For IRAS18089 lines with S/N$\gtrsim$1 have been used.}
         \label{fig:rtd_c2h5oh}
\end{centering}
   \end{figure}
\begin{figure}
\begin{centering}
   \includegraphics[width=0.45\textwidth]{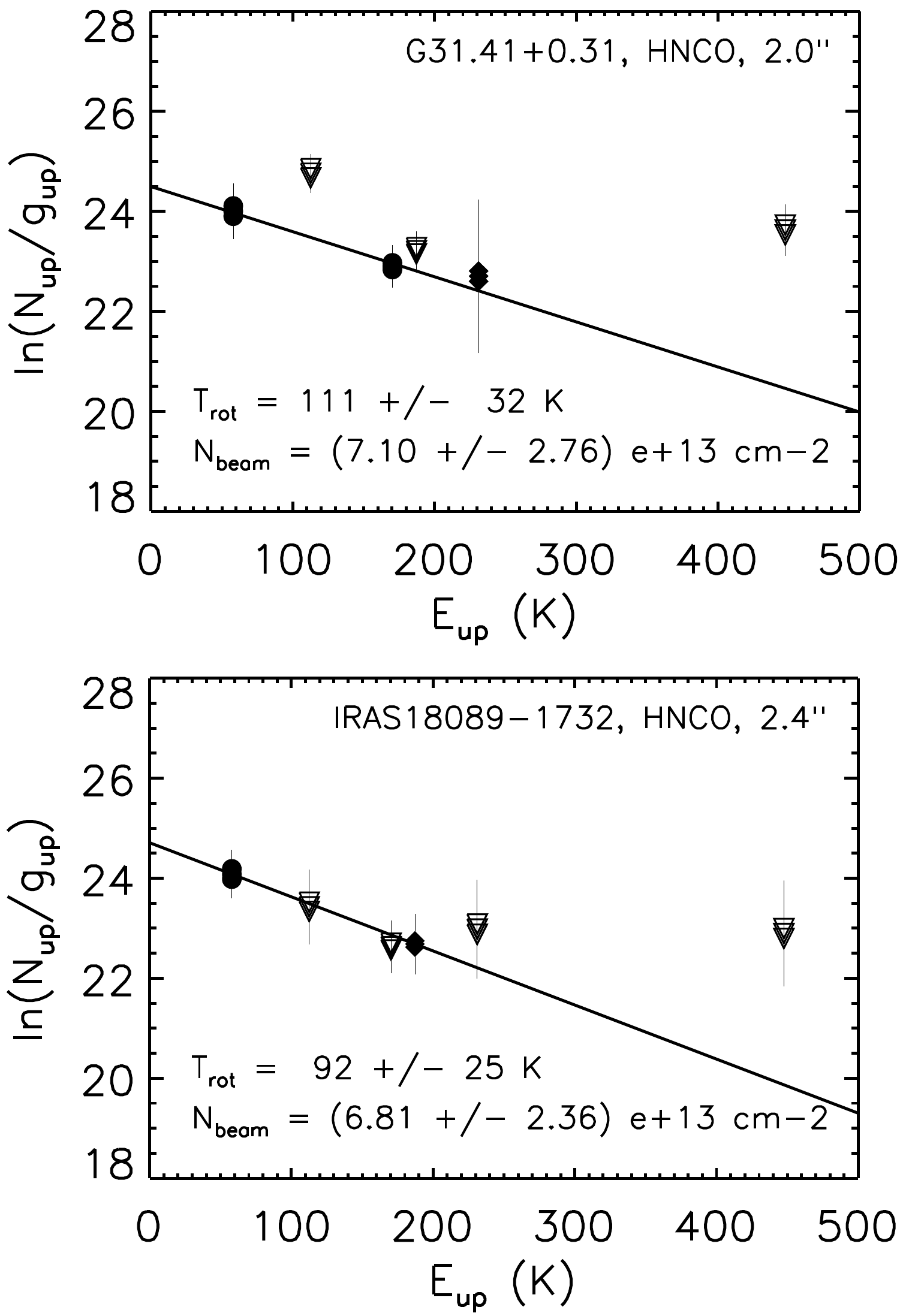}
      \caption{RTD fits for \isocyanicacid. No \isocyanicacid~was detected in IRAS20126+4104.}
         \label{fig:rtd_hnco}
\end{centering}
   \end{figure}
\begin{figure}
\begin{centering}
   \includegraphics[width=0.45\textwidth]{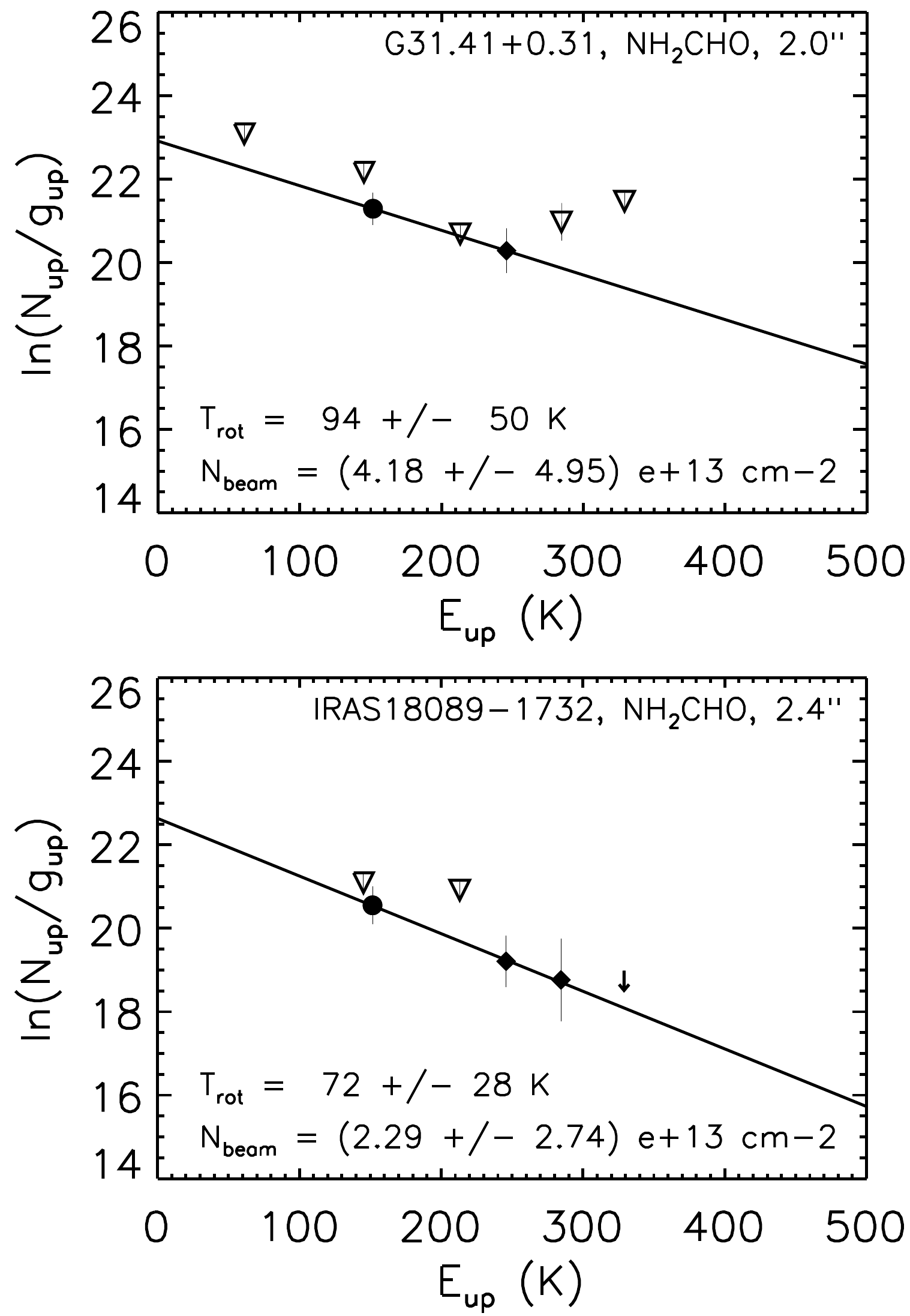}
      \caption{RTD fits for \formamide. No \formamide~was detected in IRAS20126+4104.}
         \label{fig:rtd_nh2cho}
\end{centering}
   \end{figure}
\begin{figure}
\begin{centering}
   \includegraphics[width=0.45\textwidth]{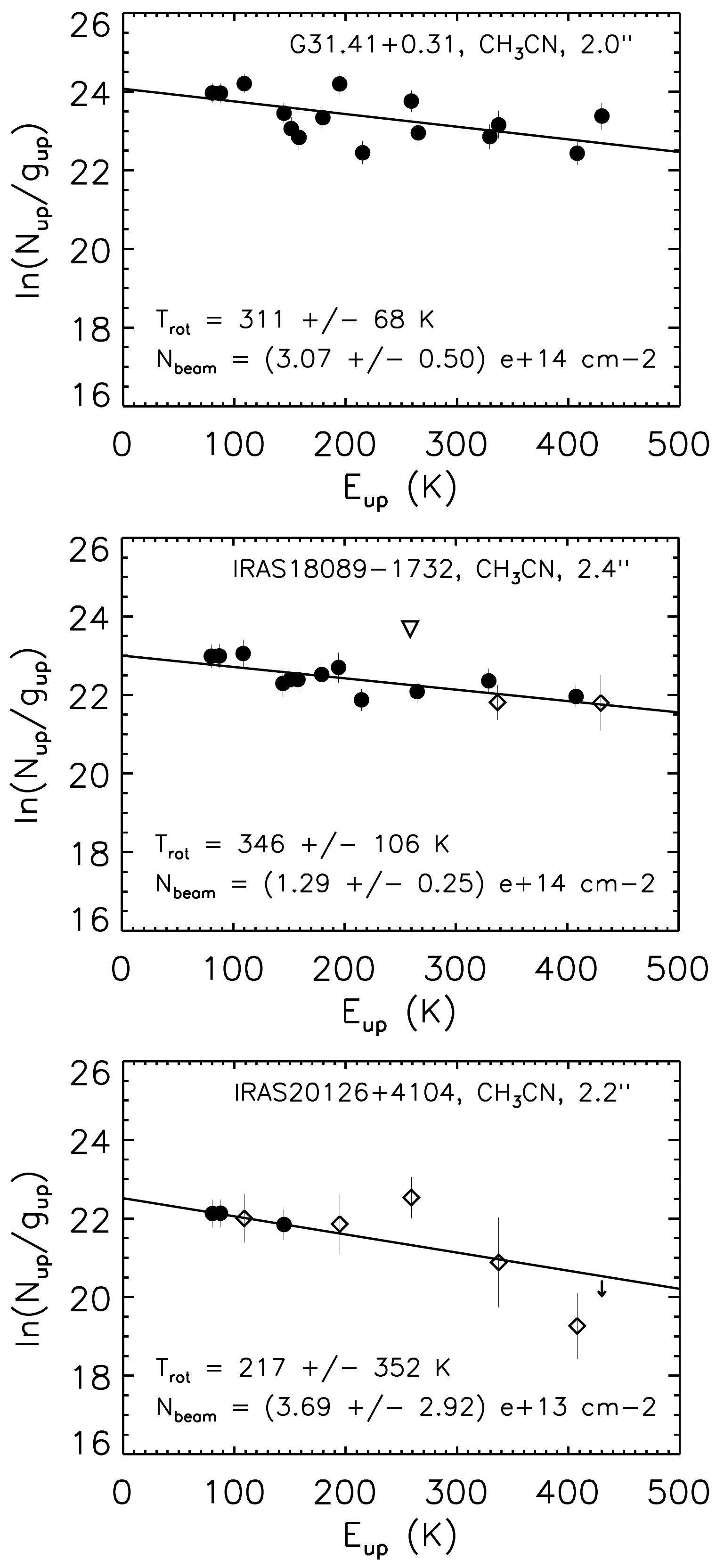}
      \caption{RTD fits for \acetonitrile.}
         \label{fig:rtd_ch3cn}
\end{centering}
   \end{figure}
\begin{figure}
\begin{centering}
   \includegraphics[width=0.45\textwidth]{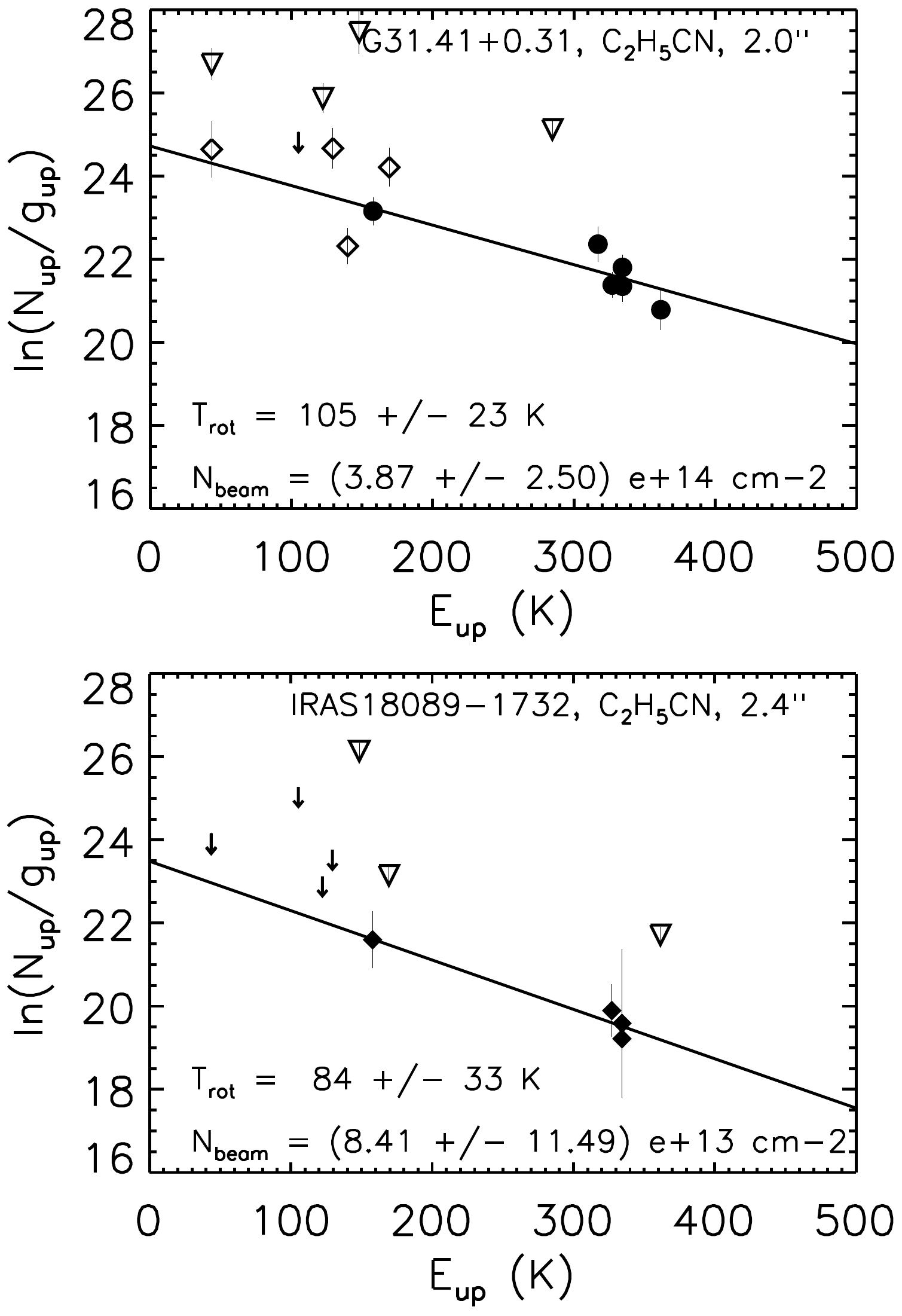}
      \caption{RTD fits for \propionitrile. No \propionitrile~was detected in IRAS20126+4104. For IRAS18089 lines with $S/N\lesssim2$ have been included in the fit.}
         \label{fig:rtd_c2h5cn}
\end{centering}
   \end{figure}
\begin{figure}
\begin{centering}
   \includegraphics[width=0.45\textwidth]{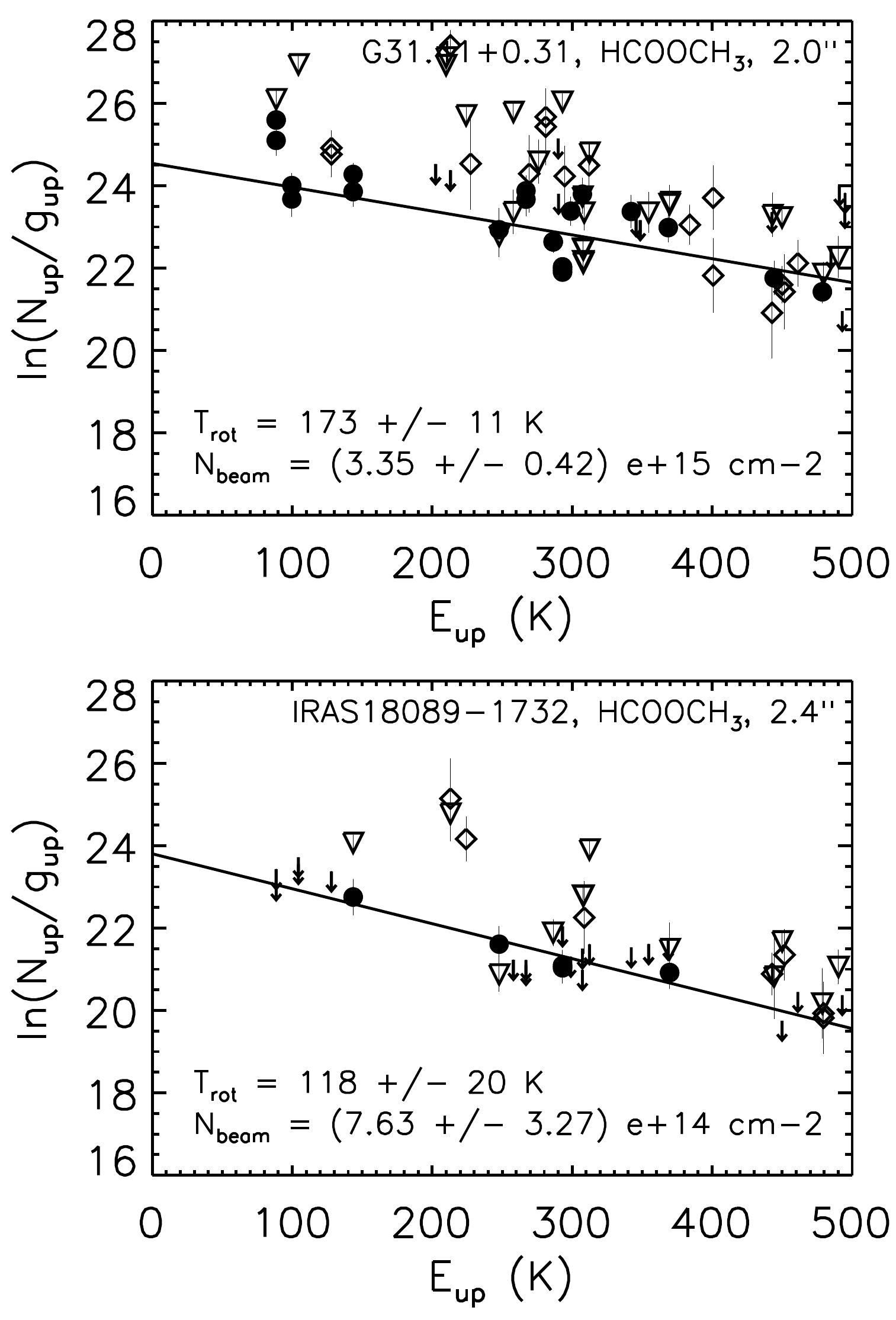}
      \caption{RTD fits for \methylformate. No \methylformate~was detected in IRAS20126+4104.}
         \label{fig:rtd_ch3ocho}
\end{centering}
   \end{figure}
\begin{figure}
\begin{centering}
   \includegraphics[width=0.45\textwidth]{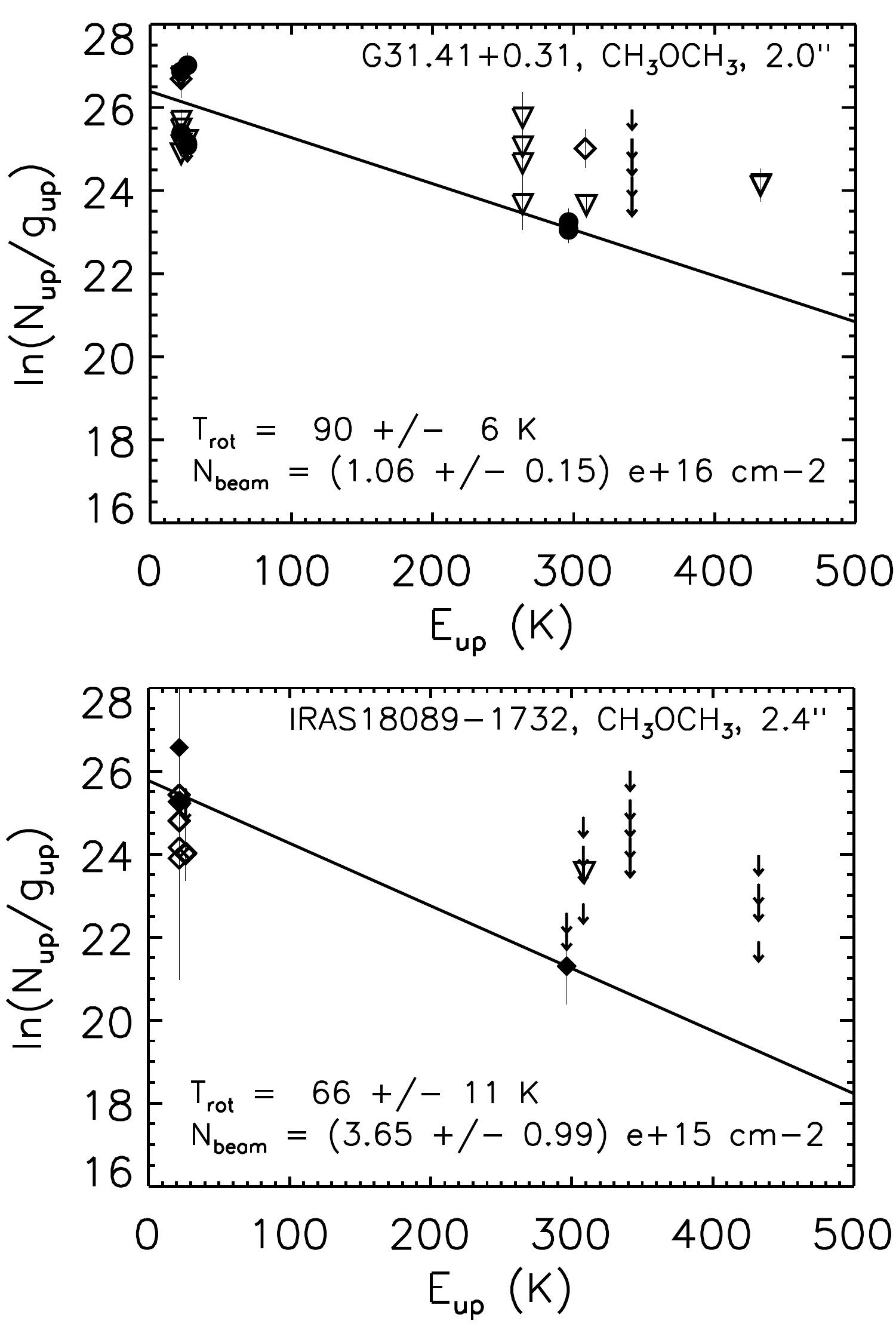}
      \caption{RTD fits for \methylether. No \methylether~was detected in IRAS20126+4104.}
         \label{fig:rtd_ch3och3}
\end{centering}
   \end{figure}
\begin{figure}
\begin{centering}
   \includegraphics[width=0.45\textwidth]{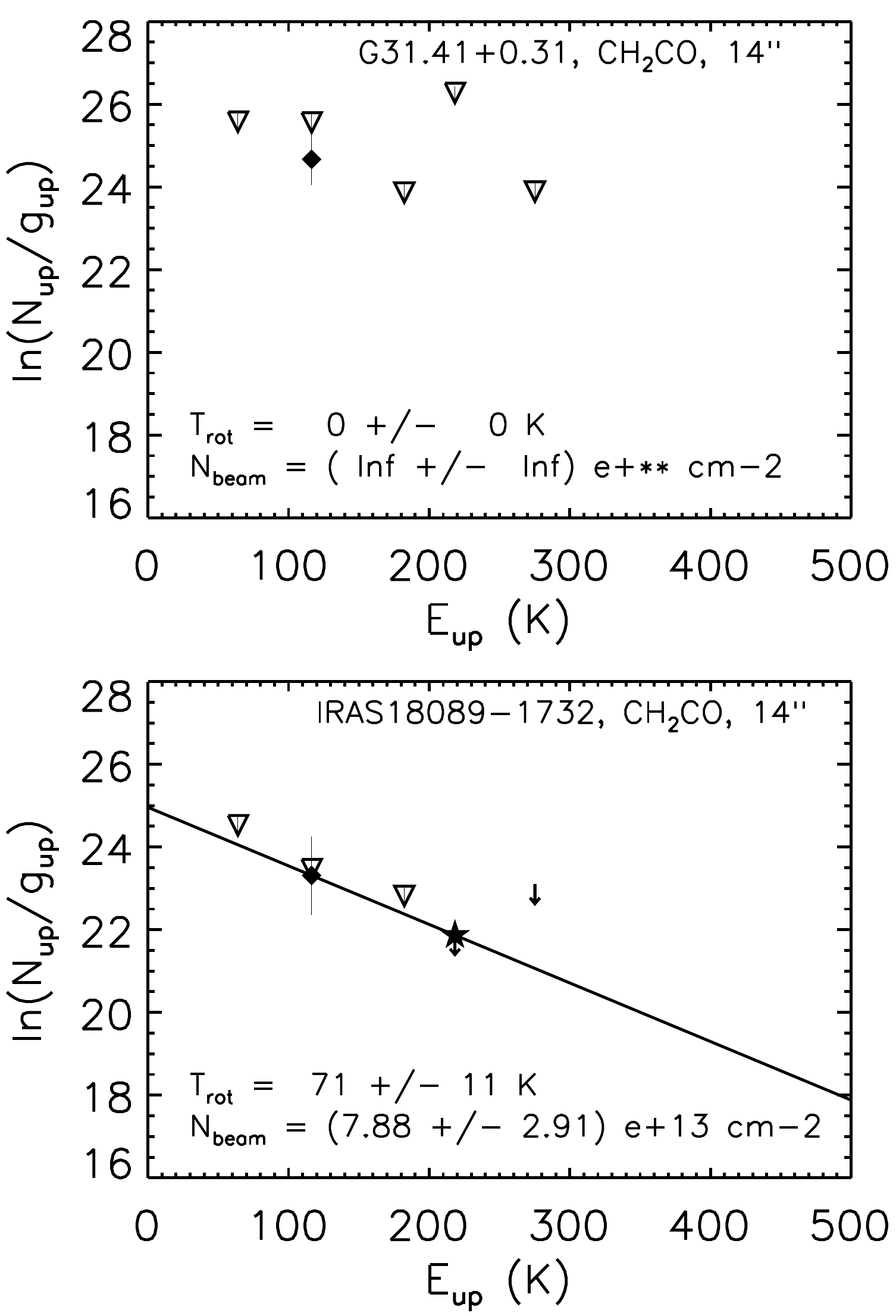}
      \caption{RTD fits for \ketene. No \ketene~was detected in IRAS20126+4104.}
         \label{fig:rtd_ch2co}
\end{centering}
   \end{figure}
\begin{figure}
   \includegraphics[width=0.45\textwidth]{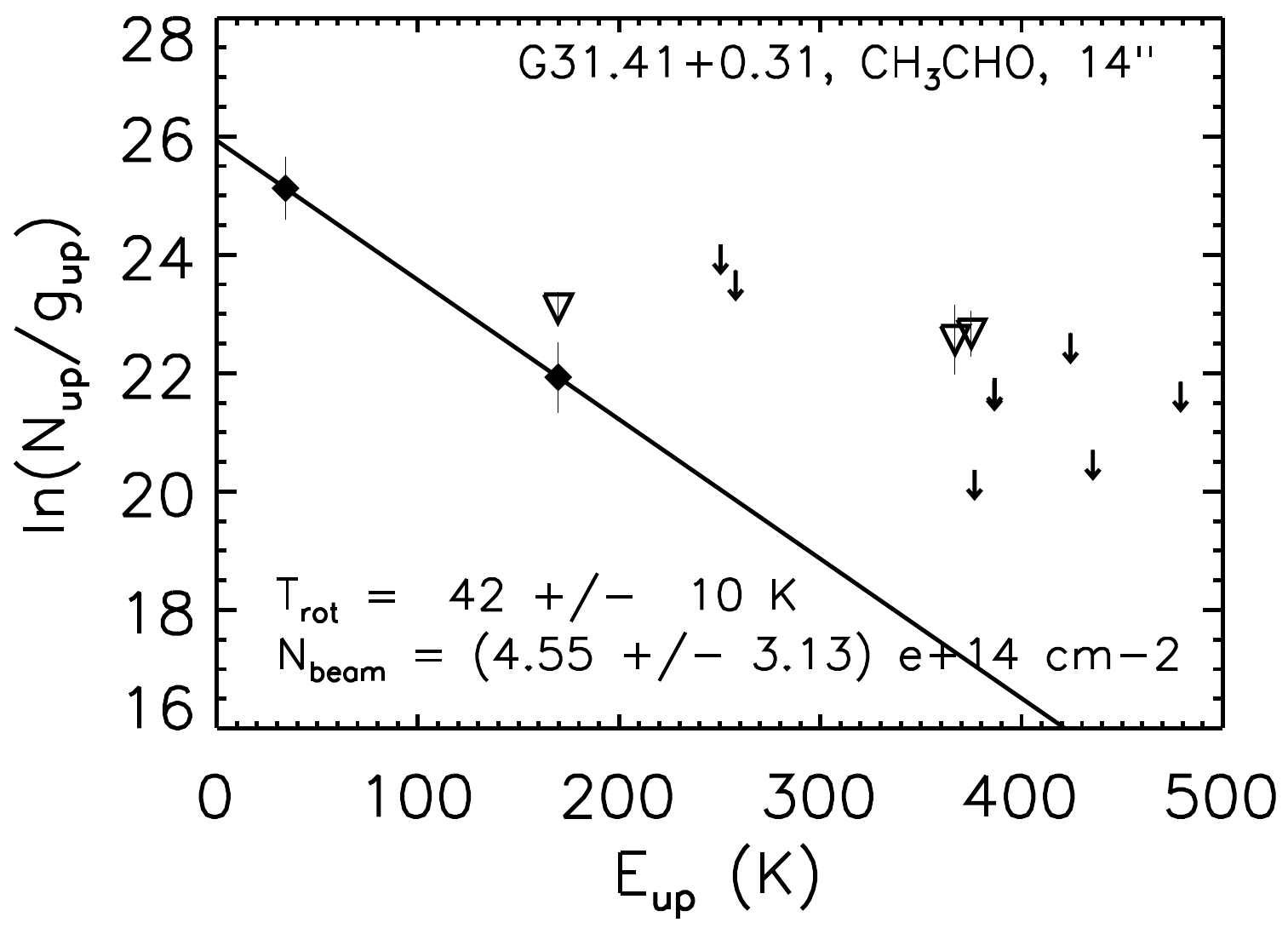}
      \caption{RTD fits for \acetaldehyde.}
         \label{fig:rtd_ch3cho}
   \end{figure}
\begin{figure}
   \includegraphics[width=0.5\textwidth]{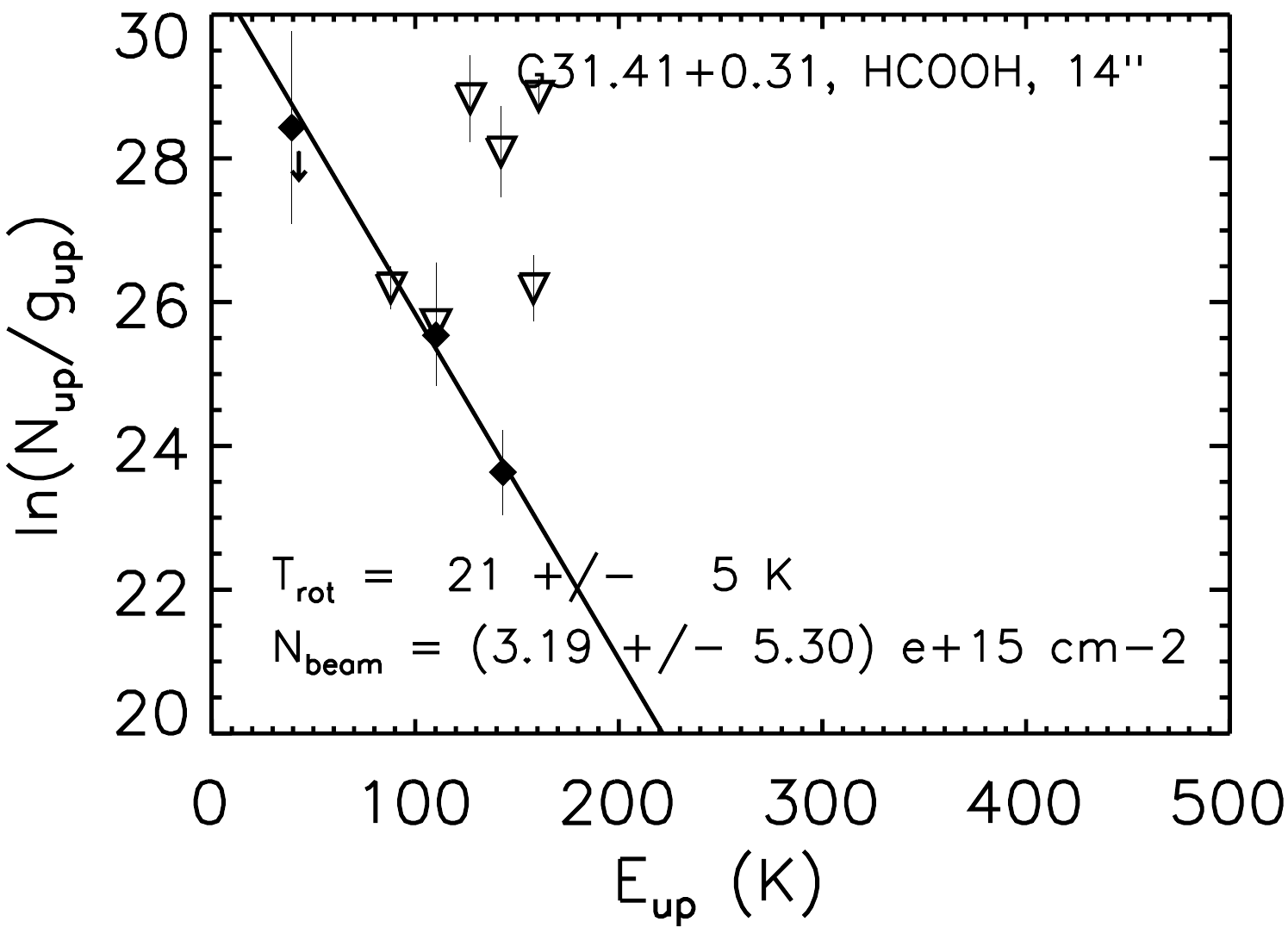}
      \caption{RTD fits for HCOOH.}
         \label{fig:rtd_hcooh}
   \end{figure}
\begin{figure}
\begin{centering}
   \includegraphics[width=0.45\textwidth]{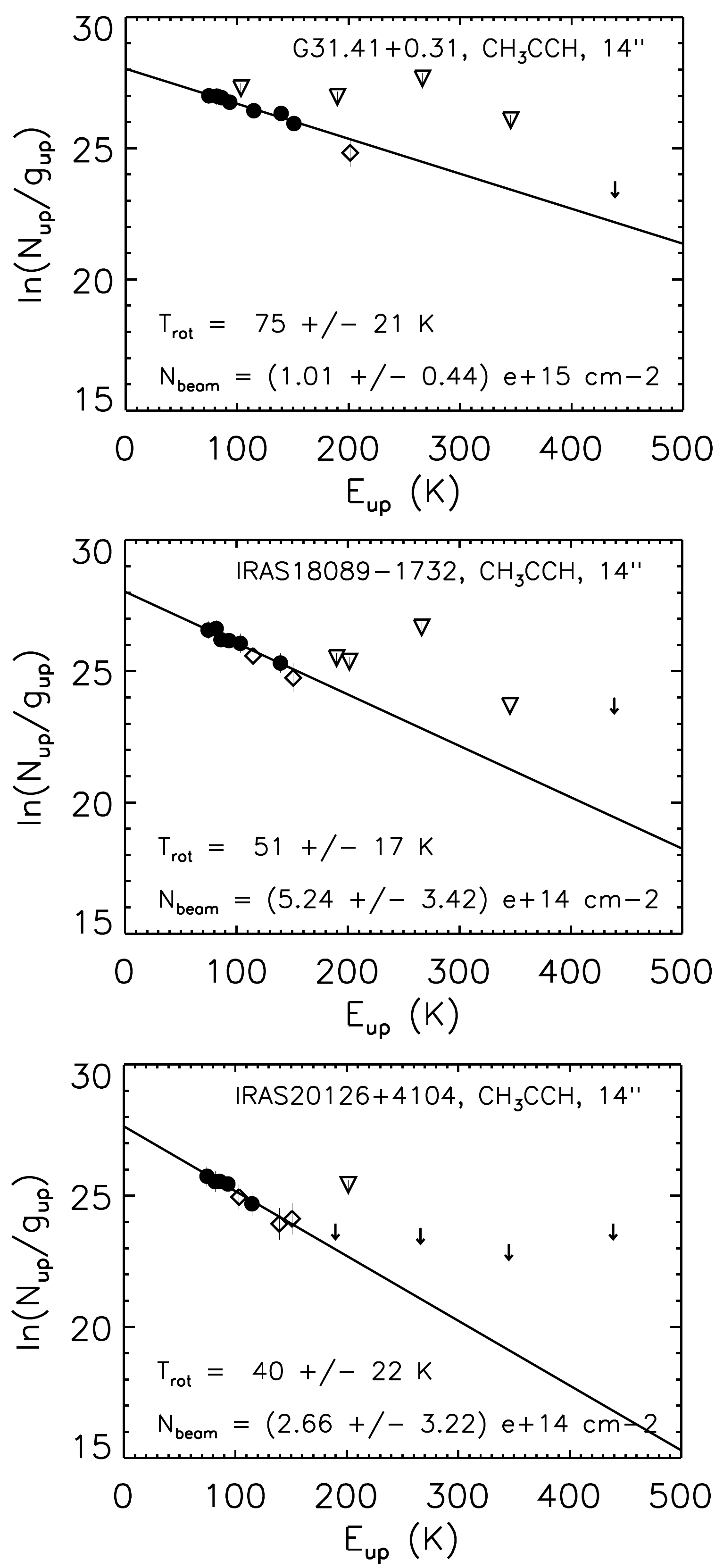}
      \caption{RTD fits for \methylacetylene.}
         \label{fig:rtd_ch3cch}
\end{centering}
   \end{figure}

\section{Weeds model parameters}

\begin{table}
\begin{center}
\caption{Weeds model parameters for IRAS20126+4104}
\label{tab:weeds_i20}
\begin{tabular}{l | c c c c}
\hline \hline
Species  & N$_\mathrm{S}$ & T$_\mathrm{ex}$ & $\theta_\mathrm{S}$ & $\Delta v$ \\
 & [cm$^{-2}$] & [K] & [''] & [km s$^{-1}$] \\
\hline
\methanol & 1.5E+17 & 300 & 2.2 & 6 \\
\methanol & 4.8E+16 & 20 & 3.0 & 5 \\
\cmethanol & -- & -- & -- & -- \\
\formaldehyde & 6.0E+15 & 150 & 2.2 & 6 \\
\cformaldehyde & $<$2.0E+15 & 150 & 2.2 & 6 \\
\acetonitrile~$^1$ & 1.3E+15  & 200 & 2.2 & 6 \\
\acetonitrilec &$<$2.0E+13 & 200 & 2.2 & 6 \\
\isocyanicacid & 1.0E+15  & 200 & 2.2 & 6 \\
\isocyanicacidc & $<$1.0E+14  & 200 & 2.2 & 6 \\
\ethanol & $<$3.0E+15  & 100 & 2.2 & 6 \\
\propionitrile & $<$5.0E+14  & 80 & 2.2 & 5 \\
\methylether & $<$1.0E+16  & 100 & 2.2 & 6 \\
\methylformate & $<$2.0E+15 & 200 & 2.2 & 6 \\
\formamide & 3.0E+14  & 300 & 2.2 & 6 \\
\ketene & $<$5.0E+13 & 50 & 14.0 & 6 \\
\acetaldehyde & $<$1.0E+14  & 50 & 8.0 & 6 \\
\formicacid & $<$5.0E+13  & 40 & 8.0 & 6 \\
\methylacetylene & 7.0E+14 & 35 & 14.0 & 2 \\
\hline
\multicolumn{5}{l}{Additional species} \\
\hline
CH$_3$COCH$_3$ & $<$0.1E+17 & 300 & 2.2 & 5\\
HCCCN & 1E+13 & 100 & 14 & 7\\
SO & 0.3E+15 & 50 & 14 & 6\\
$^{34}$SO & 0.2E+14 & 50 & 14.0 & 6\\
SO$_2$ & 0.3E+15 & 50 & 14.0 & 6\\
$^{33}$SO2 & 0.05E+15 & 50 & 14.0 & 5\\
HCN & 0.4E+14 & 50 & 14.0 & 20\\
HCN & 4E+13 & 50 & 14.0 & 5\\
H$^{13}$CN & 0.1E+13 & 50 & 14.0 & 6\\
CN, v = 0, 1 & 0.2E+15 & 50 & 14 & 2\\
\hline
\end{tabular}
\end{center}
\cmethanol~is not available in the JPL database used for Weeds modeling. \\
$^1$ based on 21'' beam spectra. \\
\end{table}

\begin{table}
\begin{center}
\caption{Weeds model parameters for IRAS18089-1732}
\label{tab:weeds_i18}
\begin{tabular}{l | c c c c}
\hline \hline
Species  & N$_\mathrm{S}$ & T$_\mathrm{ex}$ & $\theta_\mathrm{S}$ & $\Delta v$ \\
 & [cm$^{-2}$] & [K] & [''] & [km s$^{-1}$] \\
\hline
\methanol & 3.5E+17 & 300 & 2.4 & 5 \\
\methanol & 2.6E+17 & 20 & 3.0 & 5 \\
\cmethanol & -- & -- & -- & -- \\
\formaldehyde & 1.5E+16 & 150 & 2.4 & 6 \\
\cformaldehyde & 3.5E+15 & 150 & 2.4 & 6 \\
\acetonitrile & 3.5E+15 & 200 & 2.4 & 6 \\
\acetonitrilec & 4.0E+14 & 200 & 2.4 & 6 \\
\isocyanicacid & 4.0E+15  & 200 & 2.4 & 6 \\
\isocyanicacidc & 7.0E+14  & 200 & 2.4 & 6 \\
\ethanol & 2.5E+16  & 150 & 2.4 & 6 \\
\propionitrile & 4.0E+15  & 80 & 2.4 & 5 \\
\methylether & 1.0E+17  & 100 & 2.4 & 6 \\
\methylformate & 3.0E+16  & 200 & 2.4 & 6 \\
\formamide & 5.0E+14  & 100 & 2.4 & 6 \\
\ketene & 1.5E+14 & 50 & 14.0 & 5 \\
\acetaldehyde & $<$1.0E+14  & 50 & 8.0 & 6 \\
\formicacid & $<$5.0E+13  & 40 & 8.0 & 6 \\
\methylacetylene & 1.8E+15 & 40 & 14.0 & 3.5 \\
\hline
\multicolumn{5}{l}{Additional species} \\
\hline
CH$_3$COCH$_3$ & 0.5E+17 & 300 & 2.4 & 5\\
HCCCN & 3E+13 & 150 & 14 & 6\\
SO & 1E+15 & 50 & 14 & 6\\
$^{34}$SO & 0.5E+14 & 50 & 14.0 & 6\\
SO$_2$ & 0.5E+15 & 50 & 14.0 & 6\\
$^{33}$SO$_2$ & 0.5E+15 & 50 & 14.0 & 5\\
HCN & 2E+14 & 50 & 14.0 & 7\\
HCN & -5E+13 & 50 & 14.0 & 4\\
H$^{13}$CN & 1.5E+13 & 50 & 14.0 & 6\\
CN, v = 0, 1 & 0.4E+15 & 50 & 14 & 5\\
NH$_2$CN & 0.1E+14 & 50 & 14 & 7\\
OC$^{34}$S & 0.5E+15 & 50 & 14.0 & 7\\
CP & 0.5E+15 & 50 & 14.0 & 4\\
C$^{34}$S & 1.4E+14 & 50 & 14.0 & 5\\
HCOCH$_2$OH & 0.05E+17 & 300 & 2.0 & 6\\
\hline
\end{tabular}
\end{center}
\cmethanol~is not available in the JPL database used for Weeds modeling. \\
\end{table}

\begin{table}
\begin{center}
\caption{Weeds model parameters for G31.41+0.31}
\label{tab:weeds_g31}
\begin{tabular}{l | c c c c}
\hline \hline
Species  & N$_\mathrm{S}$ & T$_\mathrm{ex}$ & $\theta_\mathrm{S}$ & $\Delta v$ \\
 & [cm$^{-2}$] & [K] & [''] & [km s$^{-1}$] \\
\hline
\methanol & 2.0E+18 & 300 & 2.0 & 7 \\
\methanol & 2.7E+17 & 20 & 5.0 & 7 \\
\cmethanol & -- & -- & -- & -- \\
\formaldehyde & 6.0E+16 & 150 & 2.0 & 6 \\
\cformaldehyde & 7.0E+15 & 150 & 2.0 & 6 \\
\acetonitrile~$^1$ & 2.0E+16  & 300 & 2.0 & 7 \\
\acetonitrilec & 2.0E+15 & 300 & 2.0 & 7 \\
\isocyanicacid & 6.0E+16  & 200 & 2.0 & 7 \\
\isocyanicacidc & 2.0E+15  & 200 & 2.0 & 7 \\
\ethanol & 2.0E+17  & 100 & 2.0 & 7 \\
\propionitrile & 2.0E+16  & 80 & 2.0 & 5 \\
\methylether & 1.0E+18  & 100 & 2.0 & 5 \\
\methylformate~$^2$ & 1.0E+18 & 300 & 2.0 & 7 \\
\formamide & 4.0E+15  & 300 & 2.0 & 7 \\
\ketene & 6.5E+14 & 50 & 14.0 & 6 \\
\acetaldehyde & 1.0E+15  & 50 & 8.0 & 6 \\
\formicacid & 0.2E+16  & 40 & 8.0 & 6 \\
\methylacetylene & 2.8E+15 & 60 & 14.0 & 4 \\
\hline
\multicolumn{5}{l}{Additional species} \\
\hline
CH$_3$COCH$_3$ & 1E+17 & 100 & 2.0 & 7\\
HCCCN & 7E+13 & 150 & 14 & 7\\
SO & 7E+14 & 50 & 14 & 8\\
$^{34}$SO & 1E+14 & 50 & 14.0 & 6\\
SO$_2$ & 0.1E+16 & 50 & 14.0 & 7\\
$^{33}$SO$_2$ & 1E+15 & 50 & 14.0 & 6\\
HCN & 2E+14 & 50 & 14.0 & 12\\
HCN & -6E+13 & 50 & 14.0 & 6\\
H$^{13}$CN & 2E+13 & 50 & 14.0 & 6\\
CN, v = 0, 1 & 0.8E+15 & 50 & 14 & 6 \\
NH$_2$CN & 0.3E+14 & 50 & 14 & 7\\
O$^{13}$CS & 5E+14 & 50 & 14 & 5\\
OC$^{34}$S & 1E+15 & 50 & 14.0 & 7\\
CP & 1E+15 & 50 & 14.0 & 6\\
C$^{34}$S & 2E+14 & 50 & 14.0 & 7\\
HCOCH$_2$OH & 0.2E+17 & 300 & 2.0 & 7\\
S$_3$ & 1.5E+16 & 15 & 14 & 6\\
CH$_3$NH$_2$ & 1E+17 & 150 & 2.0 & 7\\
\hline
\end{tabular}
\end{center}
\cmethanol~is not available in the JPL database used for Weeds modeling. \\
$^1$ based on 21'' beam spectra. \\
$^2$ based on 21'' beam spectra, 1.5E+17 in the 14'' beam spectra. \\
\end{table}

 \section{Additional detections}

\begin{figure}
   \includegraphics[width=0.45\textwidth]{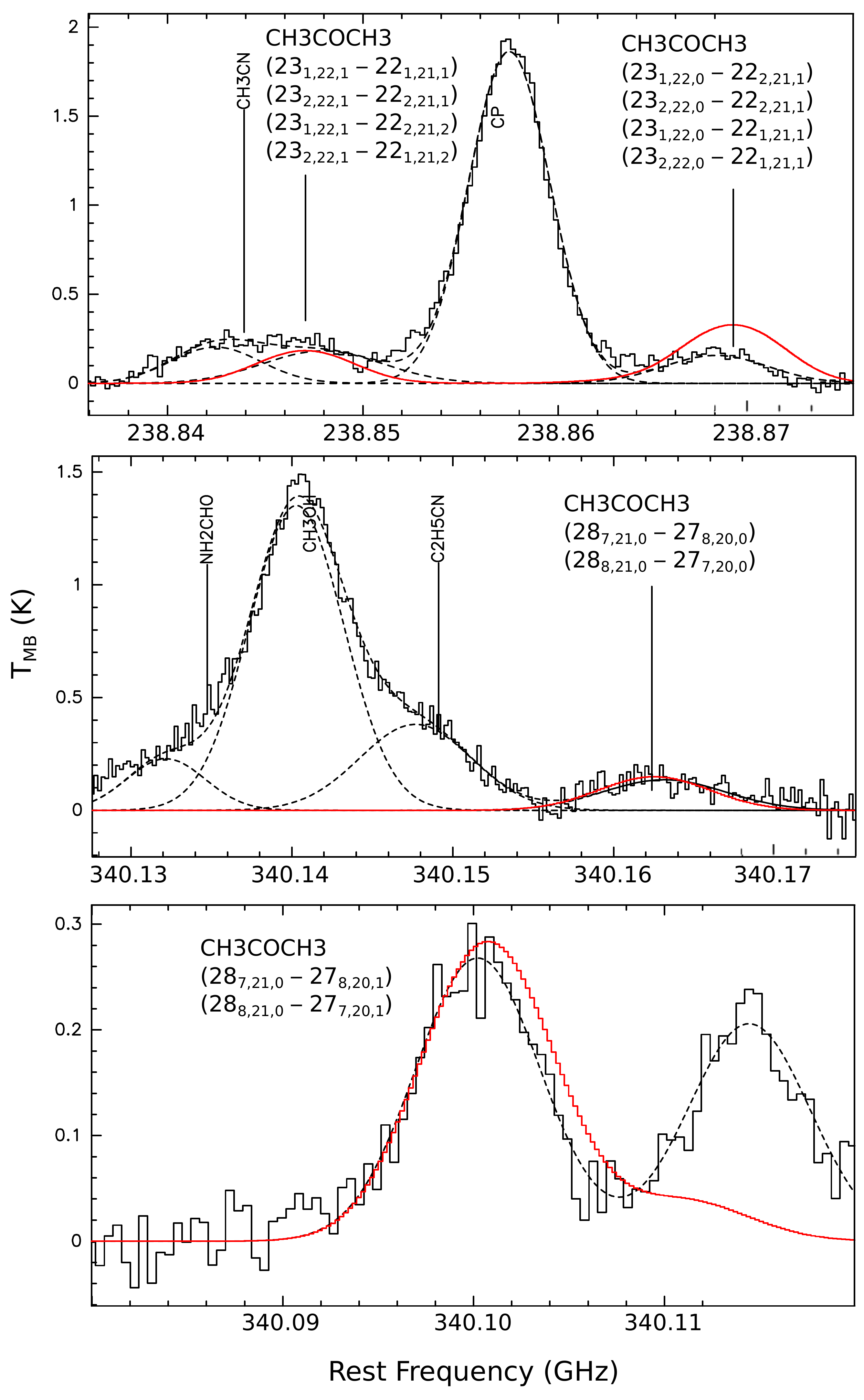}
      \caption{Beam-averaged spectra of \acetone~(23$_{1/2, 22, 1}$--22$_{1/2, 21,1}$), (23$_{1/2, 22, 0}$--22$_{1/2, 21,1}$), (28$_{7/8, 21, 0}$--27$_{8/7, 20, 0}$) and (28$_{7/8, 21, 0}$--27$_{8/7, 20, 1}$) towards G31.41+0.31 hot core. Dashed lines indicate Gaussian fits to the lines and red solid line show the Weeds model on the \acetone~lines at 100~K.}
         \label{fig:acetone}
   \end{figure}

Several lines of other species were found in the observed frequency
ranges, particularly in the line-rich G31.41+0.31. Table
\ref{tab:additional_detections} lists additional detections with
column densities obtained from the Weeds analysis with the best \Tex~when
this could be derived. Several transitions of acetone, CH$_3$COCH$_3$,
are detected. In G31, the best agreement is obtained at 100~K, whereas for
IRAS18089, this is \Tex$=$300~K. No CH$_3$COCH$_3$ was
detected in IRAS20126, and the tabulated value is an upper limit at
300~K. Figure \ref{fig:acetone} shows the strongest acetone lines in G31 together with the Weeds model.

Glycoaldehyde, HCOCH$_2$OH, has previously been detected in G31 \citep{Beltran2009}, has several lines in the covered ranges. All lines are however blended with other transitions, and only
upper limit could therefore be derived reliably. The tabulated column
density is for a temperature of 300~K, constrained by non-detection of
lines with low \Eup.

Two transitions of HC$_3$N are detected at 218.325 GHz (\Eup 131.0~K)
($J = 24 \rightarrow 23$) and 354.697 GHz (\Eup 340.5 K) ($J = 39
\rightarrow 38$). The Weeds model on the two lines, well separated in
\Eup, gives a beam averaged column density of $7\times10^{13}$~cm$^{-2}$
and a temperature of 150~K for G31. The HC$_3$N emission cannot be matched with
emission contained within a 2.0$''$ volume as the low-\Eup transition becomes
optically thick. 
For IRAS20126, the tabulated column density is at
100~K, constrained by the two transitions. Only the high
\Eup~transition was covered for IRAS18089 giving a column density of
$3\times10^{13}$~cm$^{-2}$ at 150~K and $6\times10^{13}$~cm$^{-2}$ at 100~K.

Several CH$_3$NH$_2$ transitions are observed with a \Tex~of
150~K. Blends with other species make the rotation temperature and
column density inaccurate, however.  Several unidentified lines
coincide with NH$_2$CH$_2$CH$_2$OH, but only upper limits of 1.0$\times10^{16}$~cm$^{-2}$ at 100--300~K could be derived reliably using Weeds.

Several weak lines of NH$_2$CN are detected, with a \Tex~of 50~K
derived from the Weeds model, but line blends make the rotation
temperature and column density inaccurate. 
For CN, SO, $^{34}$SO, O$^{13}$CS, OC$^{34}$S, SO$_2$, $^{33}$SO$_2$,
C$^{34}$S, and CP not enough lines were observed to derive
\Tex~values, and tabulated column densities are assuming a temperature of 50 K. 

One line of HCN, $J$=4--3, was observed. HCN emission
in G31 is composed of a broad (12 km~s$^{-1}$)
and a narrow (6 km~s$^{-1}$) component (redshifted by
2.5 km~s$^{-1}$) causing self-absorption. For IRAS18089, the broad
component has a width of 7 km~s$^{-1}$ and narrow 4~km~s$^{-1}$
(redshifted by 1 km~s$^{-1}$). For IRAS20126 the HCN emission has a very broad 20
km~s$^{-1}$ component blueshifted by 2 km~s$^{-1}$ and a narrow (5 km~s$^{-1}$) component redshifted by 1 km~s$^{-1}$. 
The tabulated column densities are those derived from the H$^{13}$CN column density.

\begin{table*}
\begin{center}
\caption{Additional detections. Column densities given are source averaged for species with \Tex$>$100~K and beam averaged for species with \Tex$<$100~K (except for HCCCN, the emission of which is warm and arises from extended volume). The tabulated values are those obtained from Weeds analysis with the best \Tex.}
\label{tab:additional_detections}
\begin{tabular}{ l | c c c }
\hline
\hline
Species & IRAS20126+4104 & IRAS18089-1732 & G31.41+0.31 \\
\hline
\hline
\multicolumn{4}{l}{Source-averaged column densities} \\
\hline
CH3COCH3 & 	$<$0.1E+17 & 0.5E+17 &	1.0E+17 \\
HCOCH2OH & 	-- & $<$0.5E+16 & 	$<$0.2E+17 \\
CH$_3$NH$_2$ & 		-- & -- & 	2.4E+14   \\
NH$_2$CH$_2$CH$_2$OH & 		-- & -- & 	$<$1.0E+16 \\
\hline
\multicolumn{4}{l}{Beam-averaged column densities} \\
\hline
NH$_2$CN & 	-- & $<$0.1E+14 & 	7.0E+14 \\
OC-13-S & 	-- & -- & 	5.0E+14 \\
OC-34-S & 	-- & 0.5E+15 & 	1.0E+15 \\
HCCCN & 	1.0E+13 & 3.0E+13  & 	7.0E+13 \\
S$_3$ & 		-- & -- & 	1.5E+16  \\
CN, v = 0, 1 & 		0.2E+15 & 0.4E+15 & 	0.8E+15   \\
SO$_2$ & 		0.3E+15 & 0.5E+15 & 	0.1E+16 \\
SO & 		0.3E+15 & 1.0E+15 & 	7.0E+14  \\
CS-34 & 		-- & 1.4E+14 & 	2.4E+14 \\
CP & 		-- & -- & 	1.0E+15  \\
S-33-O$_2$ & 		$<$0.5E+14 & 0.5E+15 & 	1.0E+15 \\
S-34-O & 		0.2E+14 & 0.5E+14 & 	1.0E+14  \\
HCN & 		8E+13 & 2.0E+14 & 	2.0E+14 \\
HC-13-N & 		0.1E+13 & 1.5E+13 & 	2.0E+13 \\
\hline
\end{tabular}
\end{center}
\end{table*}

\end{appendix}

\end{document}